\documentclass[manuscript=article]{achemso}
\setkeys{acs}{articletitle = true}
\usepackage[T1]{fontenc}
\usepackage[utf8]{inputenc}
\usepackage{graphicx}
\usepackage{amsmath}
\usepackage{color}
\usepackage{comment}
\usepackage{hyperref}
\usepackage{multirow}
\usepackage{subfig}
\usepackage{amsmath,mathtools}
\usepackage[version=3]{mhchem}
\usepackage{longtable}
\usepackage{siunitx}
\usepackage[normalem]{ulem}
\usepackage[para,online]{threeparttable}

\definecolor{ao}{rgb}{0.0, 0.5, 0.0}
\definecolor{bs}{rgb}{1.0, 0.44, 0.37}

\title{
Efficient adiabatic connection approach for strongly correlated systems.
Application to singlet-triplet gaps of biradicals}

\author{Daria Drwal}
\affiliation{Institute of Physics, Lodz University of Technology, \mbox{ul. Wolczanska 219, 90-924 Lodz, Poland}}
\altaffiliation{Contributed equally.}

\author{Pavel Beran}
\affiliation{J. Heyrovsk\'{y} Institute of Physical Chemistry, Academy of Sciences of the Czech \mbox{Republic, v.v.i.}, Dolej\v{s}kova 3, 18223 Prague 8, Czech Republic}
\alsoaffiliation{Faculty of Mathematics and Physics, Charles University, Prague, Czech Republic}
\altaffiliation{Contributed equally.}

\author{Micha{\l} Hapka}
\affiliation{Faculty of Chemistry, University of Warsaw, ul.\ L.\ Pasteura 1, 02-093 Warsaw, Poland}
\alsoaffiliation{Institute of Physics, Lodz University of Technology, \mbox{ul. Wolczanska 219, 90-924 Lodz, Poland}}

\author{Marcin Modrzejewski}
\affiliation{Faculty of Chemistry, University of Warsaw, ul.\ L.\ Pasteura 1, 02-093 Warsaw, Poland}

\author{Adam Sok{\'o}{\l}}
\affiliation{Institute of Physics, Lodz University of Technology, \mbox{ul. Wolczanska 219, 90-924 Lodz, Poland}}

\author{Libor Veis}
\email{libor.veis@jh-inst.cas.cz}
\affiliation{J. Heyrovsk\'{y} Institute of Physical Chemistry, Academy of Sciences of the Czech \mbox{Republic, v.v.i.}, Dolej\v{s}kova 3, 18223 Prague 8, Czech Republic}

\author{Katarzyna Pernal}
\email{pernalk@gmail.com}
\affiliation{Institute of Physics, Lodz University of Technology, \mbox{ul. Wolczanska 219, 90-924 Lodz, Poland}}


\begin{document}

\begin{abstract}
Strong correlation can be essentially captured with multireference wavefunction methods such as complete active space self-consistent field (CASSCF) or  density matrix renormalization group (DMRG). Still, an accurate description of the electronic structure of strongly correlated systems requires accounting for the dynamic electron correlation, which CASSCF and DMRG largely miss. In this work a new approach for the correlation energy based on the adiabatic connection (AC) is proposed. The AC$_{\rm n}$ method accounts for terms up to the desired order n in the coupling constant, is rigorously size-consistent, free from instabilities and intruder states. It employs the particle-hole multireference random phase approximation and the Cholesky decomposition technique, which leads to a computational cost growing with the fifth power of the system size. Thanks to AC$_{\rm n}$ depending solely on one- and two-electron CAS reduced density matrix, the method is much more efficient than existing \textit{ab initio} dynamic correlation methods for strong correlation. AC$_{\rm n}$ affords excellent results for singlet-triplet gaps of challenging organic biradicals. 
Development presented in this work opens new perspectives for accurate calculations of systems with dozens of strongly correlated electrons.
\end{abstract}

\maketitle


Electron correlation energy is  defined with respect to the energy of a model (a reference) used to describe a given system. In other words, given a Hamiltonian $\hat{H}$, if $\Psi^{\rm ref}$ is the reference wavefunction  and $E^{\rm ref}$ the corresponding energy i.e.
\begin{equation}
E^{\rm ref} = \left\langle \Psi^{\rm ref}|\hat{H}|\Psi^{\rm ref} \right\rangle
\ \ \ ,\label{eref}
\end{equation}
then electron correlation comprises all electron interaction effects not
accounted for by the chosen model, and the correlation energy pertains to the energy error
\begin{equation}
E_{\rm corr}\equiv E_{\rm exact}-E^{\rm ref}\ \ \ ,\label{ecorr}%
\end{equation}
computed with respect to the exact energy $E_{\rm exact}$ (an eigenvalue of the Hamiltonian $\hat{H}$).
Strongly correlated molecular systems require model wavefunctions consisting of multiple configurations to capture static correlation effects.  The complete active space (CAS) method assumes selecting a number of (active) electrons and orbitals crucial for the static correlation and performing exact diagonalization in the active orbital subspace.~\cite{Roos1987,olsen2011casscf} The CAS model is a base of CASSCF-wavefunction and is frequently employed also in density matrix renormalization group (DMRG) calculations. The  DMRG method is one of the most promising tools for strongly-correlated molecules\cite{chan_review, Szalay2015, olivares2015ab,reiherDMRG,cheng2022post} due to its  favourable scaling, which enables handling of much more extensive active spaces than CASSCF allows. The reference energy, $E^{\rm ref}$ in Eq.~(\ref{eref}), of all CAS-based methods does not include a substantial portion of electron correlation, called dynamic correlation, $E_{\rm corr}$ in Eq.~(\ref{ecorr}). Even inclusion of dozens of orbitals in the active space is not sufficient to achieve a reliable description and the necessity to recover dynamic correlation remains the major challenge of DMRG.~\cite{reiherDMRG}
Although there exists many post-CAS methods aimed at including  dynamic correlation, see e.g.\ Ref.\citenum{cheng2022post}, none is satisfactory due to the limitations both in accuracy and efficiency. In particular, perturbation theory-based approximations may suffer from the lack of size-consistency, intruder states, or unbalanced treatment of closed- and open-shell systems, which must be cured by level-shifting.~\cite{roca2012multiconfiguration} The limitation of PT2 when combined with DMRG is the high scaling with the number of active orbitals resulting from treatment of 3- and 4-electron reduced density matrices (RDMs). Efforts to reduce the cost of handling high-order RDMs in NEVPT2 are worth noticing. These include the stochastic strongly contracted scheme\cite{Mahajan2019, Blunt2020}, employing the cumulant expansion\cite{kurashige2014complete} or pre-screening techniques.~\cite{guo1}  However, the improved efficiency may come at a cost of additional intruder states.~\cite{guo2}

The goal of this work is to address the challenge of recovering dynamic correlation and proposing an efficient and reliable computational method applicable to large active spaces. The presented approach builds upon the adiabatic connection formalism first introduced in the framework of Kohn-Sham DFT\cite{harris,perdew,gunnarsson,helgaker_ac} and recently formulated for CAS  models.\cite{ac_prl,Pernal:18b}

Although the following discussion will
pertain to a ground state energy, the presented formalism is general and can be directly applied to higher states. Derivation of the formula for the correlation energy in the adiabatic connection (AC) formalism begins with assuming a model Hamiltonian $\hat{H}^{(0)}$ (typically electron-electron interaction is either reduced or removed from $\hat{H}^{(0)}$), such that the reference function $\Psi^{\rm ref}$ is its eigenfunction
\begin{equation}
\hat{H}^{(0)}\left\vert \Psi^{\rm ref}\right\rangle = E^{(0)}\left\vert \Psi^{\rm ref}\right\rangle \ \ \ .
\end{equation}
The AC Hamiltonian $\hat{H}^{\alpha}$ is introduced as a combination of
$\hat{H}^{(0)}$ and a scaled complementary operator $\hat{H}^{\prime}$
\begin{align}
\forall_{\alpha\in\lbrack0,1]}\ \ \ \hat{H}^{\alpha} &  =\hat{H}^{(0)}%
+\alpha\hat{H}^{\prime}\label{HAC}\\
\hat{H}^{\prime} &  =\hat{H}-\hat{H}^{(0)}\ \ \ ,
\end{align}
The eigenequation for $\hat{H}^{\alpha}$ reads
\begin{equation}
\hat{H}^{\alpha}\left\vert \Psi_{\nu}^{\alpha}\right\rangle=E_{\nu}^{\alpha}\left\vert \Psi_{\nu}^{\alpha}\right\rangle\ \ \ .
\end{equation}
where index $\nu$ pertains to the $\nu$-th electronic state. The role of the coupling parameter $\alpha$ is to adiabatically turn on full electron correlation by varying $\alpha$ from $0$ to $1$. Namely, at $\alpha=0$ electron interaction is reduced according to the assumed $\hat{H}^{(0)}$ model and the  reference wavefunction is obtained as $\Psi_{0}^{\alpha}$
\begin{equation}
\left\vert \Psi_{0}^{\alpha=0}\right\rangle = \left\vert \Psi^{\rm ref}
\right\rangle \ \ \ .
\end{equation}
The $\alpha=1$ limit corresponds to electrons interacting at their full strength, so that both the exact energy and wavefunction are obtained
\begin{align}
E_{0}^{\alpha=1}  & = E^{\rm exact}\ \ \ ,\\
\left\vert \Psi_{0}^{\alpha=1}\right\rangle & =\left\vert \Psi^{\rm exact}
\right\rangle \ \ \ \ .
\end{align}
Exploiting the Hellmann-Feynman theorem $\frac{\partial E_{0}^{\alpha}}{\partial\alpha} = \left\langle \Psi_{0}^{\alpha}|\hat{H}^{\prime}|\Psi_{0}^{\alpha}\right\rangle $, satisfied for $\alpha\in\lbrack0,1]$, it is straightforward to show that the correlation energy, Eq.(\ref{ecorr}), is given exactly as
\begin{equation}
E_{\rm corr} = \int_{0}^{1}\left(  \left\langle \Psi_{0}^{\alpha}|\hat{H}^{\prime}|\Psi_{0}^{\alpha}\right\rangle 
-\left\langle \Psi^{\rm ref}|\hat{H}^{\prime}|\Psi^{\rm ref}\right\rangle \right)  \ \mathrm{d}\alpha\ \ \ .
\label{ecorr1}
\end{equation}

The choice for the $\hat{H}^{(0)}$ Hamiltonian depends on the reference wavefunction. Our interest is in multireference CAS-based models which assume partitioning orbitals into sets of inactive (fully occupied), active (fractionally occupied) and virtual (unoccupied) orbitals and constructing
$\Psi^{\rm ref}$ as an antisymmetrized product of a single determinant comprising inactive orbitals and a multiconfigurational function utilizing active orbitals. Thus, we represent $\hat{H}^{(0)}$ as a sum of group Hamiltonians $\hat{H}_{I}$\cite{rosta,ac_prl}
\begin{equation}
\hat{H}^{(0)}=\sum_{I}\hat{H}_{I}
\end{equation}
where $I$ corresponds to inactive, active or virtual group and $\hat{H}_{I}$ consists of one- and two-particle operators
\begin{align}
\hat{H}_{I} &  =\sum_{pq\in I} h_{pq}^{\rm eff}\ \hat{a}_{p}^{\dagger}\hat{a}_{q} + \frac{1}{2}\sum_{pqrs\in I}\hat{a}_{r}^{\dagger}\hat{a}_{s}^{\dagger}
\hat{a}_{q}\hat{a}_{p}\left\langle rs|pq\right\rangle \ \ \ \label{HII}\\
\forall_{pq\in I}\ \ \ h_{pq}^{\rm eff} &  =h_{pq}+\sum_{J\neq I} \sum_{r\in J} n_{r} \left[  \left\langle pr|qr\right\rangle -\left\langle
pr|rq\right\rangle \right]  \ \ \ .\label{h_eff}%
\end{align}
Notice that $\left\langle rs|pq\right\rangle $ denotes a two-electron
integral in the $x_1x_2x_1x_2$ convention, and the effective one-electron Hamiltonian $h^{\rm eff}$ is a sum of
kinetic and electron-nuclei operators, and the self-consistent field
interaction of orbitals in a group $I$ with the other groups [see the second
term in Eq.(\ref{h_eff})]. Throughout the paper it is assumed that indices
$p,q,r,s$ denote natural spinorbitals of the reference ($\alpha=0$)
model and $\left\{  n_{p}\right\}$ are the corresponding natural occupation
numbers. For this choice of $\hat{H}^{(0)}$, the $\alpha$-dependent integrand
in the correlation energy expression, Eq.~\eqref{ecorr1}, includes, among
others, one electron terms depending on the difference between 1-RDM at
$\alpha=1$ and the reference one, $\gamma^{\alpha
}-\gamma^{\alpha=0}=\gamma^{\alpha}-\gamma^{\rm ref}$. Such terms are set to $0$, under the assumption
that for the properly chosen multireference wavefunction for a strongly
correlated system, variation of $\gamma^{\alpha}$ with $\alpha$ can be ignored. 

As it has been shown in Refs.~\citenum{ac_prl,Pernal:18b}, see also the Supporting Information (SI), 
choosing $\hat{H}^{(0)}$ as a group Hamiltonian and assuming that the 1-RDM\ stays constant with
$\alpha$, turn Eq.~\eqref{ecorr1} into the following AC correlation energy expression 
\begin{equation}
E_{\rm corr}^{\rm AC}=\frac{1}{2}\sum_{pqrs}{}^{\prime}\left(  \int_{0}^{1}\sum
_{\nu\neq0}\gamma_{pr}^{\alpha,0\nu}\gamma_{qs}^{\alpha,\nu0}\ \text{d}%
\alpha-\sum_{\nu\neq0}\gamma_{pr}^{\alpha=0,0\nu}\gamma_{qs}^{\alpha=0,\nu
0}\right)  \left\langle rs|pq\right\rangle \ \ \ \ ,\label{EAC}%
\end{equation}
where $\gamma^{0\nu,\alpha}$ are one-electron transition reduced density
matrices (1-TRDM) 
\begin{equation}
\gamma_{pq}^{0\nu,\alpha}=\left\langle \Psi_{0}^{\alpha}|\hat{a}_{q}^{\dagger
}\hat{a}_{p}|\Psi_{\nu}^{\alpha}\right\rangle \ \ \ .\label{TRDM}%
\end{equation}
It is important to notice a prime in the AC formula in Eq.(\ref{EAC}), which
indicates that terms pertaining to $pqrs$ belonging to the same group are
excluded. This implies that electron correlation already accounted for by the
active-orbitals-component of the reference wavefunction is not doubly counted
in $E_{\rm corr}^{\rm AC}$.

We now briefly recapitulate developments presented in our earlier works\cite{ac_prl,Pernal:18b,Pastorczak:18a,Pastorczak:18b}
leading to approximate correlation energy methods called AC and AC0. To
formulate a working expression for the AC correlation energy, we have used Rowe's equation of motion\cite{rowe,erpa1} in the particle-hole RPA\ approximation,
where the excitation operator $\hat{O}_{\nu}^{\dag}$
generating a state $\nu$, $\hat{O}_{\nu}^{\dag}\left\vert 0\right\rangle
=\left\vert \nu\right\rangle $, is approximated by single excitation operators
as $\hat{O}_{\nu}^{\dag}=\sum_{p>q}\left(  X_{pq}\hat{a}_{p}^{\dag}\hat{a}%
_{q}+Y_{pq}\hat{a}_{q}^{\dag}\hat{a}_{p}\right)  $. To distinguish this
approximation from the conventional RPA,\cite{rowe,rpa_eshuis,rpa_rev_furche,rpa_rinke} which assumes a single determinant as a reference, we called the ph-RPA\ equations
\begin{equation}
\left(
\begin{array}
[c]{cc}%
\mathbf{\mathcal{A}}_{-}^{\alpha} & 0\\
0 & \mathbf{\mathcal{A}}_{+}^{\alpha}%
\end{array}
\right)  \left(
\begin{array}
[c]{c}%
\mathbf{\tilde{Y}}_{\nu}^{\alpha}\\
\mathbf{\tilde{X}}_{\nu}^{\alpha}%
\end{array}
\right)  =\omega_{\nu}^{\alpha}\left(
\begin{array}
[c]{cc}%
\mathbf{0} & \mathbf{1}\\
\mathbf{1} & \mathbf{0}%
\end{array}
\right)  \left(
\begin{array}
[c]{c}%
\mathbf{\tilde{Y}}_{\nu}^{\alpha}\\
\mathbf{\tilde{X}}_{\nu}^{\alpha}%
\end{array}
\right)  \ \ \ ,\label{ERPA}%
\end{equation}
which are introduced for a general, multiconfigurational reference, the
extended RPA\ (ERPA).\cite{erpa1,erpa2} The ERPA\ equations have been written for the AC Hamiltonian, see Eq.~\eqref{HAC}, assuming the 1- and 2-RDM reference density matrices, leading to $\mathbf{\mathcal{A}}_{\pm}^{\alpha}$ defined as
\begin{equation}
\forall_{\substack{p>q\\r>s}}\ \ \ \left[  \mathbf{\mathcal{A}}_{\pm}^{\alpha
}\right]  _{pq,rs}=\frac{\left\langle \Psi^{\rm ref}|[\hat{a}_{p}^{\dag}\hat
{a}_{q},[\hat{H}^{\alpha},\hat{a}_{s}^{\dag}\hat{a}_{r}]]\pm\lbrack\hat{a}%
_{p}^{\dag}\hat{a}_{q},[\hat{H}^{\alpha},\hat{a}_{r}^{\dag}\hat{a}_{s}%
]]|\Psi^{\rm ref}\right\rangle }{(n_{p}-n_{q})(n_{r}-n_{s})}\ \ \ ,
\end{equation}
Explicit expressions of $\mathbf{\mathcal{A}}_{\pm}^{\alpha}$
in terms of 1- and 2-RDMs are presented in SI. Both $\mathbf{\mathcal{A}}_{+}^{\alpha}$ and
$\mathbf{\mathcal{A}}_{-}^{\alpha}$ are symmetric and positive definite at
$\alpha=0$ and $\alpha=1$ for the Hellmann-Feynman reference wavefunction
$\Psi^{\rm ref}$. Since the coupling constant dependence is passed to ERPA
equations only via AC Hamiltonian $\hat{H}^{\alpha}$, the matrices
$\mathbf{\mathcal{A}}_{\pm}^{\alpha}$ are linear in $\alpha$, i.e.
\begin{equation}
\mathbf{\mathcal{A}}_{\pm}^{\alpha}=\mathbf{\mathcal{A}}_{\pm}^{(0)}%
+\alpha\mathbf{\mathcal{A}}_{\pm}^{(1)}\ \ \ .\label{Alinear}%
\end{equation}
In the ERPA\ model,\cite{erpa_int} the $\alpha$-dependent 1-TRDMs, Eq.~\eqref{TRDM}, are given by the eigenvectors $\mathbf{\tilde{Y}}_{\nu}^{\alpha}$ as $\forall
_{p>q}\ \left(  n_{p}^{1/2}+n_{q}^{1/2}\right)  \left[ \mathbf{\tilde{Y}}_{\nu}^{\alpha}\right]_{pq} = \gamma_{qp}^{\alpha,0\nu}+\gamma_{pq}^{\alpha,0\nu}$, which allows one to turn Eq.~\eqref{EAC} into a spin-free
formula\cite{Pastorczak:18a}
\begin{equation}
E_{\rm corr}^{\rm AC} = 2 \sum_{\substack{p>q\\r>s}}{}^{\prime}\ (n_{p}^{1/2}+n_{q}^{1/2})(n_{r}^{1/2}+n_{s}^{1/2})\left(  \int_{0}^{1}\sum_{\nu}{}\left[
\mathbf{\tilde{Y}}_{\nu}^{\alpha}\right]  _{pq}\left[  \mathbf{\tilde{Y}}%
_{\nu}^{\alpha}\right]  _{rs} \text{d}\alpha-\left[  \mathbf{\tilde{Y}}_{\nu
}^{(0)}\right]_{pq}\left[  \mathbf{\tilde{Y}}_{\nu}^{(0)}\right]
_{rs}\right)   \left\langle pr|qs\right\rangle \ \ \ ,\label{ecorr2}
\end{equation}
where $\mathbf{\tilde{Y}}_{\nu}^{(0)}=\mathbf{\tilde{Y}}_{\nu}^{\alpha=0}$.
Eqs.~\eqref{ERPA} and \eqref{ecorr2} form ground for practical correlation energy calculation. This, however, requires solving the ERPA\ problem which formally scales with the $6$th power of the system size. In addition, using the reference wavefunction in which the choice of the active orbitals is not optimal could
lead to developing instability in the ERPA\ problem, for $\alpha \gg 0$.\cite{Pastorczak:18a} To lower the computational cost and avoid potential instabilities, we introduced an AC0
variant, assuming linearization of the integrand in Eq.~\eqref{ecorr2}, namely using $\mathbf{\tilde{Y}}_{\nu}^{\alpha}=\mathbf{\tilde{Y}}_{\nu}^{(0)}+\mathbf{\tilde{Y}}_{\nu}^{(1)}\alpha$, keeping the linear terms in $\alpha$ and carrying out the $\alpha$ integration,\cite{Pastorczak:18a}
\begin{equation}
E_{\rm corr}^{\rm AC0} = 2\sum_{\substack{p>q\\r>s}}{}^{\prime}\ (n_{p}^{1/2}%
+n_{q}^{1/2})(n_{r}^{1/2}+n_{s}^{1/2})\sum_{\nu}{}\left[  \mathbf{\tilde{Y}%
}_{\nu}^{(0)}\right]  _{pq}\left[  \mathbf{\tilde{Y}}_{\nu}^{(1)}\right]
_{rs}\ \left\langle pr|qs\right\rangle \ \ \ .\label{AC0}%
\end{equation}
The low computational cost of AC0 stems from the fact that ERPA equations must be only solved at $\alpha=0$ and for this value of the coupling constant the $\mathbf{\mathcal{A}}_{\pm}$ matrices are block diagonal. The largest block is
of the dimension $\rm N_{act}^{2}\times N_{act}^{2}\ $($\rm N_{act}$ denotes a number of active orbitals), so the cost of its diagonalization is marginal even for dozens of active orbitals. 

Despite the fact that encouraging results have been obtained with AC0 when combined with CASSCF\cite{Pastorczak:18a,Pastorczak:18b,pastorczak2019capturing} or DMRG,\cite{Beran:21} $\alpha$
integration should in principle account for correlation more accurately than AC0. It is thus desirable to develop an AC method which on the one hand is exact in all orders of $\alpha$, and on the other avoids solving the expensive ERPA\ problem. Ideally, such a method would be free of potential instabilities that might occur when $\alpha$ approaches $1$. A novel AC method satisfying all the requirements is  presented in this work.

Let us use the integral identity $\forall_{\operatorname{Re}a>0}%
\ \ \ 2a/\pi\int_{0}^{\infty}(\omega^{2}+a^{2})^{-1}\ $d$\omega=1$ to express
the AC correlation energy by means of the $\alpha$-dependent dynamic
density-density response matrix.~\cite{drwal2021} This
can be attained by employing the relations
\begin{equation}
\sum_{\nu}{}\left[  \mathbf{\tilde{Y}}_{\nu}^{\alpha}\right]  _{pq}\left[
\mathbf{\tilde{Y}}_{\nu}^{\alpha}\right]  _{rs}=\frac{2}{\pi}\int_{0}^{\infty
}\text{d}\omega\sum_{\nu}\left[  \mathbf{\tilde{Y}}_{\nu}^{\alpha}\right]
_{pq}\left[  \mathbf{\tilde{Y}}_{\nu}^{\alpha}\right]  _{rs}\frac{\omega_{\nu
}^{\alpha}}{\omega^{2}+\left(  \omega_{\nu}^{\alpha}\right)  ^{2}}\equiv
\frac{1}{\pi}\int_{0}^{\infty}\text{d}\omega\ \left[  \mathbf{C}^{\alpha
}(\omega)\right]  _{pq,rs}%
\end{equation}
in Eq.(\ref{ecorr2}) which results in the formula%
\begin{equation}
E_{\rm corr}^{\rm AC}=\frac{2}{\pi}\int_{0}^{1}\text{d}\alpha\int_{0}^{\infty}%
\text{d}\omega\ \text{Tr}\left\{  \left[  \mathbf{C}^{\alpha}(\omega
)-\mathbf{C}^{\alpha=0}(\omega)\right]  ^{\prime}\mathbf{g}\right\}
\ \ \ ,\label{ecorr3}%
\end{equation}
where%
\begin{equation}
\forall_{\substack{p>q\\r>s}}\ \ \ 
g_{pq,rs}=(n_{p}^{1/2}+n_{q}^{1/2})(n_{r}^{1/2}+n_{s}^{1/2})\left\langle
pr|qs\right\rangle \ \ \ ,\label{g}%
\end{equation}
and the prime in Eq.(\ref{ecorr3}) indicates that when taking a product
of the matrices $\mathbf{C}$ and $\mathbf{g}$ terms $pqrs\in active$ are excluded.
By using spectral representations of the matrices $\mathbf{\mathcal{A}}%
_{+}^{\alpha}$ and $\mathbf{\mathcal{A}}_{-}^{\alpha}$ in terms of the
ERPA\ eigenvectors,\cite{furche2001density} it is straightforward to show that the dynamic linear response matrix $\mathbf{C}^{\alpha}(\omega)$ follows from the linear equation given as (see SI\ for details)
\begin{equation}
\left[  \mathbf{\mathcal{A}}_{+}^{\alpha}\mathbf{\mathcal{A}}_{-}^{\alpha
}+\omega^{2}\mathbf{1}\right]  \mathbf{C}^{\alpha}\mathbf{(}\omega
)=\mathbf{\mathcal{A}}_{+}^{\alpha}\ \ \ .\label{Cfinal}%
\end{equation}
To reduce the computational cost of solving Eq.~\eqref{Cfinal}, we  introduce a decomposition of the modified two-electron integrals $\mathbf{g}$ 
\begin{equation}
g_{pq,rs} = \sum_{L=1}^{\rm N_{Chol}}D_{pq,L}D_{rs,L}\ \ \ ,\label{CD}
\end{equation}
where $D_{pq,L}$ are the natural-orbital transformed Cholesky vectors of the Coulomb matrix multiplied by  factors $n_{p}^{1/2}+n_{q}^{1/2}$, cf. Eq.~\eqref{g}. 
We expand $\mathbf{C}^{\alpha}(\omega)$ at $\alpha=0$
\begin{align}
\mathbf{C}^{\alpha}(\omega) &  =\sum_{n=0}\frac{1}{n!}\mathbf{C(}%
\omega\mathbf{)}^{(n)}\alpha^{n}\ \ \ ,\label{Cexpanded}\\
\mathbf{C(}\omega\mathbf{)}^{(n)} &  =\left.  \frac{\partial^{n}%
\mathbf{C}^{\alpha}(\omega)}{\partial\alpha^{n}}\right\vert _{\alpha=0}\ \ \ ,
\end{align}
and solve Eq.~\eqref{Cfinal} iteratively in the reduced space, by retrieving, in the $n$th iteration, the $n$th-order correction $\mathbf{C}^{(n)}$ projected on the space spanned by $\rm N_{Chol}$ transformed Cholesky vectors $\left\{\mathbf{D}_{L}\right\}  $. To account for the prime (exclusion of terms for
all-active indices $pqrs$) in the AC correlation energy, Eq.~\eqref{ecorr3}, define the auxiliary matrices of the transformed Cholesky vectors as
\begin{align}
\forall_{p>q}\ \ \ D_{pq,L}^{1} &  =\left\{
\begin{array}
[c]{cc}%
2D_{pq,L} & \text{if \ \ }pq\in {\rm active}\\
D_{pq,L} & \text{otherwise}%
\end{array}
\right.  \ \ \ ,\\
\forall_{p>q}\ \ \ D_{pq,L}^{2} &  =\left\{
\begin{array}
[c]{cc}%
0 & \text{if \ \ }pq\in {\rm active}\\
D_{pq,L} & \text{otherwise}%
\end{array}
\right.  \ \ \ .
\end{align}
Assuming expansion of the response matrix $\mathbf{C}^{\alpha}(\omega)$, cf.\ Eq.~\eqref{Cexpanded}, up to nth order in $\alpha$, and employing Cholesky decomposition of integrals, Eq.~\eqref{CD} together with the matrices $\mathbf{D}^{1}$ and $\mathbf{D}^{2}$ in Eq.~\eqref{ecorr3} lead to a new AC formula for the correlation energy reading
\begin{equation}
E_{\rm corr}^{\rm AC_n}=\frac{2}{\pi}\text{Tr}\left[  \left(  \int_{0}^{\infty
}\text{d}\omega\sum_{k=1}^{\rm{n}}\frac{\mathbf{\bar{C}(}%
\omega\mathbf{)}^{(k)}}{k!(k+1)}\right)  \mathbf{D}^{2}\right]
\ \ \ .\label{ecorr4}%
\end{equation}
The matrices $\mathbf{\bar{C}(}\omega\mathbf{)}^{(n)}$ defined as
\begin{equation}
\mathbf{\bar{C}(}\omega\mathbf{)}^{(n)}=\mathbf{C(}\omega\mathbf{)}%
^{(n)}\mathbf{D}^{1}\ \ \
\end{equation}
are of the dimension $\rm M^2\times \rm N_{Chol}$, which is reduced comparing to the
$\rm M^2\times M^2$ dimension of $\mathbf{C(}\omega\mathbf{)}^{(n)}$, since by
construction the number of Cholesky vectors is one order of magnitude smaller
than $\rm M^2$, i.e.\ ${\rm N_{Chol}}\sim \rm M$. Employing the linearity in $\alpha$ of the matrices $\mathbf{\mathcal{A}}_{\pm}^{\alpha}$, cf.\ Eq.~\eqref{Alinear} in Eq.~\eqref{Cfinal}, one finds the following recursive formula for the $n$th-order term $\mathbf{\bar{C}(}\omega\mathbf{)}^{(n)}$
\begin{align}
\mathbf{\bar{C}(}\omega\mathbf{)}^{(0)} &  =\mathbf{\bar{A}}%
_{+}^{(0)}\mathbf{D}^{1}\label{C0}\\
\mathbf{\bar{C}(}\omega\mathbf{)}^{(1)} &  =\mathbf{\bar{A}}%
_{+}^{(1)}\mathbf{D}^{1}-\mathbf{\bar{A}}^{(1)}\mathbf{\bar{C}(}%
\omega\mathbf{)}^{(0)}\label{C1}\\
\forall_{n\geq2}\ \ \ \mathbf{\bar{C}(}\omega\mathbf{)}^{(n)} &
=-n\mathbf{\bar{A}}^{(1)}\mathbf{\bar{C}(}\omega\mathbf{)}^{(n-1)}%
-n(n-1)\mathbf{\bar{A}}^{(2)}\mathbf{\bar{C}(}\omega\mathbf{)}^{(n-2)}%
\label{Cn}%
\end{align}
where the required matrices are given by the ERPA\ matrices
$\mathbf{\mathcal{A}}_{\pm}^{(0)}$ and $\mathbf{\mathcal{A}}_{\pm}^{(1)}$ (see SI for their explicit forms in terms of 1-, 2-RDMs)
\begin{align}
\mathbf{\bar{A}}_{+}^{(0)} &  =\Lambda(\omega)\mathbf{\mathcal{A}}_{+}%
^{(0)}\ \ \ ,\\
\mathbf{\bar{A}}_{+}^{(1)} &  =\Lambda(\omega)\mathbf{\mathcal{A}}_{+}%
^{(1)}\ \ \ ,\\
\mathbf{\bar{A}}^{(1)} &  =\Lambda(\omega)\left(  \mathbf{\mathcal{A}}%
_{+}^{(0)}\mathbf{\mathcal{A}}_{-}^{(1)}+\mathbf{\mathcal{A}}_{+}%
^{(1)}\mathbf{\mathcal{A}}_{-}^{(0)}\right)  \ \ \ ,\\
\mathbf{\bar{A}}^{(2)} &  =\Lambda(\omega)\mathbf{\mathcal{A}}_{+}%
^{(1)}\mathbf{\mathcal{A}}_{-}^{(1)}\ \ \ ,\\
\Lambda(\omega) &  =\left(  \mathbf{\mathcal{A}}_{+}^{(0)}\mathbf{\mathcal{A}%
}_{-}^{(0)}+\omega^{2}\mathbf{1}\right)  ^{-1}\ \ \ .
\label{lambda}
\end{align}

The correlation energy expression in Eq.~\eqref{ecorr4} together with the recursive relation in Eqs.~\eqref{C0}-\eqref{Cn} is the central achievement of this work. It allows one to compute the correlation energy for strongly correlated systems at the cost scaling with only the $5$th power of the system size. All matrix operations scale as $\rm M^{4} \rm N_{ Chol}$ down from $\rm M^{6}$ scaling of the original ERPA problem in Eq.~\eqref{ERPA}. Notice that the cost of computing the $\Lambda(\omega)$ matrix is marginal, since the inverted matrix is block diagonal with the largest block of the dimension $\rm N_{act}^{2}\times N_{act}^{2}$.

By setting the maximum order of expansion of the response matrix
$\mathbf{C(}\omega\mathbf{)}$ in Eq.~\eqref{ecorr4} to $1$, the correlation energy $\rm AC_n$ becomes  equivalent to the AC0 approximation, cf.\ Eq.~\eqref{AC0}. In the limit ${\rm n} \rightarrow \infty$, $ E_{\rm corr}^{ \rm AC_{n}}$ value approaches the AC energy given according to the formula in Eq.~\eqref{ecorr2}, if the Taylor series is convergent. Numerically this equality requires a sufficient accuracy both in the frequency integration and in the Cholesky decomposition of two-electron integrals.

Going beyond the 1st-order terms in the coupling constant is potentially beneficial, since higher-orders gain importance as $\alpha$ approaches $1$. 
Higher-order contributions are effectively maximized if the AC integrand $W^{\alpha}$ in Eq.~\eqref{ecorr3}, $W^{\alpha}=\int_{0}^{\infty}\text{d}\omega\ \text{Tr}\left\{  \left[
\mathbf{C}^{\alpha}(\omega)-\mathbf{C}^{\alpha=0}(\omega)\right]  ^{\prime
}\mathbf{g}\right\}$, is linearly extrapolated from $W^{\alpha=1}$ to the exact limit $W^{\alpha=0}=0$. Such an extrapolation
method leading to the approximation 
$E_{\rm corr}^{\rm AC}=\frac{2}{\pi}\int_{0}^{1}\alpha\ W^{\alpha
=1}\ \text{d}\alpha$ has already been proposed in Ref.\citenum{ac_prl}. 
If it is used together with the formula in Eq.(\ref{ecorr3}), the expansion shown in Eq.(\ref{Cexpanded}) and the Cholesky decomposition of two-electron integrals, one obtains the formula
\begin{equation}
E_{\rm corr}^{\rm AC1_n}=\frac{1}{\pi}\text{Tr}\left[  \left(  \int
_{0}^{\infty}\text{d}\omega\sum_{k=1}^{\text{n}}\frac{\mathbf{\bar{C}%
(}\omega\mathbf{)}^{(k)}}{k!}\right)  \mathbf{D}^{2}\right]  \ \ \ ,
\label{AC1}
\end{equation}
which will be denoted as AC1$\rm _n$. 
Notice that in the frequency integrated  $k$th-order term in Eq.~\eqref{AC1}
contributes to the correlation energy by the factor $(k+1)/2$ greater than its
counterpart in the expression given in Eq.~\eqref{ecorr4}. 


The Cholesky decomposition of the Coulomb integrals matrix in the AO basis was carried out
using a modified program originally used in Refs.~\citenum{Modrzejewski:20,Modrzejewski:21}.
The implementation was done according to Ref.~\citenum{Aquilante:11}.
The Cholesky vectors in the AO basis, $R_{pq,L}$, were generated until the satisfaction of
the trace condition $\sum_{p \ge q} \left( \langle pp \middle| qq \rangle - \sum_L R_{pq,L} R_{pq, L} \right)<10^{-2}$. The convergence threshold was previously tested as a part of the default set of 
numerical thresholds in Table~1 of Ref.~\citenum{Modrzejewski:20}.

For the $\omega$ integration in the AC$_n$ correlation energy, we have used a modified Gauss–Legendre quadrature as described in Ref.\citenum{ren2012resolution}. With the 18-point grid, the accuracy of the absolute value of energy achieves $10^{-2}$ mHa, which results in $10^{-2}$ eV accuracy in the singlet-triplet gaps.  

To assess the accuracy of the AC$_n$ approaches, we have applied them to two benchmark datasets of singlet-triplet (ST) energy gaps: the single-reference systems set of Schreiber et al.~\cite{Thiel:CASPT2} and multi-reference organic biradicals studied by Stoneburner et al.~\cite{Stoneburner2017} In the single-reference dataset  we employed the TZVP\cite{tzvp} basis set and compared our data against the CC3\cite{Thiel:CASPT2} results. The aug-cc-pVTZ basis and doubly electron-attached (DEA) equation-of-motion (EOM) coupled-cluster (CC) 4-particle–2-hole (4p-2h) reference\cite{Stoneburner2017} were used for the biradicals. All CASSCF calculations were performed in the Molpro\cite{Molpro:12} program. All AC methods were implemented in the GammCor program\cite{gammcor}.

\begin{figure}[h!]
\centering
\includegraphics[width=0.6\textwidth]{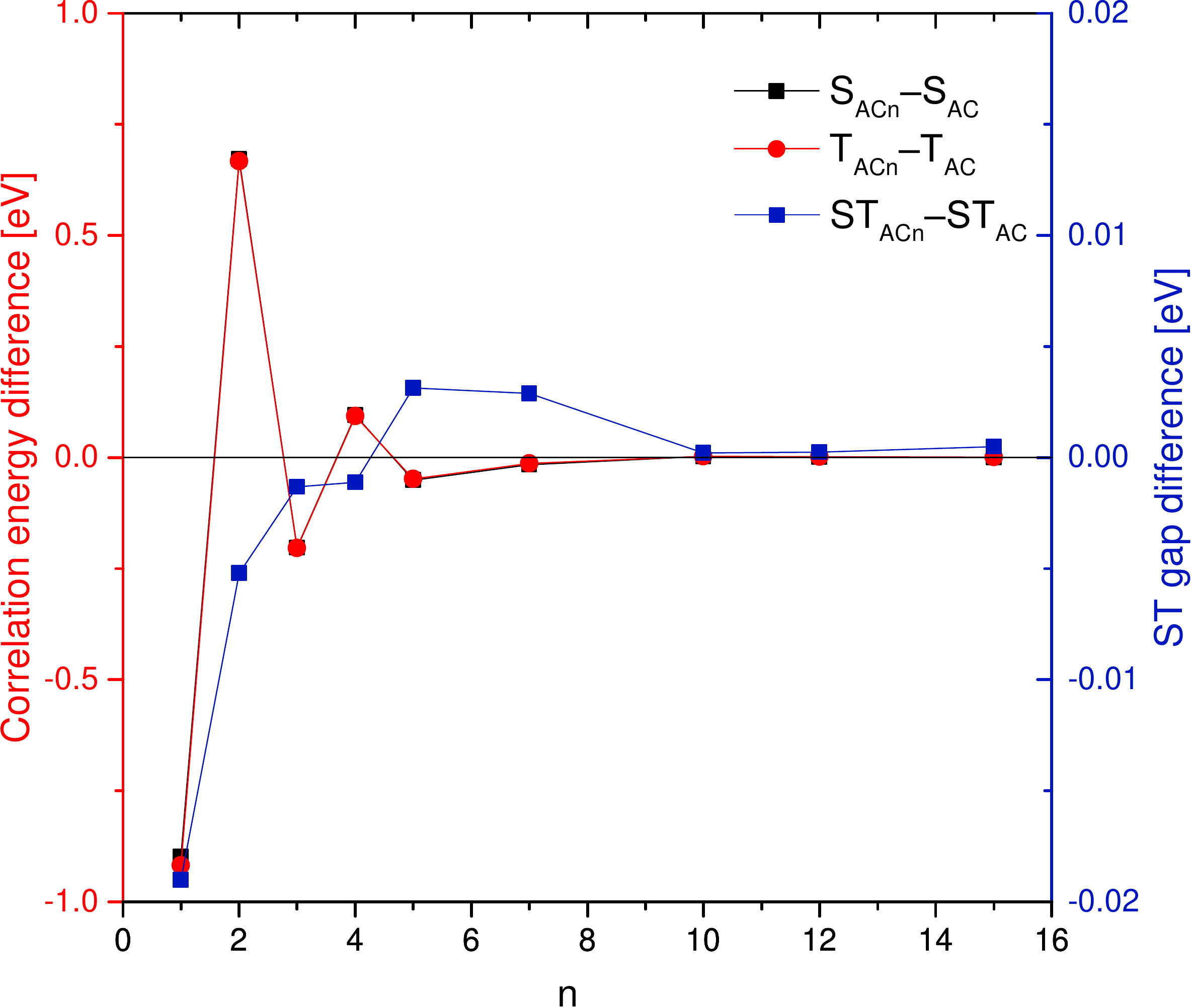}
\caption{Differences of AC$_{\rm n}$ and AC correlation energies for singlet (S) and triplet (T) states (left axis) and ST gaps (right axis) as a function of n for the C$_{4}$H$_{2}$-1,3-(CH$_{2}$)$_{2}$ biradical. Notice that black markers overlap with the red ones. \label{fig:13iso}}
\end{figure}

Computing the correlation energy with the AC${\rm_n}$ method requires either fixing the maximum order of expansion with respect to the coupling constant, $\rm n$ in Eq.(\ref{ecorr4}), or continuing the expansion until a prescribed convergence threshold is met. The advantage of the former strategy is that size-consistency is strictly preserved. For each system we found that the AC${\rm_n}$ correlation energy converges with n for the chosen active space. A typical convergence behavior for the singlet, triplet and ST energies is presented in Figure~\ref{fig:13iso}. It can be seen that already for n=3 the AC$_{\rm n}$ ST gap deviates by only 0.01~eV from the AC value, computed using Eq.~(\ref{ecorr2}). For all other biradicals and single-reference systems we found that setting $\rm n=10$ in Eq.(\ref{ecorr4}) is sufficient to converge ST gaps within 10$^{-2}$ eV, thus, n = 10 has been set for all systems.

In Table~\ref{tab:thiel} we present ST gaps for the subset of Ref.\citenum{Thiel:CASPT2} dataset. The CASSCF method predicts too narrow ST gaps with the mean error approaching $-0.4$ eV, which results from the unbalanced treatment of closed-shell singlet and open-shell triplet states. The addition of correlation energy using adiabatic connection greatly reduces the errors. The mean unsigned error (MUE) of AC0 amounts to 0.24~eV. The performance is further improved by AC$_{\rm n}$ which affords MUE of 0.13~eV. 
Maximizing the contribution from high-order terms in $\alpha$, attained in AC1$_{\rm n}$, leads to ST gaps of the same unsigned error as that of AC$_{\rm n}$. Noticeable, the signed error is reduced, which indicates that higher order terms play more important role for open-shell than for the closed-shell states. The accuracy of AC$_{\rm n}$ is on a par with NEVPT2 and only slightly worse than best CASPT2 estimations from Ref.\citenum{Thiel:CASPT2}. The standard deviation of AC0, amounting to 0.23 eV, is reduced to 0.11~eV by AC$_{\rm n}$, which parallels the standard deviation of the perturbation methods.

\begin{table}
\centering
\begin{threeparttable}
\caption{ST gaps ($\rm E_T-E_S$), mean errors (ME), mean unsigned errors (MUE), and standard deviations (Std.\ Dev.) computed with respect to CC3 reference data. All values in eV.}
\begin{tabular}{l l c c c c c  c c }
Molecule        & T state     & CASSCF\tnote{a} & AC1$\rm_n$  & AC$\rm_n$ & AC0 & NEVPT2\tnote{b} & CASPT2\tnote{c} & CC3\tnote{c} \\ \hline
Ethene          & $ 1 \prescript{3}{} B_{1u}$     & 3.78  & 4.53 & 4.56 & 4.69 & 4.60 & 4.60 & 4.48 \\
E-butadiene       & $ 1 \prescript{3}{} B_u$       & 2.77  & 3.44 & 3.43 & 3.46 & 3.38 & 3.34 & 3.32 \\
All-E-hexatriene      & $ 1 \prescript{3}{} A_g$      & 2.66  & 2.83 & 2.81 & 2.80 & 2.73 & 2.71 & 2.69 \\
All-E-octatetraene    & $ 1 \prescript{3}{} B_u$     & 2.25  & 2.46 & 2.43 & 2.39 & 2.32 & 2.33 & 2.30 \\
Cyclopropene    & $ 1 \prescript{3}{} B_2$      & 3.78  & 4.42 & 4.44 & 4.56 & 4.56 & 4.35 & 4.34 \\
Cyclopentadiene & $ 1 \prescript{3}{} B_2$      & 2.75  & 3.34 & 3.34 & 3.37 & 3.32 & 3.28 & 3.25 \\
Norbornadiene   & $ 1 \prescript{3}{} A_2$      & 3.07  & 3.92 & 3.89 & 3.86 & 3.79 & 3.75 & 3.72 \\
Benzene         & $ 1 \prescript{3}{} B_{1u}$    & 3.74  & 4.17 & 4.21 & 4.37 & 4.32 & 4.17 & 4.12 \\
Naphtalene      &  $ 1 \prescript{3}{} B_{2u}$     & 2.93  & 3.19 & 3.21 & 3.29 & 3.26 & 3.20 & 3.11 \\
Furan           & $ 1 \prescript{3}{} B_2$      &3.54  & 4.09 & 4.16 & 4.30 & 4.33 & 4.17 & 4.48 \\
Pyrrole         & $ 1 \prescript{3}{} B_2$      & 3.95  & 4.47 & 4.52 & 4.67 & 4.73 & 4.52 & 4.48 \\
Imidazole       & $ 1 \prescript{3}{} A^{\prime}$      & 4.42  & 4.70 & 4.74 & 4.85 & 4.77 & 4.65 & 4.69 \\
Pyridine        &$ 1 \prescript{3}{} A_1$      & 3.81  & 4.28 & 4.34 & 4.53 & 4.47 & 4.27 & 4.25 \\
s-Tetrazine       & $ 1 \prescript{3}{} B_{3u}$     & 2.43  & 2.27 & 2.05 & 1.51 & 1.64 & 1.56 & 1.89 \\
Formaldehyde    & $ 1 \prescript{3}{} A_2$      & 3.32  & 3.80 & 3.74 & 3.77 & 3.75 & 3.58 & 3.55 \\
Acetone         & $ 1 \prescript{3}{} A_2$      & 4.17  & 4.27 & 4.29 & 4.90 & 4.10 & 4.08 & 4.05 \\
Formamide       & $ 1 \prescript{3}{} A^{\prime\prime}$     & 4.72  & 5.31 & 5.47 & 5.60 & 5.64 & 5.40 & 5.36 \\
Acetamide       & $ 1 \prescript{3}{} A^{\prime\prime}$     & 4.77  & 5.46 & 5.57 & 5.73 & 5.52 & 5.53 & 5.42 \\
Propanamide     & $ 1 \prescript{3}{} A^{\prime\prime}$      & 4.79  & 5.51 & 5.61 & 5.80 & 5.54 & 5.44 & 5.45 \\
\hline
                \hline
               ME &        & $-0.38$ & 0.08 & 0.10 & 0.18 & 0.10 & 0.00 & - \\
                MUE&       & 0.45  & 0.13 & 0.13 & 0.24 & 0.14 & 0.07 & - \\
                Std.\ Dev.\  & & 0.35  & 0.15 & 0.11 & 0.23 & 0.13 & 0.12 & -\\
                \hline\hline

\end{tabular}
\begin{tablenotes}
\item[a] active spaces from Ref.\citenum{Thiel:CASPT2}
\item[b] results from Ref.\citenum{Neese:NEVPT2}
\item[c] results from Ref.\citenum{Thiel:CASPT2}
\end{tablenotes}
\label{tab:thiel}
\end{threeparttable}
\end{table}

In Ref.\citenum{Stoneburner2017} a systematic design of active spaces for biradicals based on the correlated participating orbital (CPO) scheme\cite{Tishchenko2008} was presented.
Here, we take a different approach and identify the most appropriate CASs by means of single-orbital entropies  and two-orbital mutual information.~\cite{Stein2016,Legeza2003,Golub2021} 

\begin{figure}[h!]
  \subfloat[C$_4$H$_4$, singlet state, CAS(20,22) \label{mut_info_c4h4_s)}]{%
    \includegraphics[width=0.45\textwidth]{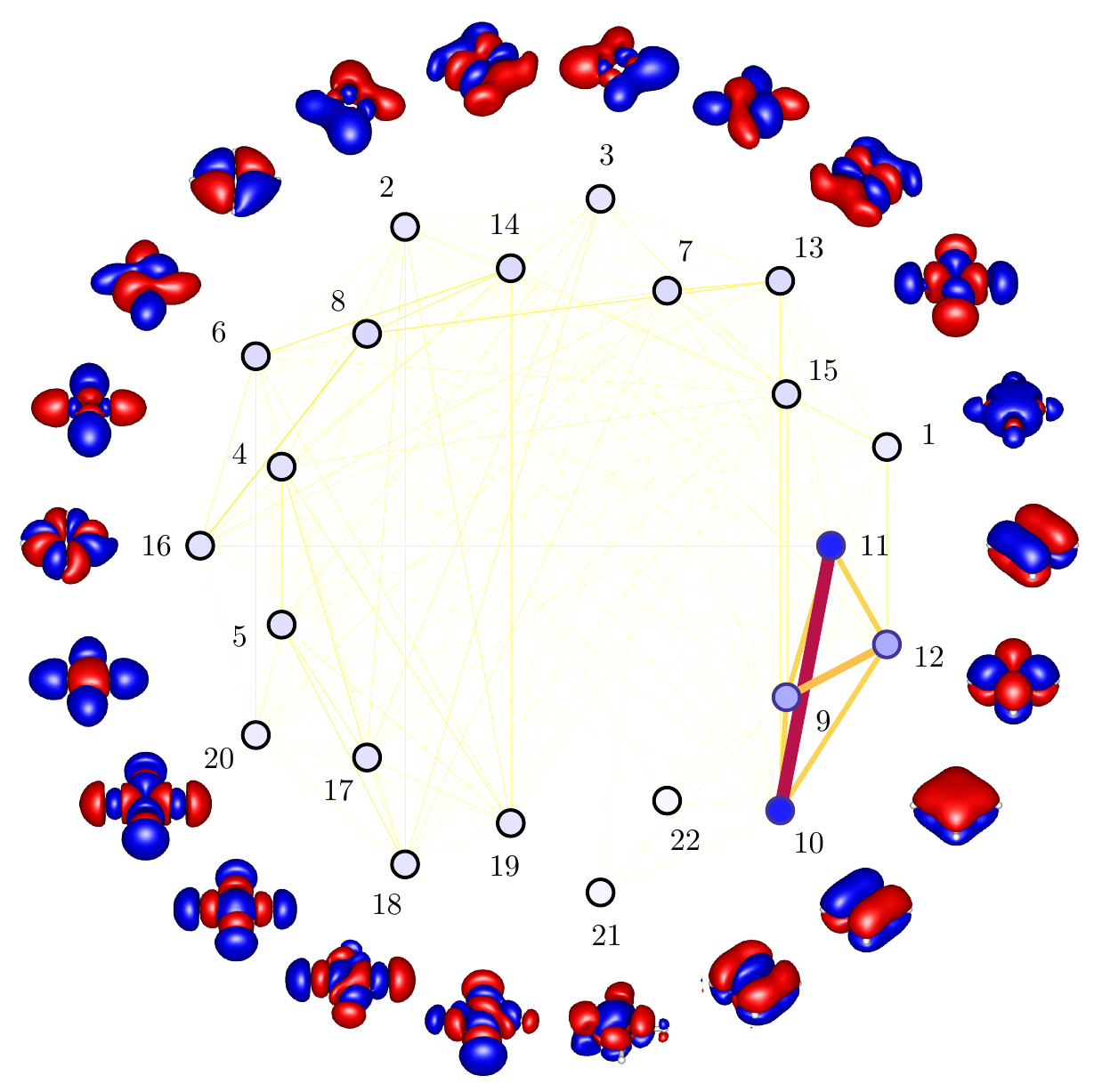}
  }
  \hfill
  \subfloat[C$_4$H$_4$, triplet state, CAS(20,22) \label{mut_info_c4h4_t}]{%
    \includegraphics[width=0.45\textwidth]{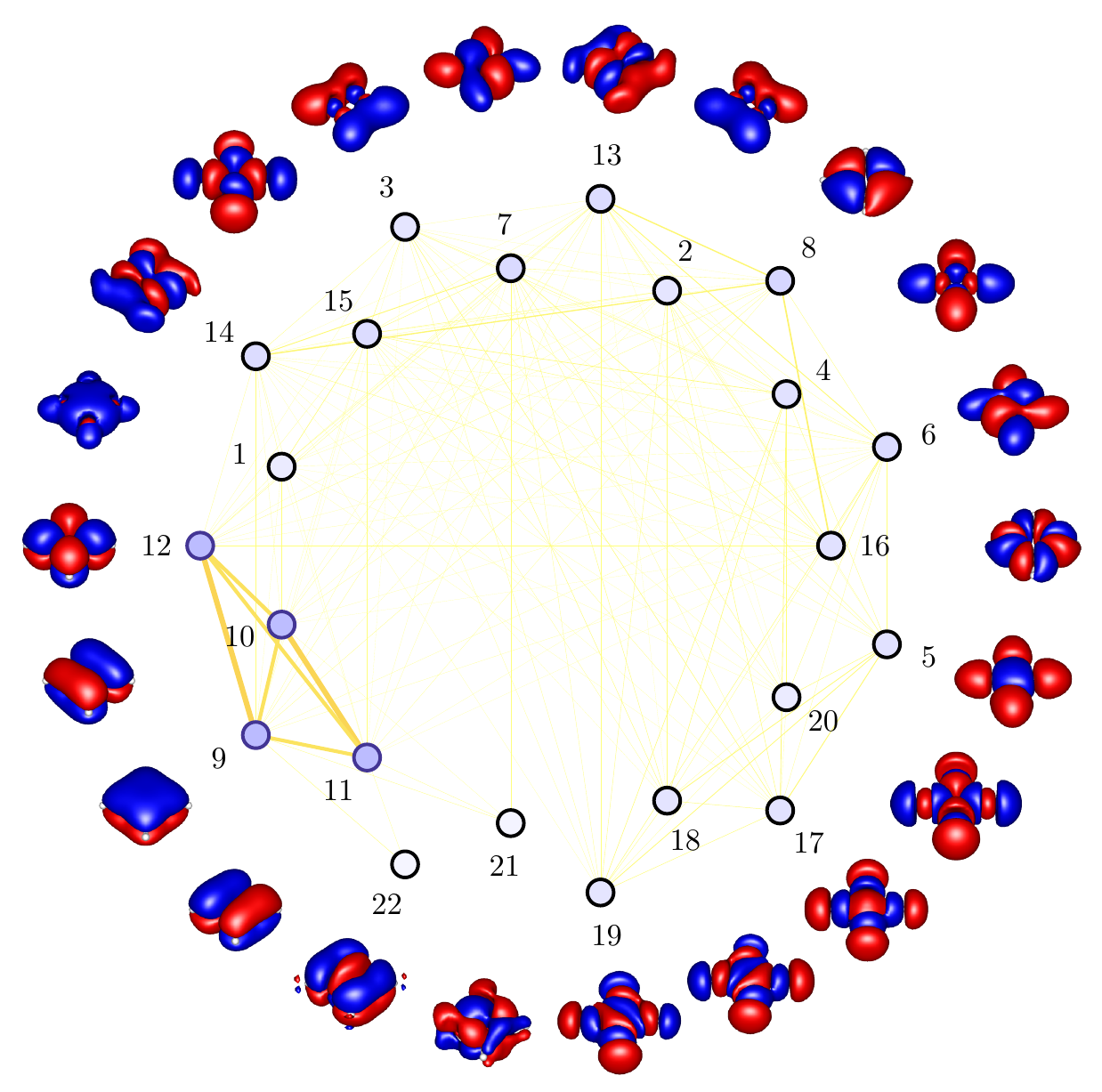}
  }
  \\
  \subfloat[C$_5$H$_5^+$, singlet state, CAS(14,16) \label{mut_info_c5h5_s}]{%
    \includegraphics[width=0.45\textwidth]{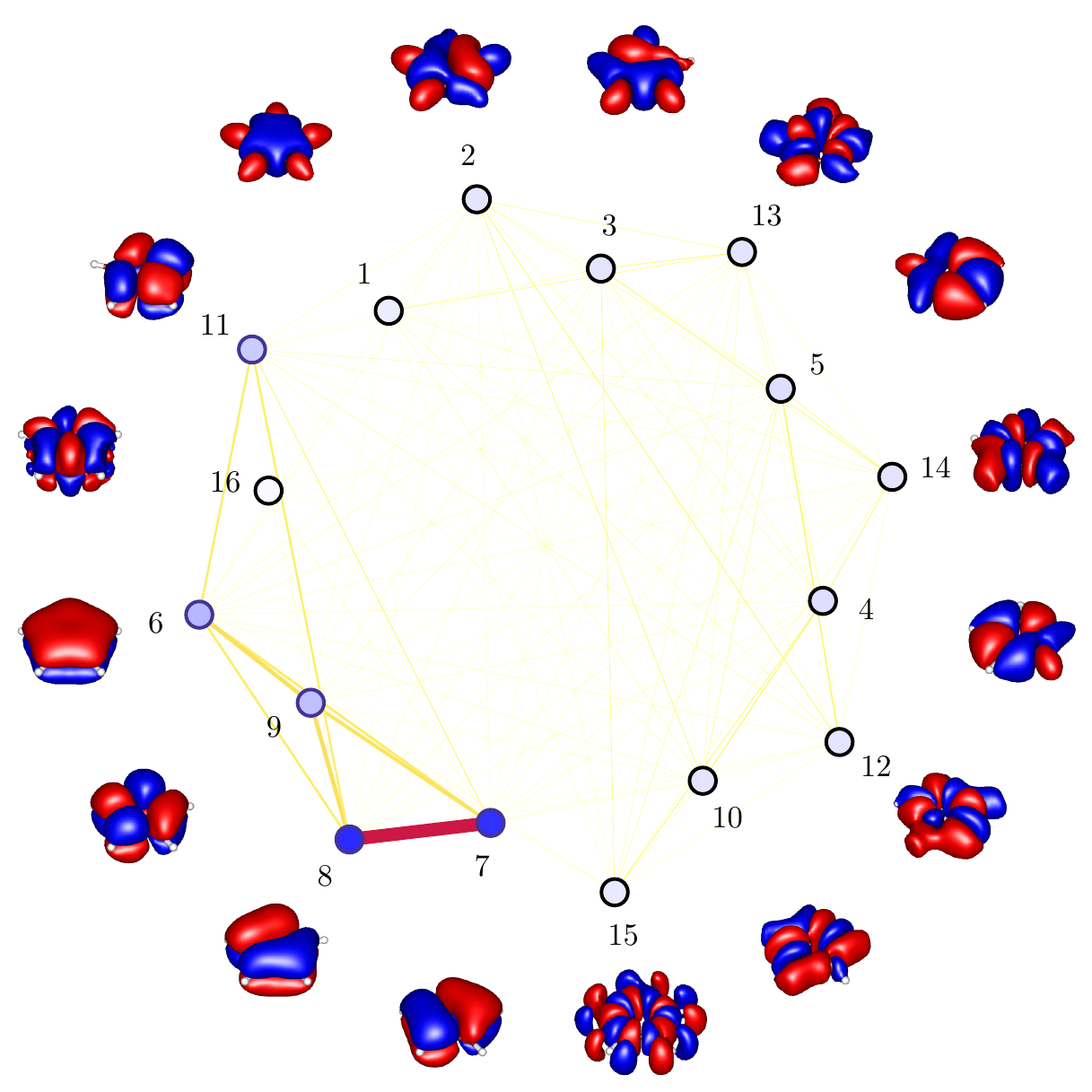}
  }
  \hfill
  \subfloat[C$_5$H$_5^+$, triplet state, CAS(14,16) \label{mut_info_c5h5_t}]{%
    \includegraphics[width=0.45\textwidth]{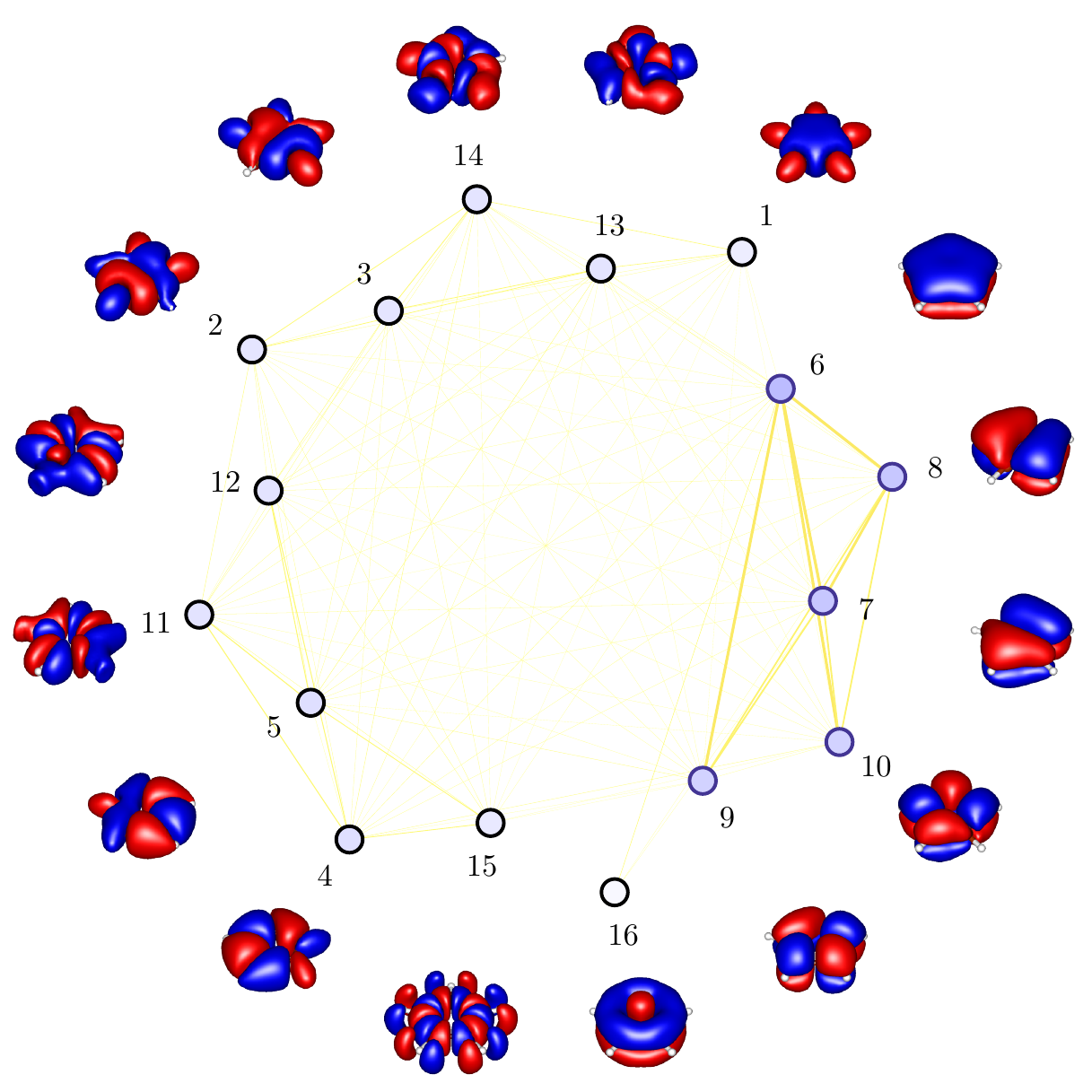}
  }
  \\
  \subfloat[Scales used for visualizing of correlations\label{scale}]{%
    \includegraphics[width=0.45\textwidth]{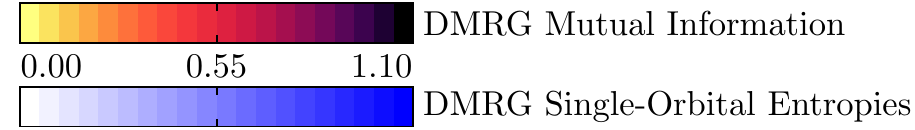}
  }
  \caption{The DMRG mutual information (colored edges) and single-orbital entropies (colored vertices) of C$_4$H$_4$ and C$_5$H$_5^+$ for the lowest singlet and triplet states. Numbers in the graphs correspond to indices of the DMRG-SCF (C$_4$H$_4$) and CASSCF (C$_5$H$_5^+$) natural orbitals presented together with their occupation numbers in SI. Blue circles represent the $\pi$ orbitals with $s_i > 0.19$. \label{mut_info}}
\end{figure}

Figure~\ref{mut_info} shows the correlation measures for singlet and triplet states of prototypical biradicals, \ce{C4H4} and C$_5$H$_5^+$, obtained with, respectively, CAS(20,22) and CAS(14,16) active spaces (cf.\ description in SI). We observe that the $\pi$ orbitals of \ce{C4H4} and C$_5$H$_5^+$ are well separated from the others in terms of their single-orbital entropies ($s_i > 0.19$, see SI) and represent a natural choice of the active space selection. The largest values of $s_i$ correspond to the singly occupied frontier orbitals in the singlet states. These orbital pairs also possess the largest values of mutual information, which stems from a strong correlation of the frontier orbitals due to the singlet type coupling of these open shells. Notice that both single-orbital entropies and mutual information of the singly occupied orbitals are much lower in the case of the triplet states. This is due to the fact that the triplet states were calculated as high-spin projections and thus can be qualitatively described with a single determinant. When analyzing the triplet states, one can however see that all the $\pi$ orbitals have similar values of their single-orbital entropies and that e.g.\ CAS(2,4) (nCPO active space in Ref.\citenum{Stoneburner2017}) is not a reasonable choice. In fact, for this imbalanced active space, we have experienced  divergence of the AC$\rm{_n}$ series, see the last entry of Table 1 in SI.

\begin{table}
\caption{ST gaps ($\rm E_T-E_S$) in eV and errors with respect to DEA-EOMCC[4p-2h]\cite{Stoneburner2017} (ref.) predictions. Labels: 1 -- \(\mathrm{C_{4}H_{4}}\), 2 -- \(\mathrm{C_{5}H_{5}^{+}}\), 3 -- \(\mathrm{C_{4}H_{3}NH_{2}}\), 4 -- \(\mathrm{C_{4}H_{3}CHO}\), 5 -- \(\mathrm{C_{4}H_{2}NH_{2}(CHO)}\), 6 -- C$_{4}$H$_{2}$-1,2-(CH$_{2}$)$_{2}$, 7 -- C$_{4}$H$_{2}$-1,3-(CH$_{2}$)$_{2}$.}
\centering
\begin{threeparttable}
\begin{tabular}{c S S S S S S S}
sys.              & \multicolumn{1}{c}{CASSCF} & \multicolumn{1}{c}{AC1$\rm_n$} & \multicolumn{1}{c}{AC$\rm_n$}  & \multicolumn{1}{c}{AC0} & \multicolumn{1}{c}{ftPBE\tnote{a}} & \multicolumn{1}{c}{RASPT2\tnote{b}} & \multicolumn{1}{c}{ref.} \\
\hline\hline
1    & 0.44   & 0.18   & 0.13     & 0.00  & 0.11   & 0.19   & 0.18      \\
2    & -0.70  & -0.71  & -0.77    & -0.86  & -0.64  & -0.65  & -0.60     \\
3    & 0.39   & 0.12   & 0.07     & -0.08  & 0.03  & 0.11   & 0.12      \\
4    & 0.41   & 0.17   & 0.12     & -0.02  & 0.09  & 0.16   & 0.16      \\
5    & 0.60   & 0.39   & 0.37     & 0.14  & 0.45    & 0.32   & 0.25      \\
6    & 3.33   & 3.44   & 3.44     & 3.46  & 3.34    & 3.27   & 3.37      \\
7    & -0.92  & -0.89  & -0.90    & -0.92  &  -0.66  & -0.80  & -0.80 \\
\hline
ME        & 0.13        & 0.00       & -0.03       & -0.14       & 0.01 & -0.01  & \multicolumn{1}{c}{-} \\
MUE       & 0.20        & 0.06        & 0.08        & 0.16       & 0.09 & 0.04   & \multicolumn{1}{c}{-} \\
MU\%E     & 102.3      & 13.7       & 26.1       & 69.6      & 37.6 & 7.2   & \multicolumn{1}{c}{-} \\
Std. Dev. & 0.20        & 0.09        & 0.10        & 0.11   &   0.12  & 0.05   & \multicolumn{1}{c}{-} \\  \hline\hline
\end{tabular}
\begin{tablenotes}
\item [a] ftPBE results taken from Ref.\citenum{stoneburner2018mc}
\item [b] RASPT2 (valence-$\pi$, $\pi$CPO active space) results taken from Ref.\citenum{Stoneburner2017}
\end{tablenotes}
\end{threeparttable}
\label{tab:biradicals}
\end{table}

The analysis of mutual information and single-orbital entropies of prototypical biradicals have allowed us to define optimal active spaces: CAS(4,4) for \ce{C4H4}, \ce{C4H3NH2}, \ce{C4H3CHO}, \ce{C4H2NH2}(CHO), CAS(4,5) for C$_{5}$H$_{5}^{+}$, and CAS(6,6) for C$_{4}$H$_{2}$-1,2-(CH$_{2}$)$_{2}$, and C$_{4}$H$_{2}$-1,3-(CH$_{2}$)$_{2}$. The choice of the orbitals in CAS is therefore such that all valence $\pi$ orbitals on or adjacent to the carbon ring are included and only the mostly correlated orbitals, of the occupancies in the range (0.05, 1.95), enter the active space. The chosen active spaces are close to the $\pi$CPO scheme considered in Ref.\citenum{Stoneburner2017}, with the difference that nearly unoccupied orbitals included in $\pi$CPO, shown to be uncorrelated according to our mutual information analysis, are excluded.

Similar to the single-reference case, the performance of the CASSCF method for the ST gaps in biradicals is seriously affected by the lack of dynamic correlation (Table~\ref{tab:biradicals}). 
Even though the CASSCF gaps of three systems (C$_5$H$_5^+$, 1,2- and 1,3-isomers) are in error of only 0.1 eV, the overall MUE is as large as 0.20~eV and the mean average unsigned percentage error (MU\%E) exceeds 100\%.
The AC0 method overcompensates the errors of CASSCF. 
For biradicals 1, 3, 4, the excessive reduction of the ST gaps by AC0  results in a wrong ordering of states. 
Both AC$_{\rm n}$ and AC1$_{\rm n}$ approaches capture correlation in high orders of $\alpha$ and greatly improve over AC0. The ordering of states is correct and the average error falls below 0.10 eV as compared to the 0.16~eV error of AC0.  AC1$_{\rm n}$ performs slightly better than AC$_{\rm n}$ in terms of MUEs, the errors are 0.06~eV and 0.08~eV, respectively, and significantly better in terms of percentage errors. The improved MU\%E of AC1$_{\rm n}$ (14\% vs.\ 26\%) is due to the good performance of this method for small gaps (systems 1, 3, and 4). These excellent results imply a crucial role of the high-order terms in AC which should enter the correlation energy with high weights.

Table~\ref{tab:biradicals} includes the ftPBE results from Ref.\citenum{stoneburner2018mc}. The latter method performs better than other MC-PDFT\cite{Gagliardi:14} approaches for ST gaps of biradicals. Similarly to AC approximations, MC-PDFT is a post-CASSCF method relying on only 1- and 2-RDMs obtained from CAS. It employs density functional exchange-correlation functionals with modified arguments to describe electron correlation.
As shown in Table~\ref{tab:biradicals}, the accuracy of ST gaps predictions by ftPBE does not match that of the AC1$\rm{_n}$ method, with the percentage error nearly tripled and amounting to 38\%. 
Comparing the computational efficiency of the adiabatic connection and MC-PDFT approximations, AC$\rm{_n}$ (or AC1$\rm{_n}$) formally scale with the 5th power of the system-size which is one order more than scaling of MC-PDFT (timings of both methods are presented in SI). It should be noticed, however, that in the case of both AC$\rm{_n}$ and MC-PDFT the major share of the total computational time is spent on CASSCF calculation. 

The accuracy achieved by AC1$_{\rm n}$ comes close to that of the RASPT2 method.
Comparison of RASPT2 (or CASPT2\cite{Stoneburner2017}) results with those of AC$_{\rm n}$ requires some care. These perturbation methods involve parameters to remove intruder states and to compensate their tendency to underestimate gap energies between closed- and open-shell states.~\cite{ghigo2004modified} The default value of the ionization potential-electron affinity shift\cite{roca2012multiconfiguration}  
used in Ref.\citenum{Stoneburner2017} improves ST gaps of biradicals predicted by CASPT2 and RASPT2 methods. In general, however, the shift may be problematic for strongly correlated systems, e.g.\ complexes with transition metals, and their tuning may be required.~\cite{kepenekian2009zeroth, lawson2012accurate}

In summary, we have proposed a computational approach to the correlation energy in complete active space models. The novel AC$_{\rm n}$ formula for the correlation energy is based on a systematic expansion with respect to the adiabatic connection coupling constant $\alpha$.  Application to singlet-triplet gaps of single- and multi-reference systems revealed the need to account for higher-order terms in the $\alpha$-expansion. The AC$_{\rm n}$/AC1$_{\rm n}$ approaches showed a systematic improvement over the first-order AC0 method. The AC1$_{\rm n}$ variant, which maximizes contributions from the high-order terms, was identified as the best-performing AC approximation. Owing to the Cholesky decomposition technique the AC$_{\rm n}$ methods achieve  $\mathcal{O}(N^5)$ scaling of the computational time with the system size. Since they involve only 1- and 2-RDMs, they are well-suited to treat large active spaces. Importantly, the formalism used to derive AC$_{\rm n}$ is not limited to a particular form of the model Hamiltonian ${\hat H}^{(0)}$, thus further improvements in accuracy could be achieved with models other than that assumed in this work.

Compared to other correlation energy methods for strong correlation, AC${\rm_n}$ emerges as having the most favorable accuracy to cost ratio. Advantages of AC${\rm_n}$ over perturbation methods, such as CASPT2 or RASPT2, include not only the ability to treat dozens of active orbitals, but also the lack of parameters and strict size-consistency.~\cite{RintelmanJCP} We believe that the presented development opens new perspectives for meeting the challenge of strong correlation, e.g., by DMRG\cite{reiherDMRG} methods.

\section*{Acknowledgment}
This work was supported by the National Science Center of Poland under grant \\ no.\ 2019/35/B/ST4/01310,
the Charles University in Prague (grant no.\ CZ.02.2.69\slash0.0\slash0.0\slash19\_073\slash0016935),
the Ministry of Education, Youth and Sports of the Czech Republic through the e-INFRA CZ (ID:90140), and by the European Centre of Excellence in Exascale Computing TREX - Targeting Real Chemical Accuracy at the Exascale. This project has received funding from the European Union’s Horizon 2020 - Research and Innovation program - under grant agreement no.\ 952165.

\bibliography{biblio_AC_CD}

\end{document}


\section{Coupling-constant-dependent ERPA\ matrices}

For a given reference wavefunction $\Psi^{\rm ref}$, introduce spin-free 1- and 2-RDMs,
$\gamma$ and $\Gamma$, respectively, as%
\begin{equation}
\gamma_{pq}=\left\langle \Psi^{\rm ref}|\hat{a}_{q_{\alpha}}^{\dag}\hat
{a}_{p_{\alpha}}|\Psi^{\rm ref}\right\rangle +\left\langle \Psi^{\rm ref}|\hat
{a}_{q_{\beta}}^{\dag}\hat{a}_{p_{\beta}}|\Psi^{\rm ref}\right\rangle
\end{equation}
and 
\begin{align}
2\Gamma_{pqrs}  & =\left\langle \Psi^{\rm ref}|\hat{a}_{r_{\alpha}}^{\dag}\hat
{a}_{s_{\alpha}}^{\dag}\hat{a}_{q_{\alpha}}\hat{a}_{p_{\alpha}}|\Psi
^{\rm ref}\right\rangle +\left\langle \Psi^{\rm ref}|\hat{a}_{r_{\alpha}}^{\dag}%
\hat{a}_{s_{\beta}}^{\dag}\hat{a}_{q_{\beta}}\hat{a}_{p_{\alpha}}|\Psi
^{\rm ref}\right\rangle \nonumber\\
& +\left\langle \Psi^{\rm ref}|\hat{a}_{r_{\beta}}^{\dag}\hat{a}_{s_{\beta}}%
^{\dag}\hat{a}_{q_{\beta}}\hat{a}_{p_{\beta}}|\Psi^{\rm ref}\right\rangle
+\left\langle \Psi^{\rm ref}|\hat{a}_{r_{\beta}}^{\dag}\hat{a}_{s_{\alpha}}^{\dag
}\hat{a}_{q_{\alpha}}\hat{a}_{p_{\beta}}|\Psi^{\rm ref}\right\rangle \ \ \ .
\end{align}
Using the representation of the natural orbitals
\begin{equation}
\gamma_{pq}=2\ \delta_{pq}n_{p}%
\end{equation}
(notice that $\forall_{p}\ 0\leq n_{p}\leq1$) the spin-free ERPA\ matrices%
\begin{align}
\mathbf{\mathcal{A}}_{+}^{\alpha}  & =\mathcal{N}^{-1/2}\left(
\mathbf{\mathcal{A}}^{\alpha}+\mathbf{\mathcal{B}}^{\alpha}\right)
\mathcal{N}^{-1/2}\label{Ap}\\
\mathbf{\mathcal{A}}_{-}^{\alpha}  & =\mathcal{N}^{-1/2}\left(
\mathbf{\mathcal{A}}^{\alpha}-\mathbf{\mathcal{B}}^{\alpha}\right)
\mathcal{N}^{-1/2}\label{Am}%
\end{align}%
\begin{equation}
\forall_{\substack{p>q\\r>s}}\ \ \ \mathcal{N}_{pq,rs}=\delta_{pr}\delta
_{qs}(n_{p}-n_{q})
\end{equation}
are given explicitly in terms of 1- and 2-RDMs and
the pertinent expression reads%
\begin{align}
\forall_{pqrs}\ \ \ \left[  \mathbf{\mathcal{A}}^{\alpha}\right]  _{pq,rs} &
=\left[  \mathbf{\mathcal{B}}^{\alpha}\right]  _{pq,sr}\nonumber\\
&  =h_{sq}^{\alpha}\delta_{pr}(n_{p}-n_{s})+h_{pr}^{\alpha}\delta_{sq}%
(n_{q}-n_{r})\nonumber\\
&  +\sum_{tu}\widetilde{\left\langle st|qu\right\rangle _{\alpha}}%
\ \Gamma_{purt}+\sum_{tu}\widetilde{\left\langle st|uq\right\rangle _{\alpha}%
}\ \Gamma_{putr}\nonumber\\
&  +\sum_{tu}\widetilde{\left\langle up|tr\right\rangle _{\alpha}}%
\ \Gamma_{stqu}+\sum_{tu}\widetilde{\left\langle up|rt\right\rangle _{\alpha}%
}\ \Gamma_{stuq}\nonumber\\
&  -\sum_{tu}\widetilde{\left\langle ps|tu\right\rangle _{\alpha}}%
\ \Gamma_{tuqr}-\sum_{tu}\widetilde{\left\langle tu|qr\right\rangle _{\alpha}%
}\ \Gamma_{sput}\nonumber\\
&  -\frac{1}{2}\delta_{sq}\sum_{twu}\widetilde{\left\langle tp|wu\right\rangle
_{\alpha}}\ \Gamma_{wutr}-\frac{1}{2}\delta_{sq}\sum_{twu}\widetilde
{\left\langle tp|uw\right\rangle _{\alpha}}\ \Gamma_{wurt}\nonumber\\
&  -\frac{1}{2}\delta_{pr}\sum_{tuw}\widetilde{\left\langle tu|wq\right\rangle
_{\alpha}}\ \Gamma_{swut}-\frac{1}{2}\delta_{pr}\sum_{tuw}\widetilde
{\left\langle tu|qw\right\rangle _{\alpha}}\ \Gamma_{swtu}\ \ \ .\label{ab}%
\end{align}
The coupling-constant-dependent modified one- and two-electron integrals are
defined as%
\begin{align}
h_{pq}^{\alpha} &  =\alpha h_{pq}+\delta_{I_{p}I_{q}}(1-\alpha)h_{pq}%
^{ \rm eff}\ \ \ ,\label{h-alpha}\\
h_{pq}^{\rm eff} &  =h_{pq}+\sum_{J\neq I}\sum_{r\in J}n_{r}\left[  2\left\langle
pr|qr\right\rangle -\left\langle pr|rq\right\rangle \right]  \ \ \ .\\
\widetilde{\left\langle pq|rs\right\rangle _{\alpha}} &  =\left[
\alpha+\delta_{I_{r}I_{s}}\delta_{I_{s}I_{p}}\delta_{I_{p}I_{q}}%
(1-\alpha)\right]  \ \left\langle pq|rs\right\rangle \ \ \ ,\label{g-alpha}%
\end{align}
the symbol $\delta_{I_{p}I_{q}}$ stands for $1$ if orbitals $p$ and $q$ are from the same orbital group (both are inactive, occupied or virtual) and $0$ otherwise.

The matrices $\mathbf{\mathcal{A}}_{\pm}^{(0)}$ and $\mathbf{\mathcal{A}}%
_{\pm}^{(1)}$ needed to find the AC correlation energy using the iterative
scheme and the Cholesky decomposition, see Eqs.(35)-(39) in the main text,
follow immediately from Eqs.(\ref{Ap}), (\ref{Am}), namely
\begin{align}
\mathbf{\mathcal{A}}_{\pm}^{(0)}  & =\mathbf{\mathcal{A}}^{(0)}\pm
\mathbf{\mathcal{B}}^{(0)}\ \ \ ,\\
\mathbf{\mathcal{A}}_{\pm}^{(1)}  & =\mathbf{\mathcal{A}}^{(1)}\pm
\mathbf{\mathcal{B}}^{(1)}\ \ \ ,
\end{align}
where the matrices $\mathbf{\mathcal{A}}^{(0)}$, $\mathbf{\mathcal{B}}^{(0)}$
are obtained from Eq.(\ref{ab}) be replacing $h_{pq}^{\alpha}$ and
$\widetilde{\left\langle pq|rs\right\rangle _{\alpha}}$ integrals with their
zero-order counterparts $h_{pq}^{(0)}$, $\widetilde{\left\langle
pq|rs\right\rangle^{(0)}}$ reading%
\begin{align}
h_{pq}^{(0)}  & =\delta_{I_{p}I_{q}}h_{pq}^{\rm eff}\ \ \ ,\\
\widetilde{\left\langle pq|rs\right\rangle^{(0)}} & =\delta_{I_{r}I_{s}%
}\delta_{I_{s}I_{p}}\delta_{I_{p}I_{q}}\ \left\langle pq|rs\right\rangle
\ \ \ .
\end{align}
Similarly, first-order matrices $\mathbf{\mathcal{A}}^{(1)}$,
$\mathbf{\mathcal{B}}^{(1)}$ follow from Eq.(\ref{ab}) if the elements
\begin{align}
h_{pq}^{(1)}  & =\alpha h_{pq}-\delta_{I_{p}I_{q}}h_{pq}^{\rm eff}\ \ \ ,\\
\widetilde{\left\langle pq|rs\right\rangle }^{(1)}  & =\left(  1-\delta
_{I_{r}I_{s}}\delta_{I_{s}I_{p}}\delta_{I_{p}I_{q}}\right)  \ \left\langle
pq|rs\right\rangle \ \ \ .
\end{align}
replace $h_{pq}^{\alpha}$ and $\widetilde{\left\langle pq|rs\right\rangle
_{\alpha}}$, respectively.

\newpage

\section{Frequency-dependent density-density linear response function in the
ERPA approximation}

Begin with the coupling-constant dependent ERPA\ problem%
\begin{equation}
\left(
\begin{array}
[c]{cc}%
\mathbf{\mathcal{A}}_{-}^{\alpha} & 0\\
0 & \mathbf{\mathcal{A}}_{+}^{\alpha}%
\end{array}
\right)  \left(
\begin{array}
[c]{c}%
\mathbf{\tilde{Y}}_{\nu}^{\alpha}\\
\mathbf{\tilde{X}}_{\nu}^{\alpha}%
\end{array}
\right)  =\omega_{\nu}^{\alpha}\left(
\begin{array}
[c]{cc}%
\mathbf{0} & \mathbf{1}\\
\mathbf{1} & \mathbf{0}%
\end{array}
\right)  \left(
\begin{array}
[c]{c}%
\mathbf{\tilde{Y}}_{\nu}^{\alpha}\\
\mathbf{\tilde{X}}_{\nu}^{\alpha}%
\end{array}
\right)  \ \ \ ,
\end{equation}
where the eigenvectors are orthonormal in the following sense%
\begin{equation}
\forall_{\mu,\nu}\ \ \ 2\mathbf{\tilde{Y}}_{\mu}^{\alpha}\mathbf{\tilde{X}%
}_{\nu}^{\alpha}=\delta_{\mu\nu}\label{norm1}%
\end{equation}
The matrices $\mathbf{\mathcal{A}}_{\pm}^{\alpha}$, defined in Eqs.(\ref{Ap}%
)-(\ref{ab}), are assumed to be positive definite. This condition is met
if the reference wavefunction satisfies the Hellman-Feynman theorem, since in
such a case the EPRA\ matrices are equivalent to the Hessian matrix describing
variations of energy with respect to orbital rotations. It can be shown [see
theorem A.5 in Furche, JCP 114, 5982 (2001)] that spectral decomposition of
the matrices $\mathbf{\mathcal{A}}_{\pm}^{\alpha}$ is given by eigenvalues
and eigenvectors
\begin{align}
\mathbf{\bar{X}}_{\nu}^{\alpha}  & =\sqrt{2}\mathbf{\tilde{X}}_{\nu}^{\alpha
}\label{x}\\
\mathbf{\bar{Y}}_{\nu}^{\alpha}  & =\sqrt{2}\mathbf{\tilde{Y}}_{\nu}^{\alpha
}\label{y}%
\end{align}
orthonormal in the following sense
\begin{equation}
\forall_{\mu,\nu}\ \ \ \mathbf{\bar{Y}}_{\mu}^{\alpha}\mathbf{\bar{X}}_{\nu
}^{\alpha}=\delta_{\mu\nu}\label{norm2}%
\end{equation}
as%

\begin{equation}
\mathbf{\mathcal{A}}_{+}^{\alpha}=\left(
\begin{array}
[c]{ccc}%
\mathbf{\bar{Y}}_{1}^{\alpha} & \mathbf{\bar{Y}}_{2}^{\alpha} & \cdots
\end{array}
\right)  \left(
\begin{array}
[c]{ccc}%
\omega_{1}^{\alpha} & 0 & \cdots\\
0 & \omega_{2}^{\alpha} & \cdots\\
\cdots & \cdots & \cdots
\end{array}
\right)  \left(
\begin{array}
[c]{c}%
\left[  \mathbf{\bar{Y}}_{1}^{\alpha}\right]  ^{\text{T}}\\
\left[  \mathbf{\bar{Y}}_{2}^{\alpha}\right]  ^{\text{T}}\\
\cdots
\end{array}
\right)
\end{equation}
and%
\begin{equation}
\mathbf{\mathcal{A}}_{-}^{\alpha}=\left(
\begin{array}
[c]{ccc}%
\mathbf{\bar{X}}_{1}^{\alpha} & \mathbf{\bar{X}}_{2}^{\alpha} & \cdots
\end{array}
\right)  \left(
\begin{array}
[c]{ccc}%
\omega_{1}^{\alpha} & 0 & \cdots\\
0 & \omega_{2}^{\alpha} & \cdots\\
\cdots & \cdots & \cdots
\end{array}
\right)  \left(
\begin{array}
[c]{c}%
\left[  \mathbf{\bar{X}}_{1}^{\alpha}\right]  ^{\text{T}}\\
\left[  \mathbf{\bar{X}}_{2}^{\alpha}\right]  ^{\text{T}}\\
\cdots
\end{array}
\right)  \ \ \ .
\end{equation}
Using the orthonormality condition, Eq.(\ref{norm2}), and Eqs.(\ref{x}%
)-(\ref{y}), it can be checked by inspection that a matrix $\mathbf{C}%
^{\alpha}\mathbf{(}\omega)$ satisfying the following equation%
\begin{equation}
\left[  \mathbf{\mathcal{A}}_{+}^{\alpha}\mathbf{\mathcal{A}}_{-}^{\alpha
}+\omega^{2}\mathbf{1}\right]  \mathbf{C}^{\alpha}\mathbf{(}\omega
)=\mathbf{\mathcal{A}}_{+}^{\alpha}%
\end{equation}
is the dynamic linear response function, see Eq.(21) in the main text,
\begin{equation}
\forall_{\substack{p>q\\r>s}}\ \ \ C_{pq,rs}^{\alpha}(\omega)=2\sum_{\nu
}\left[  \mathbf{\tilde{Y}}_{\nu}^{\alpha}\right]  _{pq}\frac{\omega_{\nu}%
}{\left(  \omega_{\nu}\right)  ^{2}+\omega^{2}}\left[  \mathbf{\tilde{Y}}%
_{\nu}^{\alpha}\right]  _{rs}\ \ \ .
\end{equation}

\newpage

\begin{longtable}{cccccccc}
\caption{Convergence check for biradicals molecules. Energies in [Ha], ST gaps (differences between triplet and singlet energies) in [eV]. Systems (Sys.): 1 -- \(\mathrm{C_{4}H_{4}}\), 2 -- \(\mathrm{C_{5}H_{5}^{+}}\), 3 -- \(\mathrm{C_{4}H_{3}NH_{2}}\), 4 -- \(\mathrm{C_{4}H_{3}CHO}\), 5 -- \(\mathrm{C_{4}H_{2}NH_{2}(CHO)}\), 6 -- C$_{4}$H$_{2}$-1,2-(CH$_{2}$)$_{2}$, 7 -- 
C$_{4}$H$_{2}$-1,3-(CH$_{2}$)$_{2}$, 8 -- divergent case of \(\mathrm{C_{5}H_{5}^{+}}\) and CAS(2,4).}
\label{tab:convergence}\\

\endfirsthead
\hline
\toprule
Sys.              & n  & AC$\mathrm{_{n}}$(S)  & AC$\mathrm{_{n}}$(T)  & AC$\mathrm{_{n}}$(S)-AC(S)   & AC$\mathrm{_{n}}$(T)-AC(T)  & AC$\mathrm{_{n}}$(ST)  & AC$\mathrm{_{n}}$(ST)-AC(ST)  \\
  \hline\hline
\endhead
\toprule
Sys.              & n  & AC$\mathrm{_{n}}$(S)  & AC$\mathrm{_{n}}$(T)  & AC$\mathrm{_{n}}$(S)-AC(S)   & AC$\mathrm{_{n}}$(T)-AC(T)  & AC$\mathrm{_{n}}$(ST)  & AC$\mathrm{_{n}}$(ST)-AC(ST)  \\
\hline\hline
\multirow{10}{*}{1} & 1           & -154.38675       & -154.38681       & -0.01993  & -0.02478  & -0.00144     & -0.13201        \\
                    & 2           & -154.35211       & -154.36192       & 0.01471   & 0.00011   & -0.26684     & -0.39741        \\
                    & 3           & -154.37064       & -154.36790       & -0.00381  & -0.00587  & 0.07450      & -0.05607        \\
                    & 4           & -154.36520       & -154.35905       & 0.00162   & 0.00297   & 0.16744      & 0.03687         \\
                    & 5           & -154.36751       & -154.36354       & -0.00069  & -0.00151  & 0.10811      & -0.02246        \\
                    & 7           & -154.36697       & -154.36251       & -0.00015  & -0.00049  & 0.12129      & -0.00928        \\
                    & 10          & -154.36681       & -154.36192       & 0.00001   & 0.00011   & 0.13320      & 0.00263         \\
                    & 12          & -154.36682       & -154.36198       & 0.00000   & 0.00004   & 0.13174      & 0.00117         \\
                    & 15          & -154.36683       & -154.36204       & 0.00000   & -0.00002  & 0.13021      & -0.00036        \\
                    & AC & -154.36682       & -154.36202       & -   & -   & 0.13057      & -         \\
                    \hline
\multirow{10}{*}{2} & 1           & -192.78598       & -192.81767       & -0.02010  & -0.02386  & -0.86239     & -0.10249        \\
                    & 2           & -192.74597       & -192.77359       & 0.01992   & 0.02022   & -0.75165     & 0.00824         \\
                    & 3           & -192.77033       & -192.79950       & -0.00444  & -0.00569  & -0.79389     & -0.03400        \\
                    & 4           & -192.76255       & -192.79086       & 0.00333   & 0.00295   & -0.77025     & -0.01035        \\
                    & 5           & -192.76682       & -192.79519       & -0.00094  & -0.00139  & -0.77207     & -0.01217        \\
                    & 7           & -192.76610       & -192.79422       & -0.00022  & -0.00041  & -0.76512     & -0.00523        \\
                    & 10          & -192.76553       & -192.79372       & 0.00035   & 0.00009   & -0.76691     & -0.00702        \\
                    & 12          & -192.76565       & -192.79377       & 0.00023   & 0.00004   & -0.76510     & -0.00521        \\
                    & 15          & -192.76586       & -192.79381       & 0.00003   & 0.00000  & -0.76059     & -0.00069        \\
                    & AC & -192.76588       & -192.79381       & -   & -   & -0.75989     & -         \\
                    \hline
\multirow{10}{*}{3} & 1           & -209.68148       & -209.68454       & -0.03926  & -0.04462  & -0.08319     & -0.14584        \\
                    & 2           & -209.61974       & -209.61310       & 0.02248   & 0.02682   & 0.18065      & 0.11800         \\
                    & 3           & -209.64932       & -209.64934       & -0.00709  & -0.00942  & -0.00061     & -0.06326        \\
                    & 4           & -209.63933       & -209.63531       & 0.00289   & 0.00461   & 0.10950      & 0.04685         \\
                    & 5           & -209.64365       & -209.64235       & -0.00142  & -0.00243  & 0.03523      & -0.02742        \\
                    & 7           & -209.64256       & -209.64071       & -0.00033  & -0.00079  & 0.05023      & -0.01242        \\
                    & 10          & -209.64216       & -209.63970       & 0.00006   & 0.00022   & 0.06702      & 0.00437         \\
                    & 12          & -209.64219       & -209.63981       & 0.00003   & 0.00011   & 0.06479      & 0.00214         \\
                    & 15          & -209.64220       & -209.63993       & 0.00002   & 0.00000   & 0.06193      & -0.00072        \\
                    & AC & -209.64222       & -209.63992       & -   & -   & 0.06265      & -         \\
                    \hline
\multirow{10}{*}{4} & 1           & -267.57835       & -267.57912       & -0.06117  & -0.06606  & -0.02111     & -0.13292        \\
                    & 2           & -267.48330       & -267.47536       & 0.03387   & 0.03771   & 0.21615      & 0.10434         \\
                    & 3           & -267.53097       & -267.52893       & -0.01379  & -0.01587  & 0.05536      & -0.05646        \\
                    & 4           & -267.51019       & -267.50456       & 0.00699   & 0.00851   & 0.15309      & 0.04128         \\
                    & 5           & -267.52136       & -267.51816       & -0.00418  & -0.00509  & 0.08705      & -0.02476        \\
                    & 7           & -267.51880       & -267.51512       & -0.00162  & -0.00205  & 0.09996      & -0.01185        \\
                    & 10          & -267.51662       & -267.51233       & 0.00056   & 0.00074   & 0.11661      & 0.00480         \\
                    & 12          & -267.51686       & -267.51265       & 0.00032   & 0.00042   & 0.11454      & 0.00273         \\
                    & 15          & -267.51724       & -267.51318       & -0.00006  & -0.00011  & 0.11055      & -0.00126        \\
                    & AC & -267.51718       & -267.51307       & -   & -   & 0.11181      & -        \\
                    \hline
\multirow{10}{*}{5} & 1           & -322.88405       & -322.87894       & -0.07863  & -0.08765  & 0.13894      & -0.24570        \\
                    & 2           & -322.75998       & -322.74413       & 0.04545   & 0.04716   & 0.43122      & 0.04658         \\
                    & 3           & -322.82401       & -322.81120       & -0.01858  & -0.01991  & 0.34854      & -0.03610        \\
                    & 4           & -322.79486       & -322.78070       & 0.01057   & 0.01059   & 0.38516      & 0.00052         \\
                    & 5           & -322.81221       & -322.79752       & -0.00679  & -0.00624  & 0.39960      & 0.01496         \\
                    & 7           & -322.80884       & -322.79377       & -0.00341  & -0.00248  & 0.41001      & 0.02537         \\
                    & 10          & -322.80384       & -322.79036       & 0.00159   & 0.00093   & 0.36678      & -0.01786        \\
                    & 12          & -322.80431       & -322.79075       & 0.00111   & 0.00054   & 0.36910      & -0.01553        \\
                    & 15          & -322.80606       & -322.79141       & -0.00064  & -0.00012  & 0.39881      & 0.01417         \\
                    & AC & -322.80542       & -322.79129       & -   & -   & 0.38464      & -        \\
                    \hline
\multirow{10}{*}{6} & 1           & -231.74407       & -231.61704       & -0.03381  & -0.03307  & 3.45658      & 0.02014         \\
                    & 2           & -231.68434       & -231.56013       & 0.02591   & 0.02384   & 3.38011      & -0.05632        \\
                    & 3           & -231.71815       & -231.59113       & -0.00790  & -0.00716  & 3.45641      & 0.01998         \\
                    & 4           & -231.70636       & -231.58071       & 0.00389   & 0.00326   & 3.41918      & -0.01725        \\
                    & 5           & -231.71231       & -231.58559       & -0.00206  & -0.00163  & 3.44816      & 0.01173         \\
                    & 7           & -231.71092       & -231.58440       & -0.00067  & -0.00043  & 3.44276      & 0.00632         \\
                    & 10          & -231.71012       & -231.58386       & 0.00014   & 0.00011   & 3.43570      & -0.00074        \\
                    & 12          & -231.71018       & -231.58390       & 0.00007   & 0.00007   & 3.43621      & -0.00022        \\
                    & 15          & -231.71024       & -231.58394       & 0.00001   & 0.00003   & 3.43697      & 0.00054         \\
                    & AC & -231.71026       & -231.58397       & -   & -   & 3.43643      & -         \\
                    \hline
\multirow{10}{*}{7} & 1           & -231.64530       & -231.67913       & -0.03302  & -0.03372  & -0.92055     & -0.01902        \\
                    & 2           & -231.58760       & -231.62092       & 0.02469   & 0.02450   & -0.90673     & -0.00520        \\
                    & 3           & -231.61972       & -231.65290       & -0.00744  & -0.00749  & -0.90286     & -0.00133        \\
                    & 4           & -231.60880       & -231.64197       & 0.00348   & 0.00344   & -0.90265     & -0.00112        \\
                    & 5           & -231.61416       & -231.64717       & -0.00187  & -0.00176  & -0.89839     & 0.00314         \\
                    & 7           & -231.61288       & -231.64591       & -0.00060  & -0.00049  & -0.89866     & 0.00288         \\
                    & 10          & -231.61219       & -231.64531       & 0.00010   & 0.00011   & -0.90132     & 0.00021         \\
                    & 12          & -231.61224       & -231.64536       & 0.00005   & 0.00006   & -0.90128     & 0.00025         \\
                    & 15          & -231.61228       & -231.64540       & 0.00000   & 0.00002   & -0.90104     & 0.00050         \\
                    & AC & -231.61228       & -231.64541       & -   & -   & -0.90153     & - \\
                    \hline
\multirow{11}{*}{8} & 1           & -192.79829       & -192.82560       & -0.04506   & -0.03862   & -0.74314     &  0.17552        \\
                    & 2           & -192.72167       & -192.75934       &  0.03156   &  0.02764   & -1.02505     & -0.10639        \\
                    & 3           & -192.77238       & -192.79638       & -0.01916   & -0.00940   & -0.65307     &  0.26559        \\
                    & 4           & -192.74718       & -192.78110       &  0.00605   &  0.00589   & -0.92301     & -0.00435        \\
                    & 5           & -192.76827       & -192.78992       & -0.01505   & -0.00294   & -0.58913     &  0.32953        \\
                    & 6           & -192.75531       & -192.78477       & -0.00208   &  0.00222   & -0.80165     &  0.11701        \\
                    & 7           & -192.77100       & -192.78820       & -0.01778   & -0.00121   & -0.46776     &  0.45090        \\
                    & 8           & -192.76309       & -192.78593       & -0.00986   &  0.00106   & -0.62151     &  0.29715        \\
                    & 9           & -192.77912       & -192.78757       & -0.02589   & -0.00058   & -0.23021     &  0.68845        \\
                    & 10          & -192.77532       & -192.78642       & -0.02209   &  0.00057   & -0.30205     & 0.61661         \\
                    & AC          & -192.75323       & -192.78699       & -          & -          & -0.91866     & -               \\
                    \hline\hline
\end{longtable}

\newpage

\section{Timings}

Table \ref{tab:timings} shows timings for AC, AC$_{n}$ and AC0 correlation energy calculations for singlet energy of C$_{4}$H$_{2}$-1,3-(CH$_{2}$)$_{2}$ biradical in two basis sets: aug-cc-pVTZ (aTZ) and aug-cc-pVQZ (aQZ). 

\begin{table}[h!]
\caption{CPU times of computing correlation energy for  C$_{4}$H$_{2}$-1,3-(CH$_{2}$)$_{2}$ in two basis sets: aug-cc-pVTZ (aTZ) and aug-cc-pVQZ (aQZ). 
AC$\mathrm{_{n}}$ corresponds to n=10. 
The timings were measured on a 4 core Intel(R) Xeon(R) Gold 6240 CPU  machine clocked at 2.60 GHz using a threaded MKL BLAS library.}
\begin{tabular}{ccccccc}
             & \multicolumn{2}{c}{AC}        & \multicolumn{2}{c}{AC$\mathrm{_{n}}$}  & \multicolumn{2}{c}{AC0} \\
             \hline\hline
active space & aTZ          & aQZ          & aTZ      & aQZ          & aTZ     & aQZ         \\
\hline
(4,4)          & 3h 14m 3s & 19h 55m 18s & 42m  9s   & 4h  1m 17s  & 22s     & 1m 26s     \\
(6,6)          & 3h 54m 36s & 21h 38m 33s & 46m 46s   & 4h  9m 57s  & 28s     & 1m 36s     \\
(8,8)          & 3h 56m 39s & 24h 52m 45s & 50m 15s   & 4h 28m  1s  & 37s     & 2m 22s        \\
\hline\hline
\end{tabular}
\label{tab:timings}
\end{table}


\newpage

\section{DMRG single-orbital entropies and mutual information for selected biradicals}

For all studied biradicals we have used geometries from Ref.\citenum{saito2011symmetry} and  aug-cc-pVTZ basis set.\cite{augDunning} For the representative examples of C$_4$H$_4$ and  C$_5$H$_5^+$  we have performed the CASSCF orbital optimizations in the medium size active spaces of Ref.\citenum{Stoneburner2017} (mCPO). 
Moreover, for C$_4$H$_4$ we have also performed DMRG-SCF optimizations in eCPO with fixed bond dimensions $M = 2000$, i.e.\ CAS(20,22), to verify that there is essentially no difference between correlations in mCPO and eCPO. In other words, the additional orbitals not included in mCPO are only weakly correlated. 
Finally, we have computed the DMRG($M_{\text{max}} = 4000$) single-orbital entropies ($s_i$), as well as mutual information ($I_{ij}$) \cite{Szalay2015}, which provide additional information on the correlation structure. 

\begin{figure}[!h]
  \subfloat[singlet state]{%
    \includegraphics[width=0.45\textwidth]{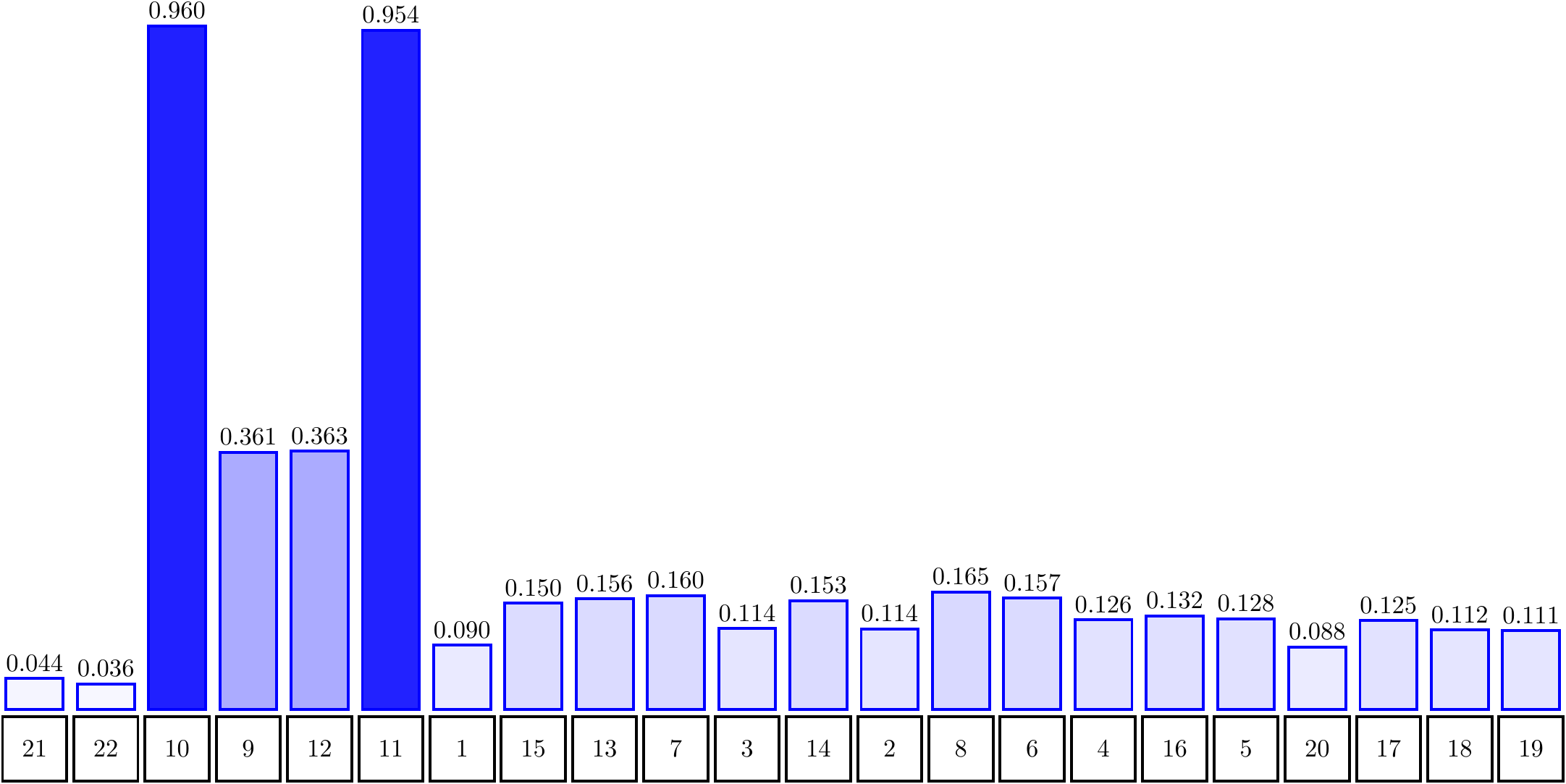}
  }
  \hfill
  \subfloat[triplet state]{%
    \includegraphics[width=0.45\textwidth]{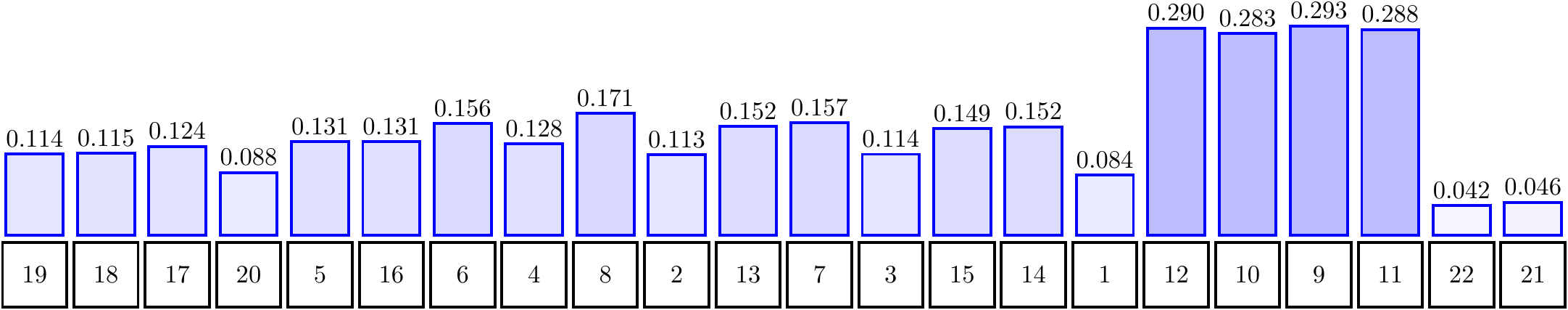}
  }
  \caption{The DMRG single-orbital entropies for C$_4$H$_4$ and eCPO, i.e.\ CAS(20,22). The orbital indices correspond to the ordering from Figures \ref{orbs_c4h4_s} and \ref{orbs_c4h4_t}. \label{c4h4_single_site_eCPO}}
\end{figure}

\begin{figure}[!h]  
  \subfloat[singlet state]{%
    \includegraphics[width=0.45\textwidth]{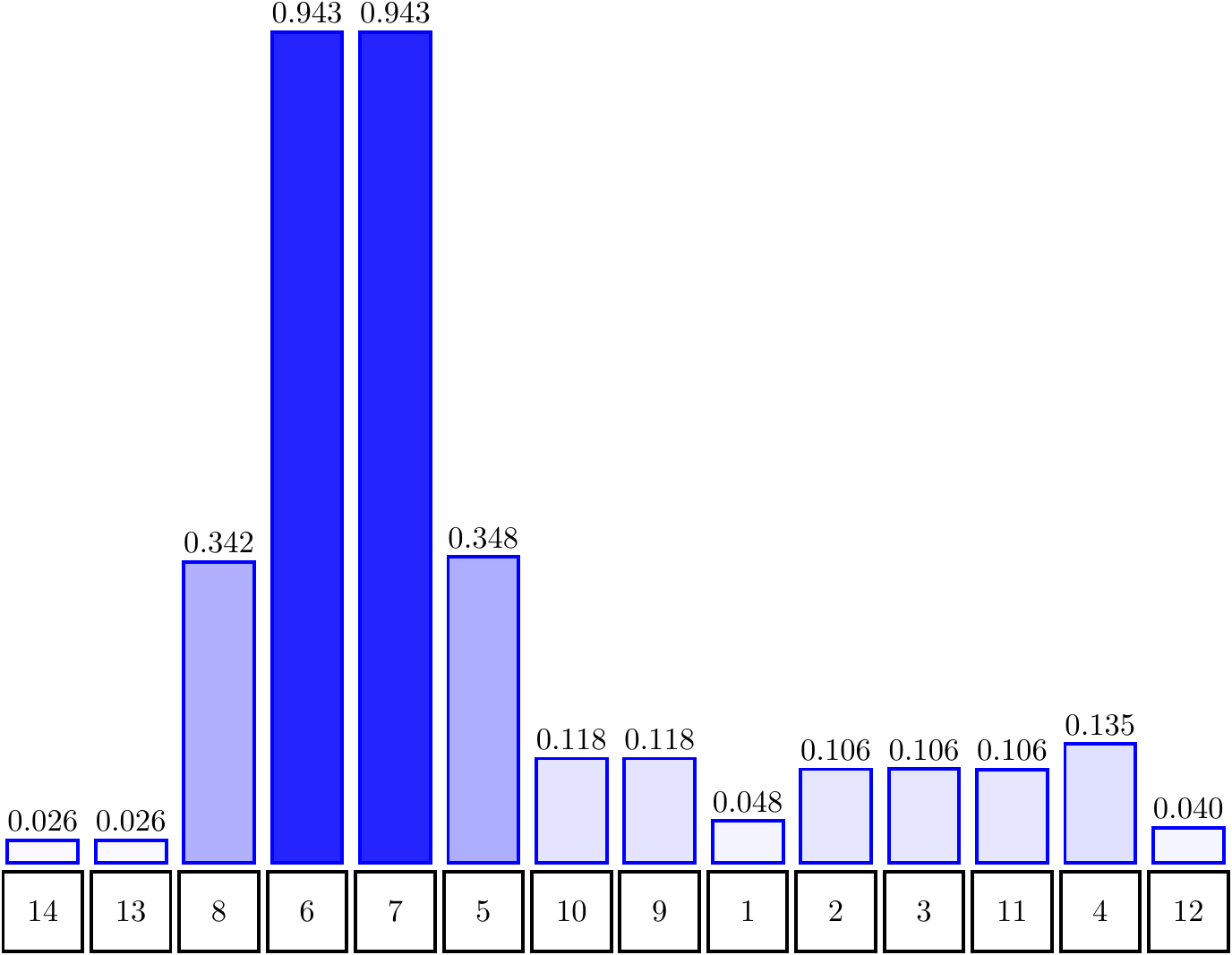}
  }
  \hfill
  \subfloat[triplet state]{%
    \includegraphics[width=0.45\textwidth]{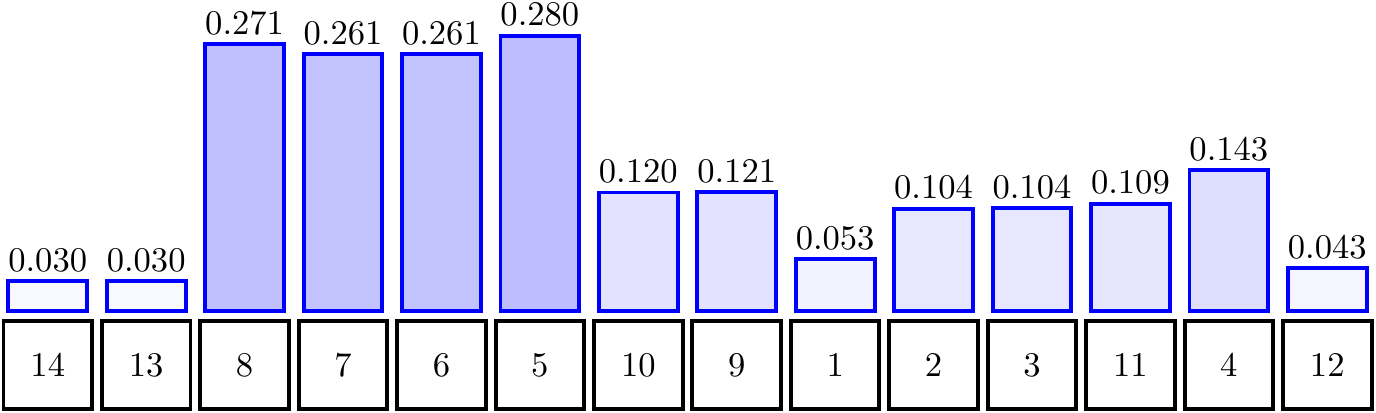}
  }
  \caption{The DMRG single-orbital entropies for C$_4$H$_4$ and mCPO, i.e.\ CAS(12,14). The orbital indices correspond to the ordering from Figures \ref{orbs_c4h4small_s} and \ref{orbs_c4h4small_t}.  \label{c4h4_single_site_mCPO}}
\end{figure}

\begin{figure}[!h]  
  \subfloat[singlet state]{%
    \includegraphics[width=0.45\textwidth]{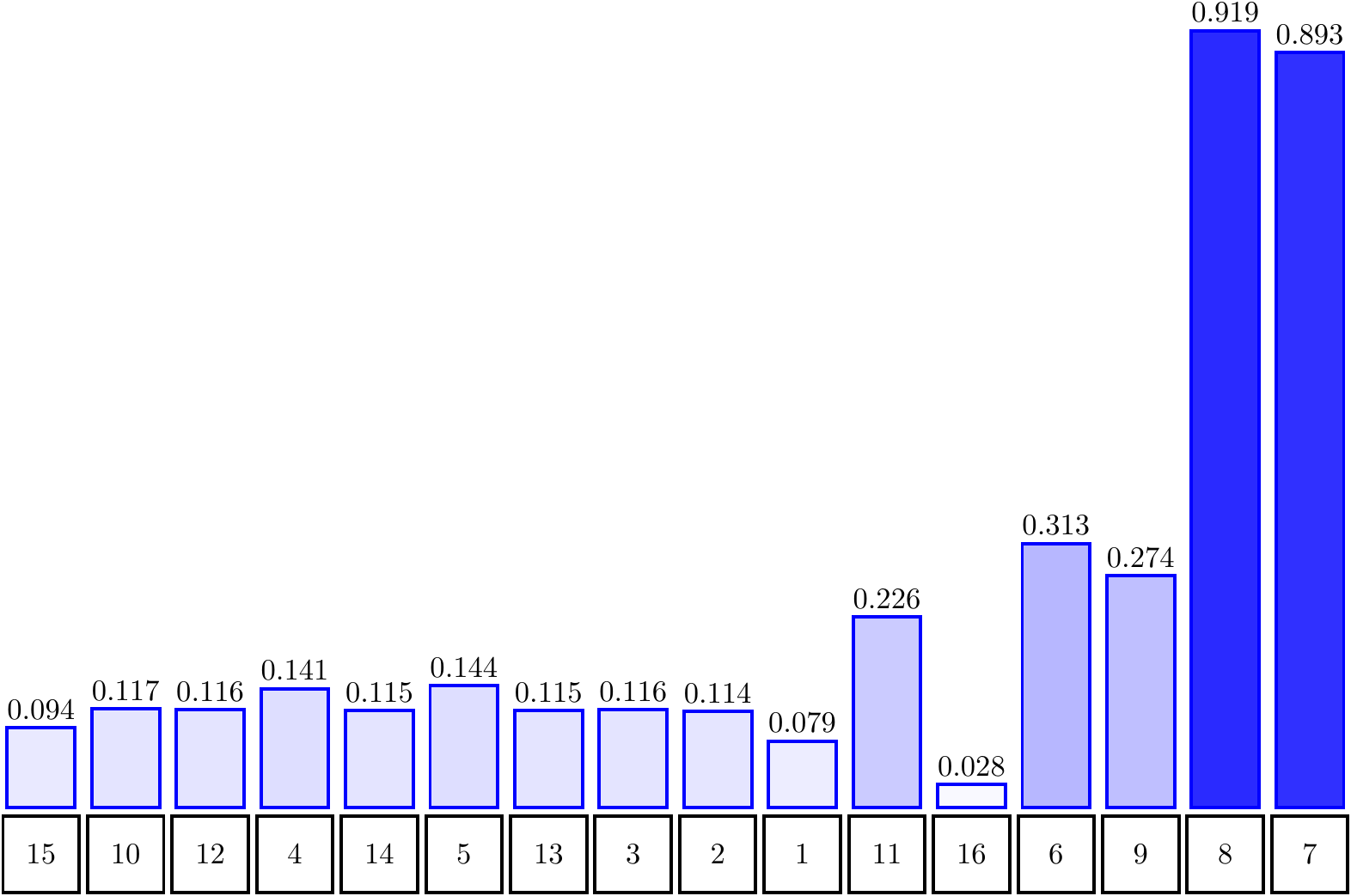}
  }
  \hfill
  \subfloat[triplet state]{%
    \includegraphics[width=0.45\textwidth]{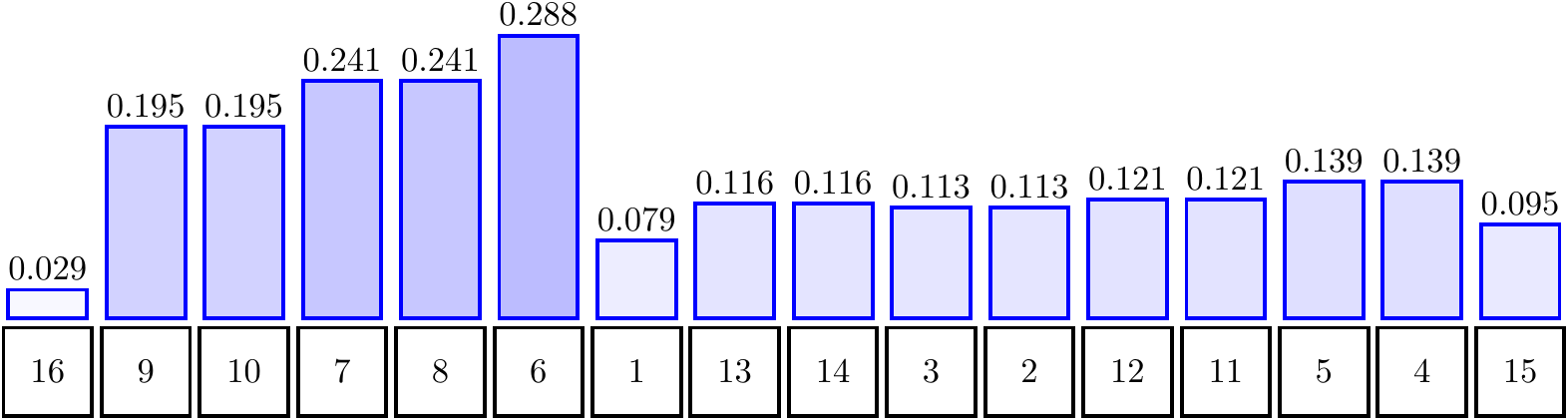}
  }
  \caption{The DMRG single-orbital entropies for C$_5$H$_5^+$ and mCPO, i.e. CAS(14,16). The orbital indices correspond to the ordering from Figures \ref{orbs_c5h5_s} and \ref{orbs_c5h5_t}. \label{c5h5p_single_site}}
\end{figure}

The DMRG single-orbital entropies for C$_4$H$_4$ (eCPO, mCPO) and C$_5$H$_5^+$ (mCPO)
are shown in Figures \ref{c4h4_single_site_eCPO} - \ref{c5h5p_single_site}.

\renewcommand{\thesubfigure}{\alph{subfigure}}



\newpage

\newpage

\section{CASSCF natural orbitals for selected biradicals}

\begin{figure}[!h]
  \subfloat[$n_{\text{occup}}$ = 1.98]{%
    \includegraphics[width=0.22\textwidth]{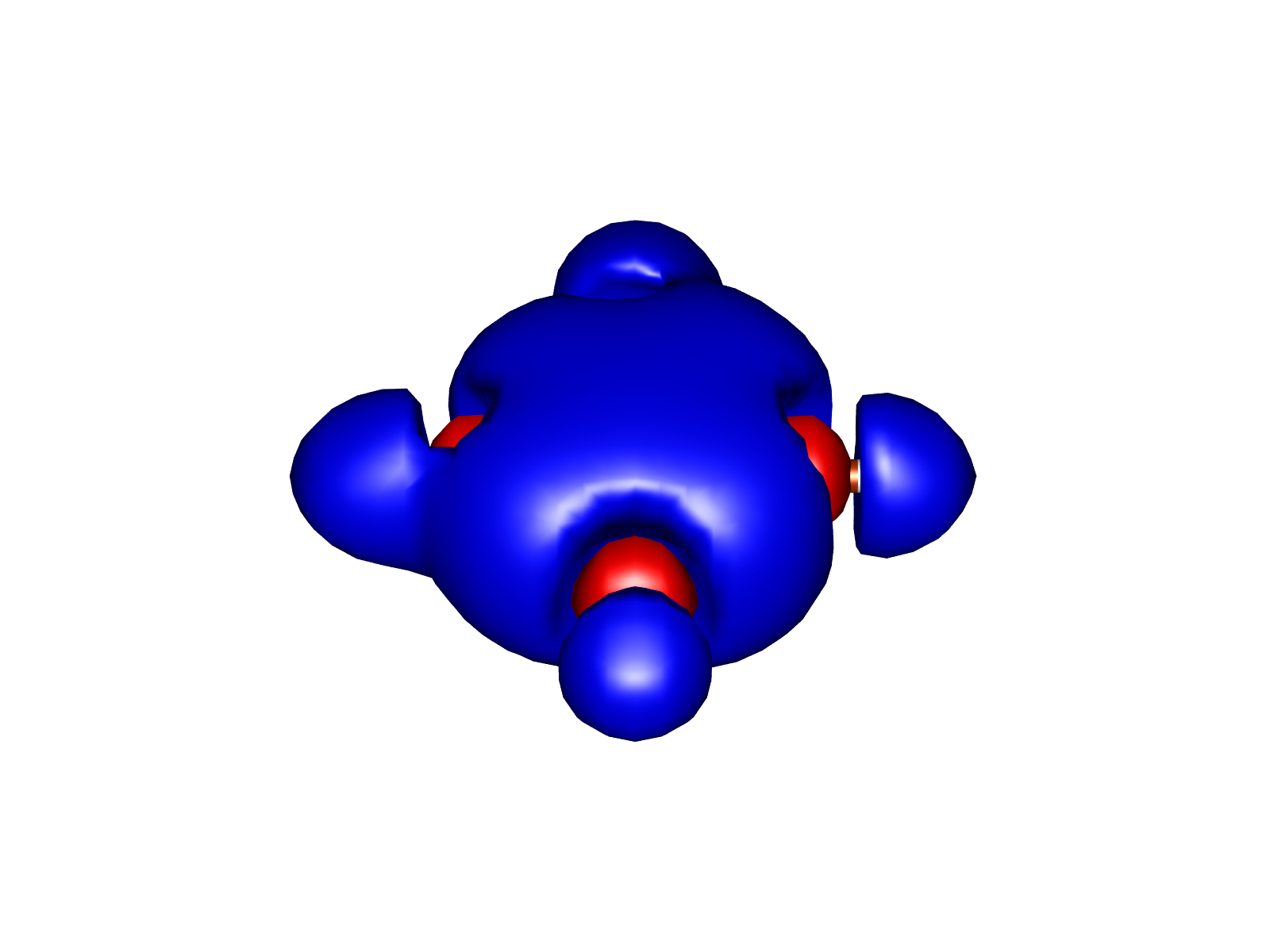}
  }
  \hfill
  \subfloat[$n_{\text{occup}}$ = 1.98]{%
    \includegraphics[width=0.22\textwidth]{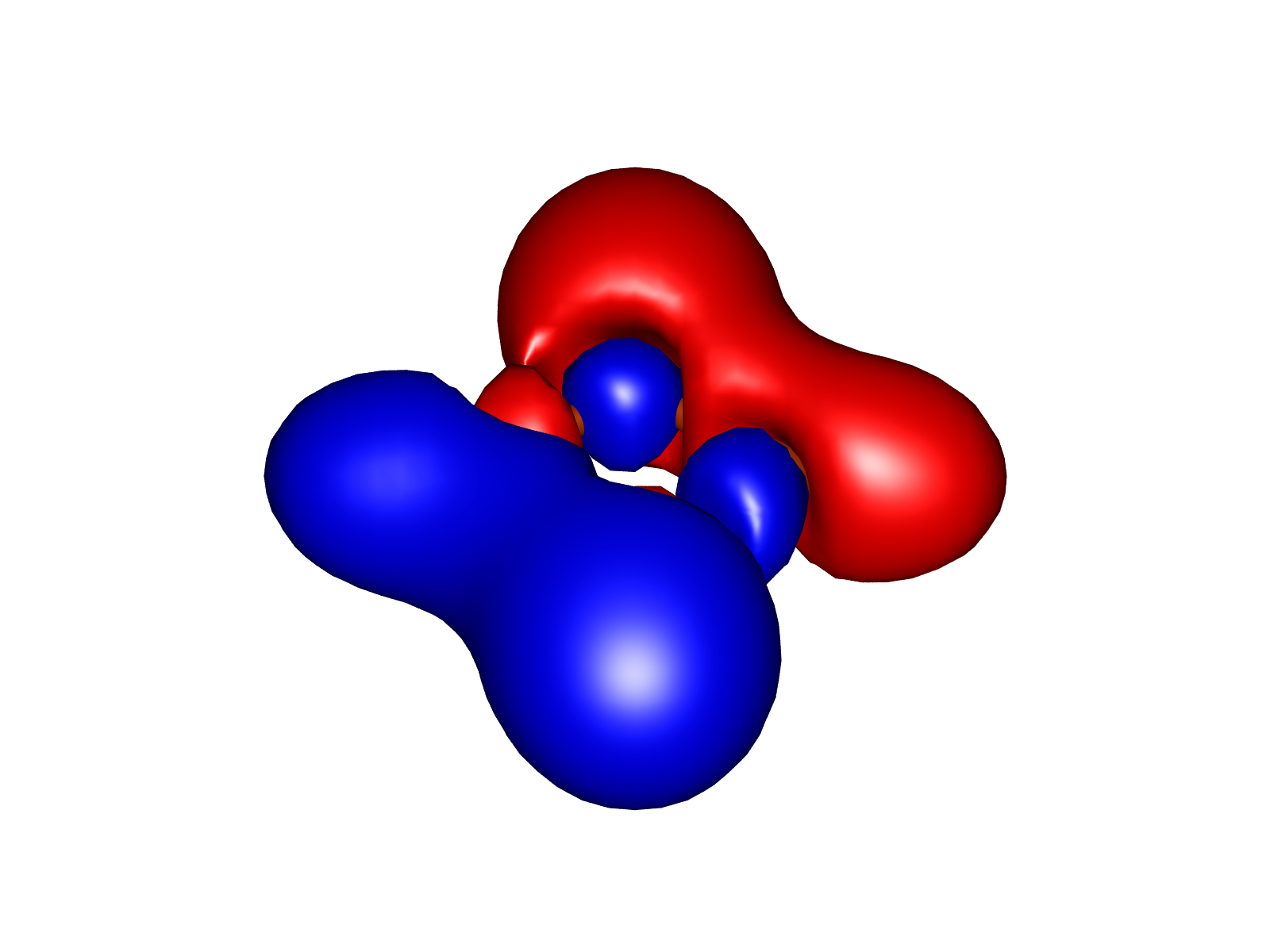}
  }
  \hfill
  \subfloat[$n_{\text{occup}}$ = 1.98]{%
    \includegraphics[width=0.22\textwidth]{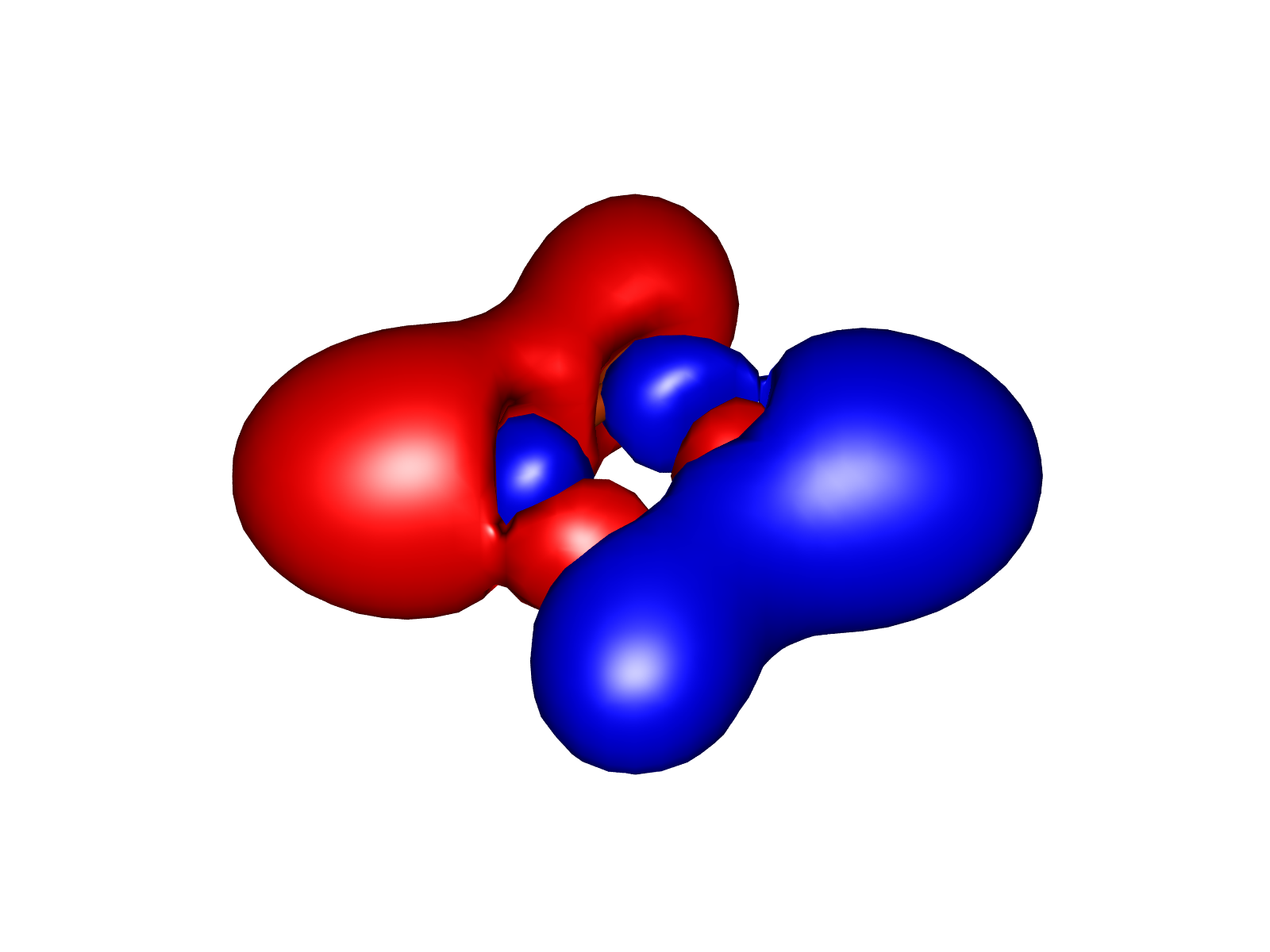}
  }
  \hfill
  \subfloat[$n_{\text{occup}}$ = 1.98]{%
    \includegraphics[width=0.22\textwidth]{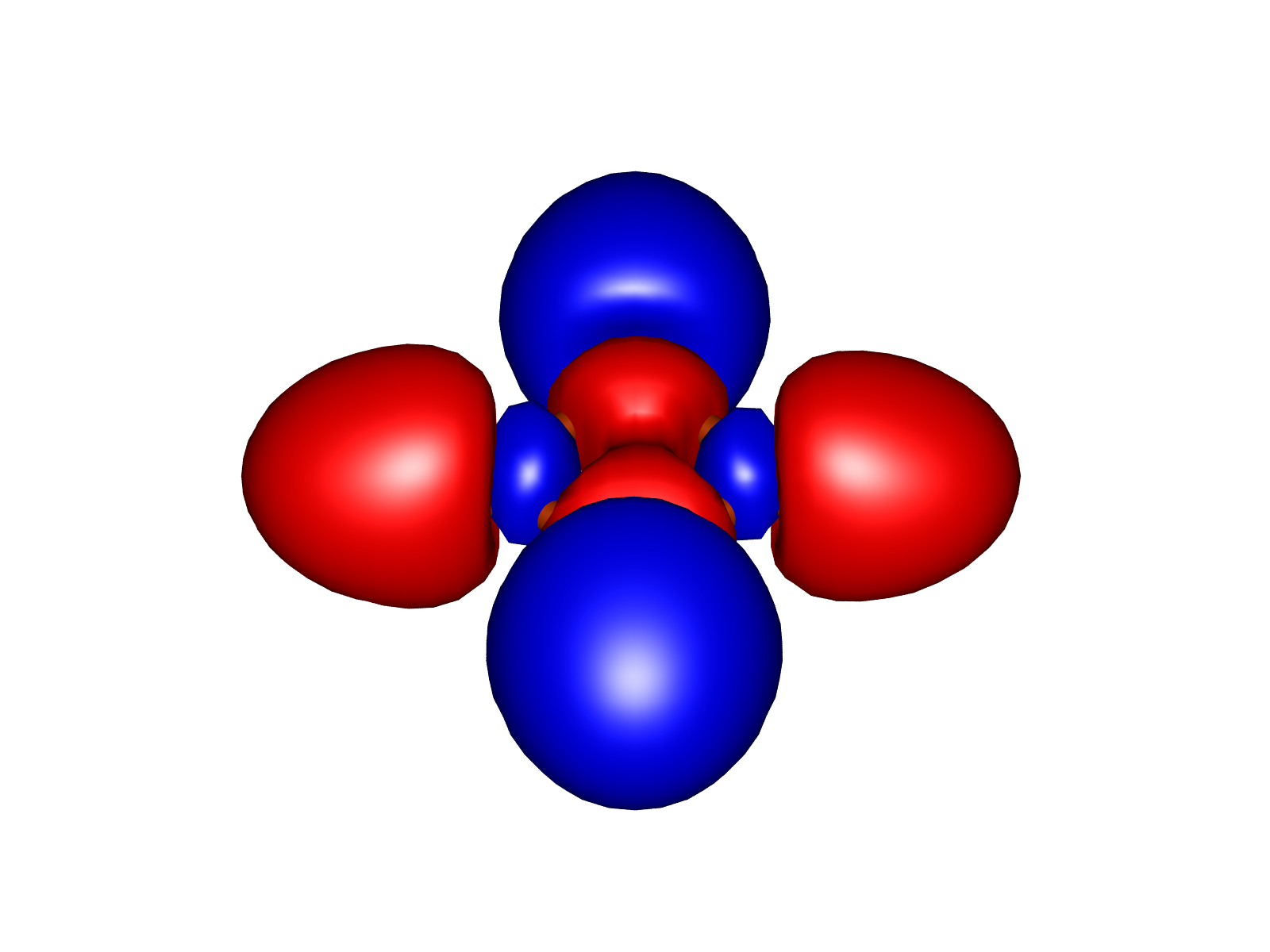}
  } \\
  \subfloat[$n_{\text{occup}}$ = 1.98]{%
    \includegraphics[width=0.22\textwidth]{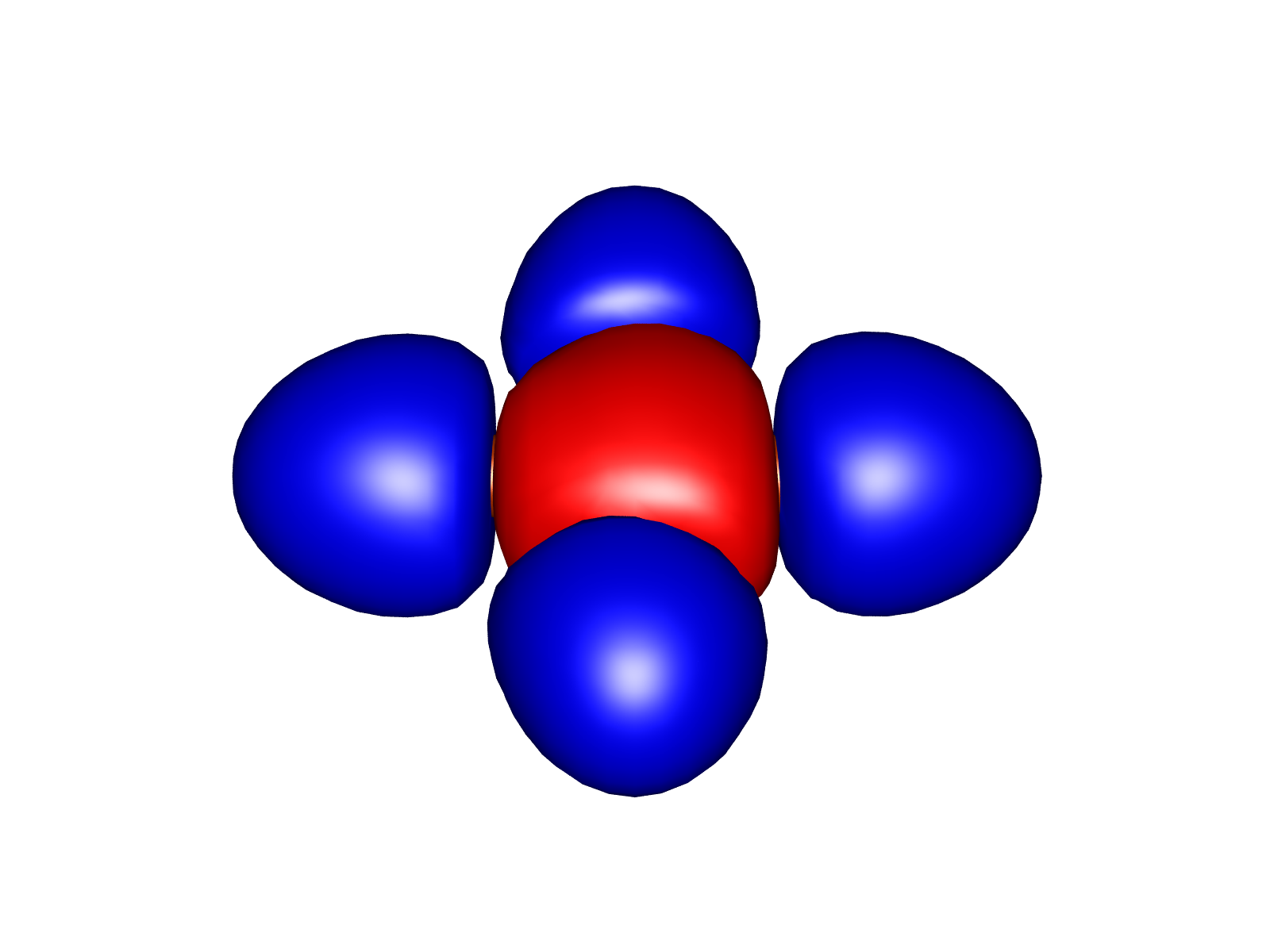}
  }
  \hfill
  \subfloat[$n_{\text{occup}}$ = 1.97]{%
    \includegraphics[width=0.22\textwidth]{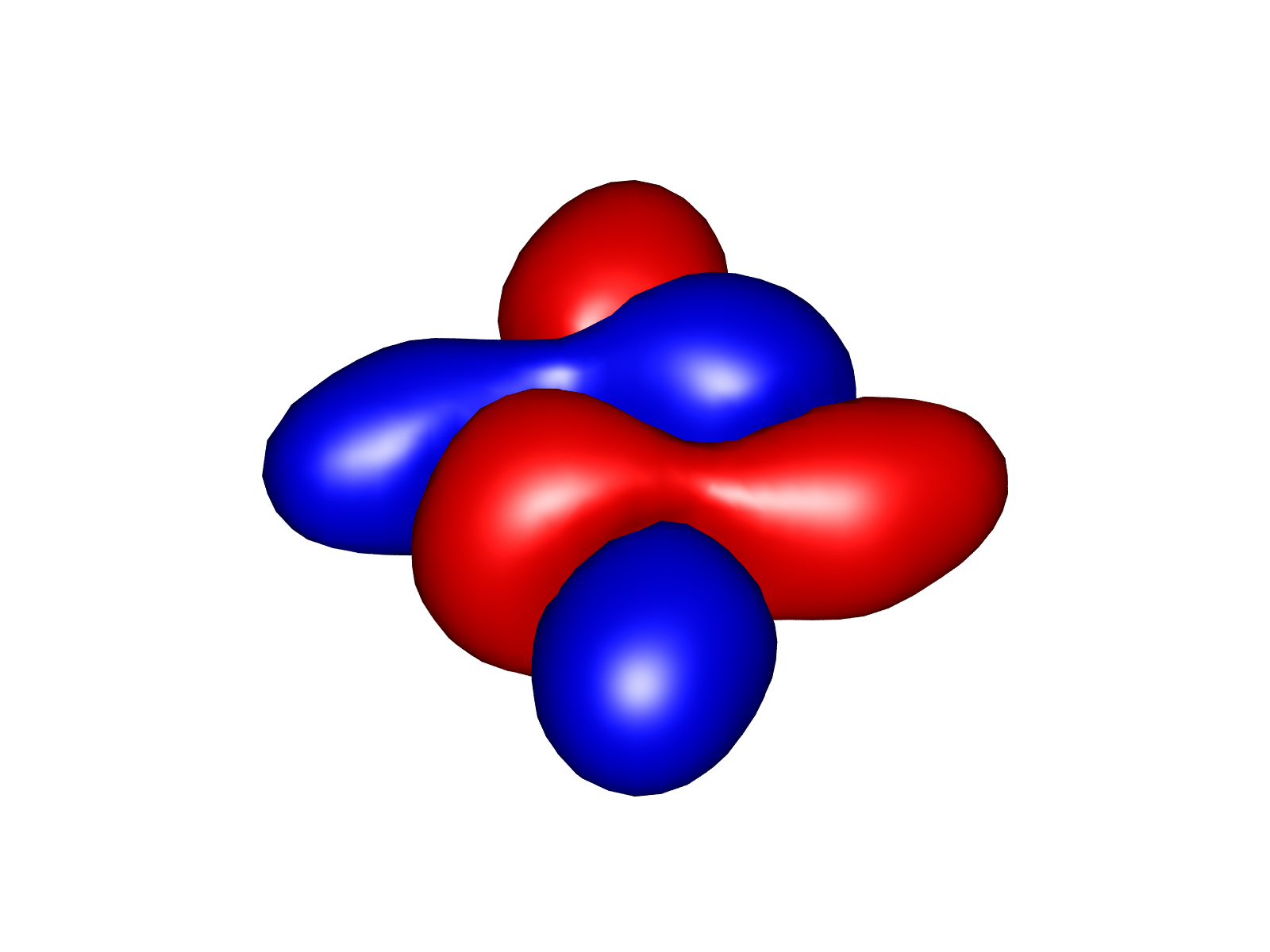}
  }
  \hfill
  \subfloat[$n_{\text{occup}}$ = 1.97]{%
    \includegraphics[width=0.22\textwidth]{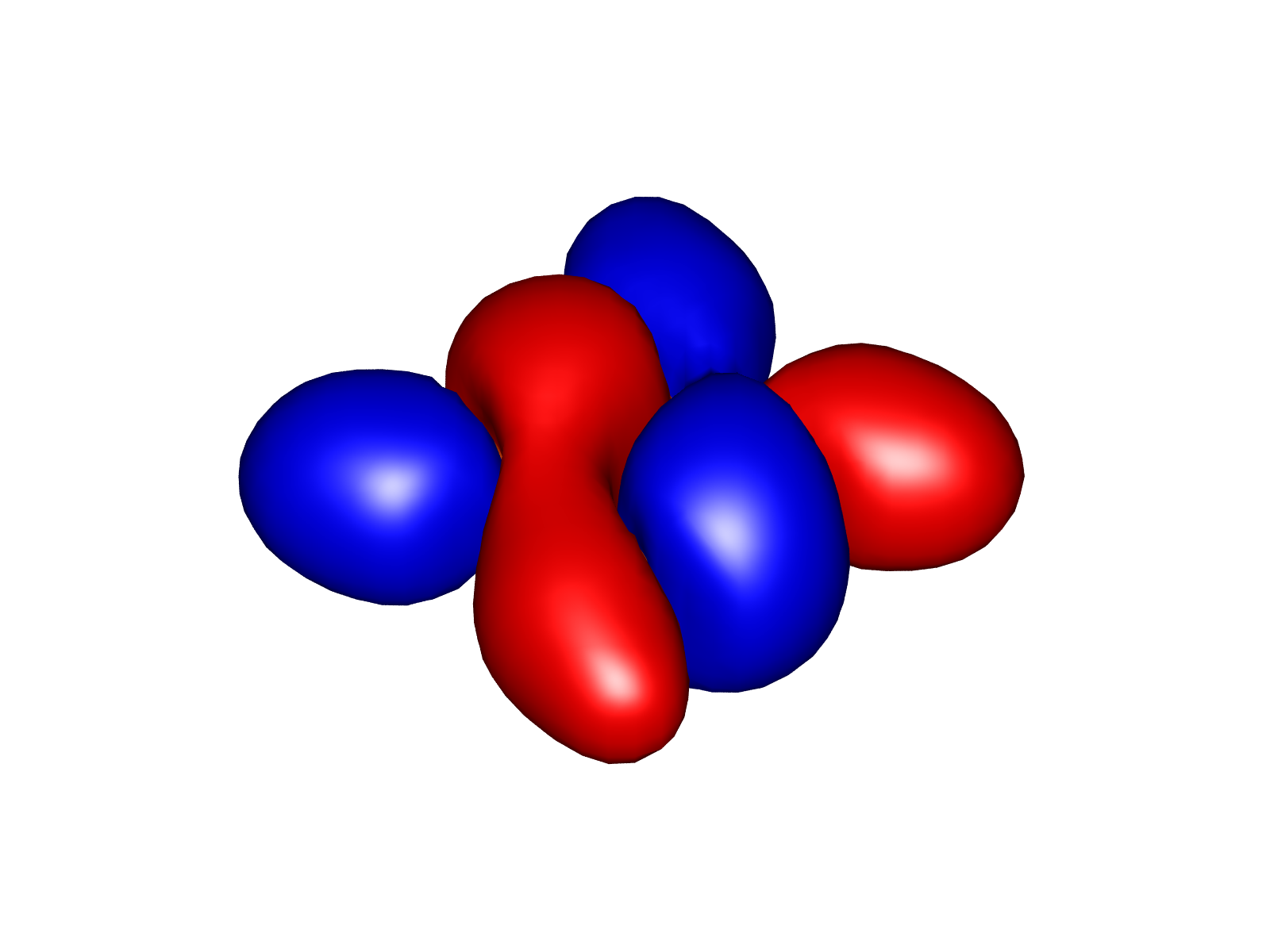}
  }
  \hfill
  \subfloat[$n_{\text{occup}}$ = 1.97]{%
    \includegraphics[width=0.22\textwidth]{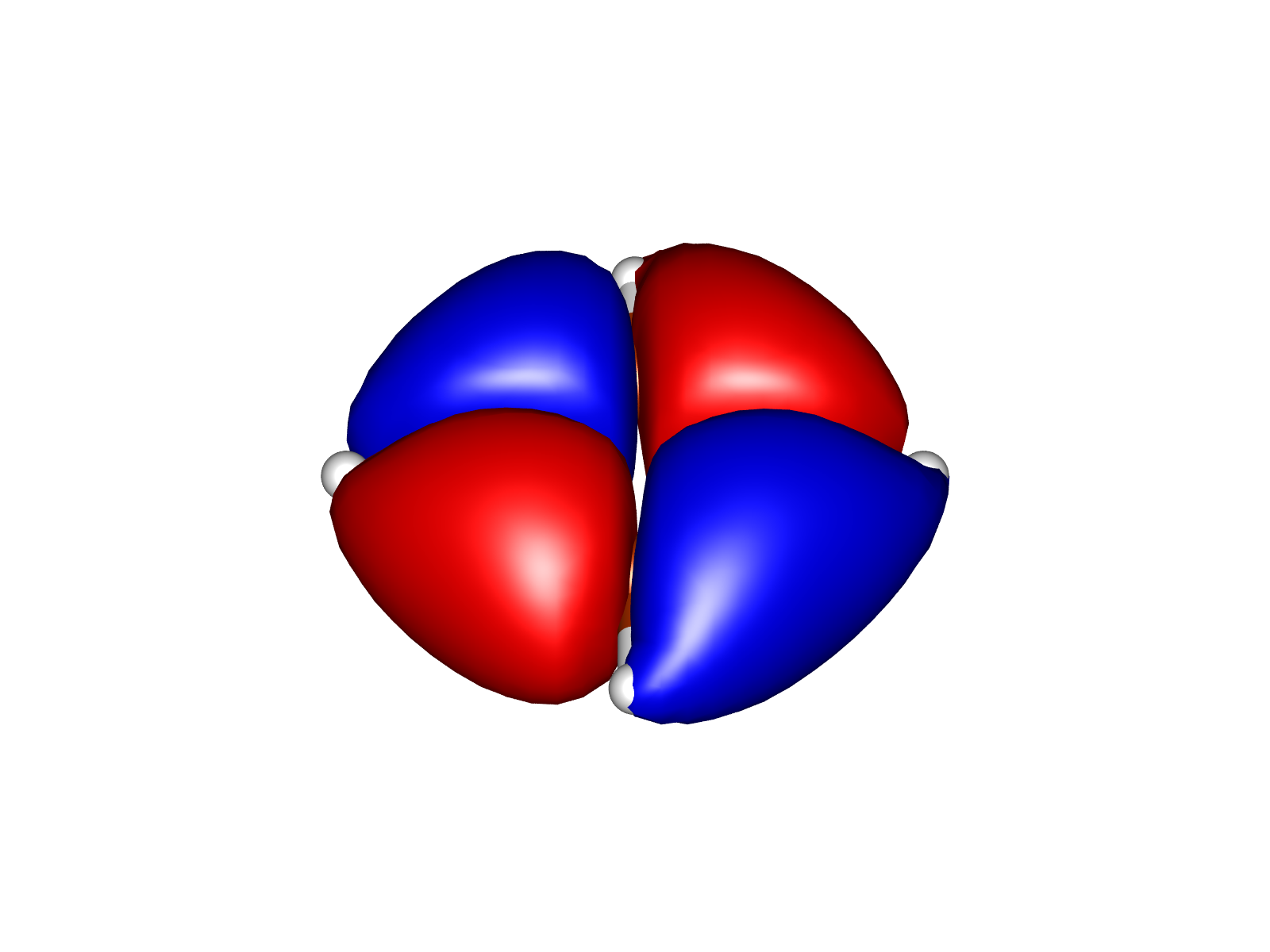}
  }
  \\
  \subfloat[$n_{\text{occup}}$ = 1.91]{%
    \includegraphics[width=0.22\textwidth]{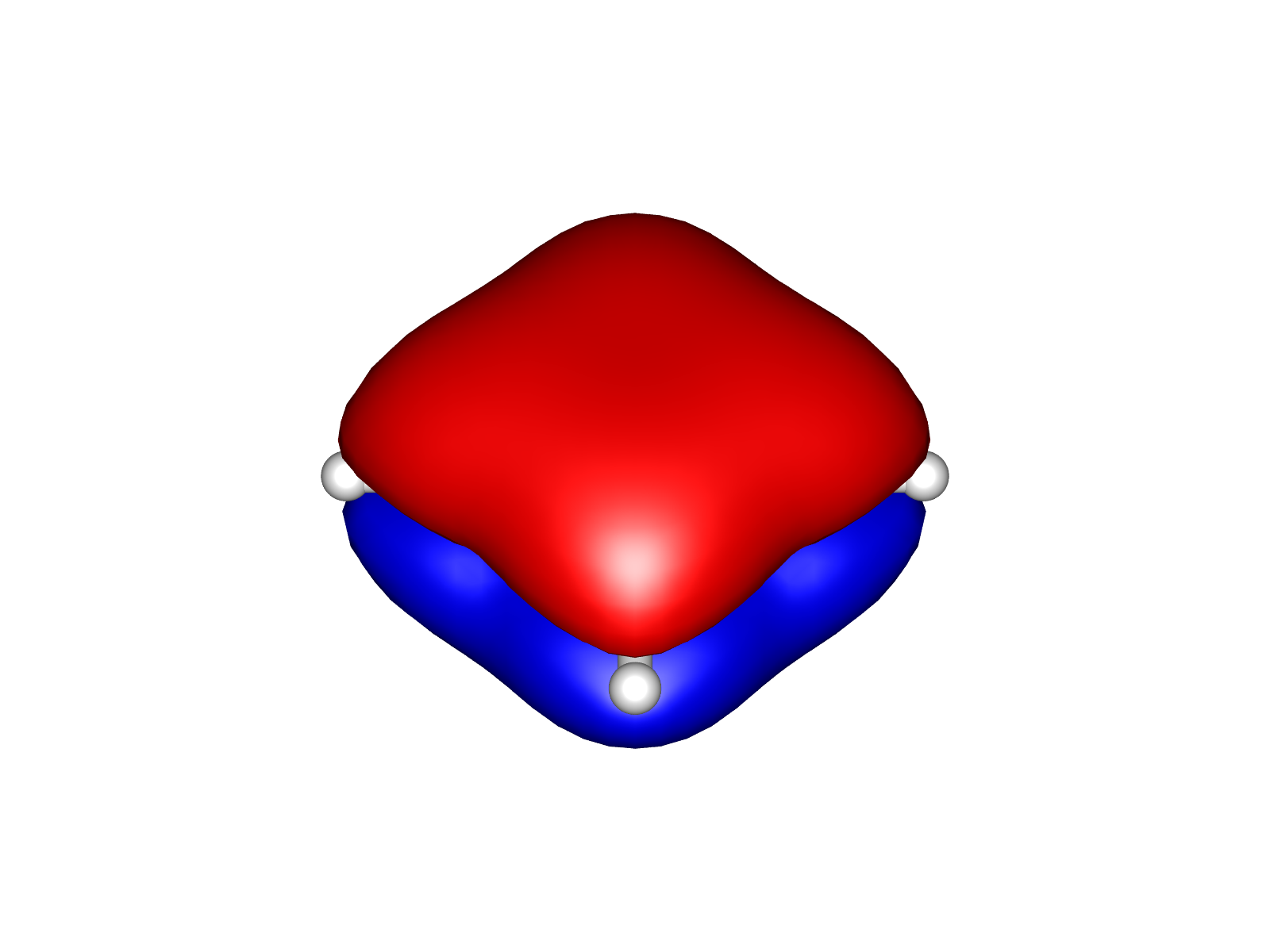}
  }
  \hfill
  \subfloat[$n_{\text{occup}}$ = 1.15]{%
    \includegraphics[width=0.22\textwidth]{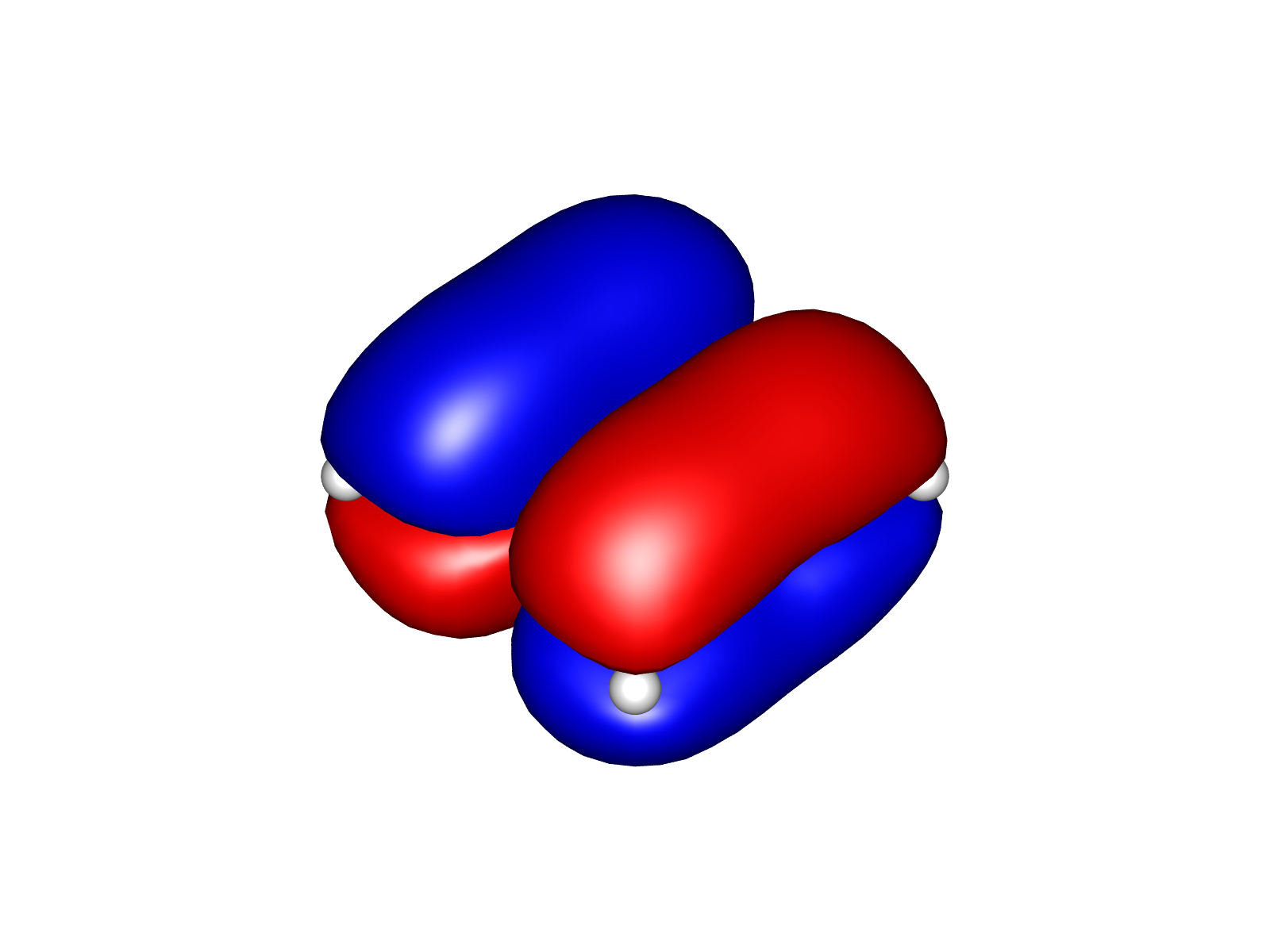}
  }
  \hfill
  \subfloat[$n_{\text{occup}}$ = 0.85]{%
    \includegraphics[width=0.22\textwidth]{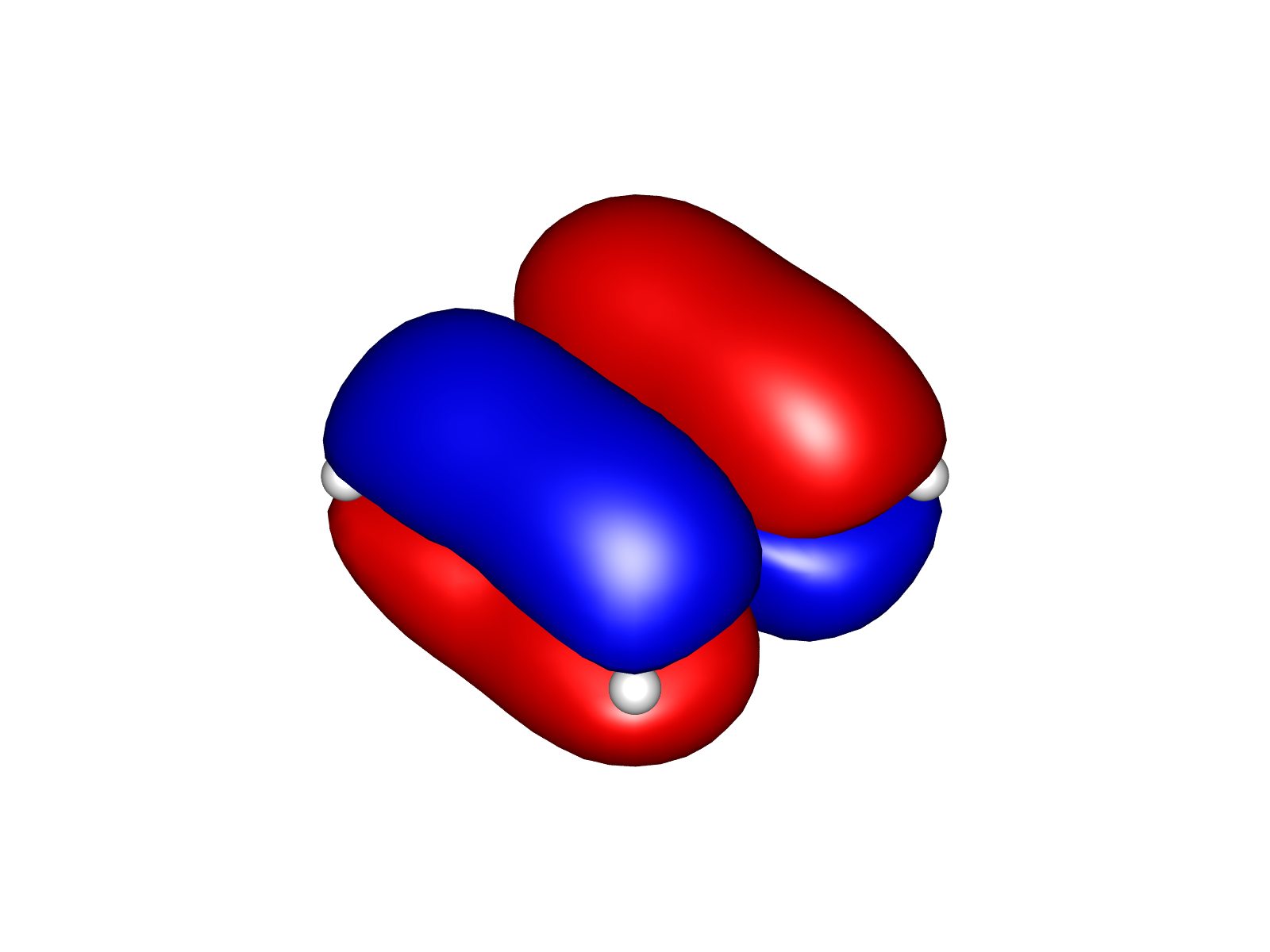}
  }
  \hfill
  \subfloat[$n_{\text{occup}}$ = 0.09]{%
    \includegraphics[width=0.22\textwidth]{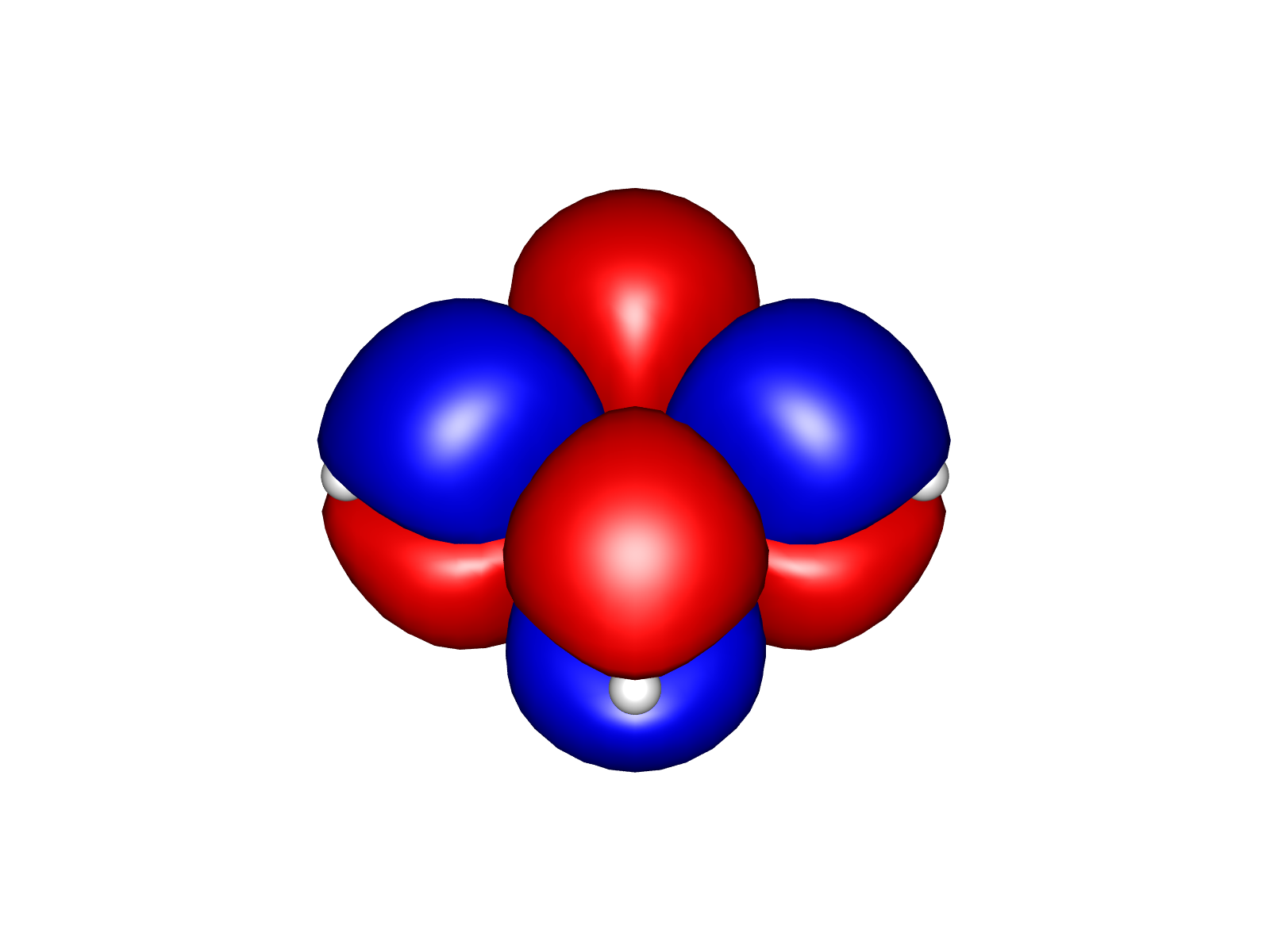}
  }
  \\
  \subfloat[$n_{\text{occup}}$ = 0.03]{%
    \includegraphics[width=0.22\textwidth]{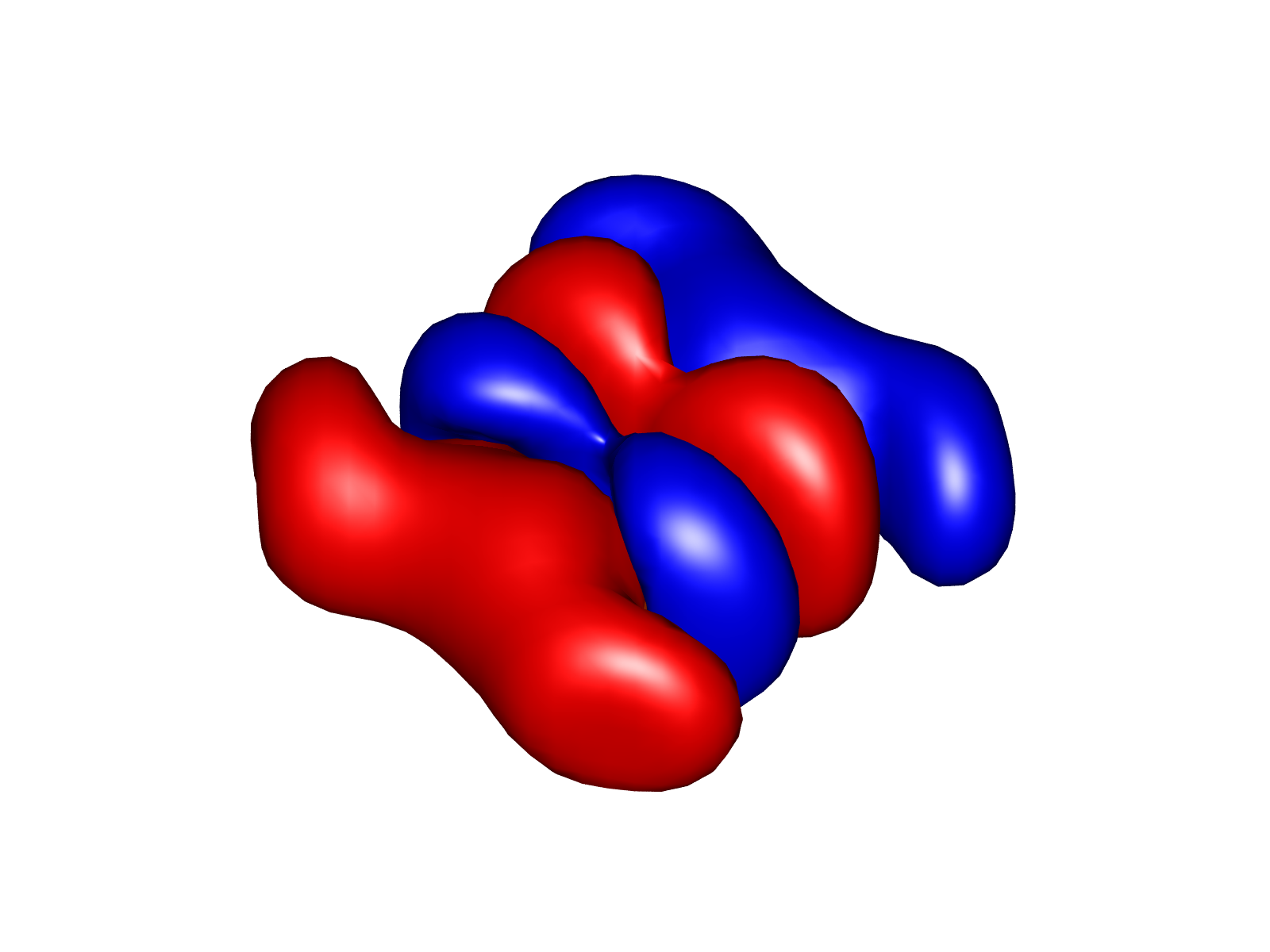}
  }
  \hfill
  \subfloat[$n_{\text{occup}}$ = 0.03]{%
    \includegraphics[width=0.22\textwidth]{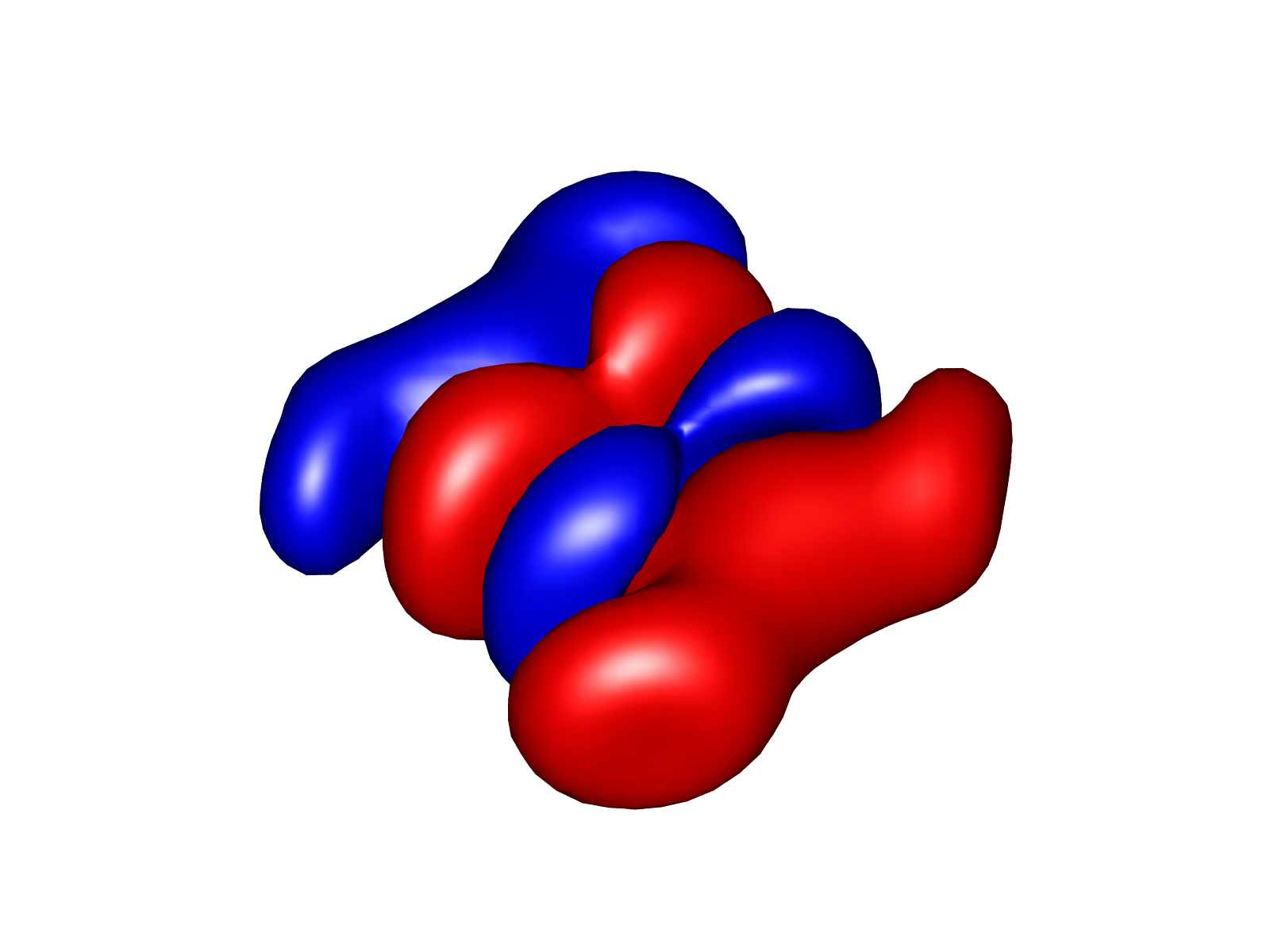}
  }
  \hfill
  \subfloat[$n_{\text{occup}}$ = 0.03]{%
    \includegraphics[width=0.22\textwidth]{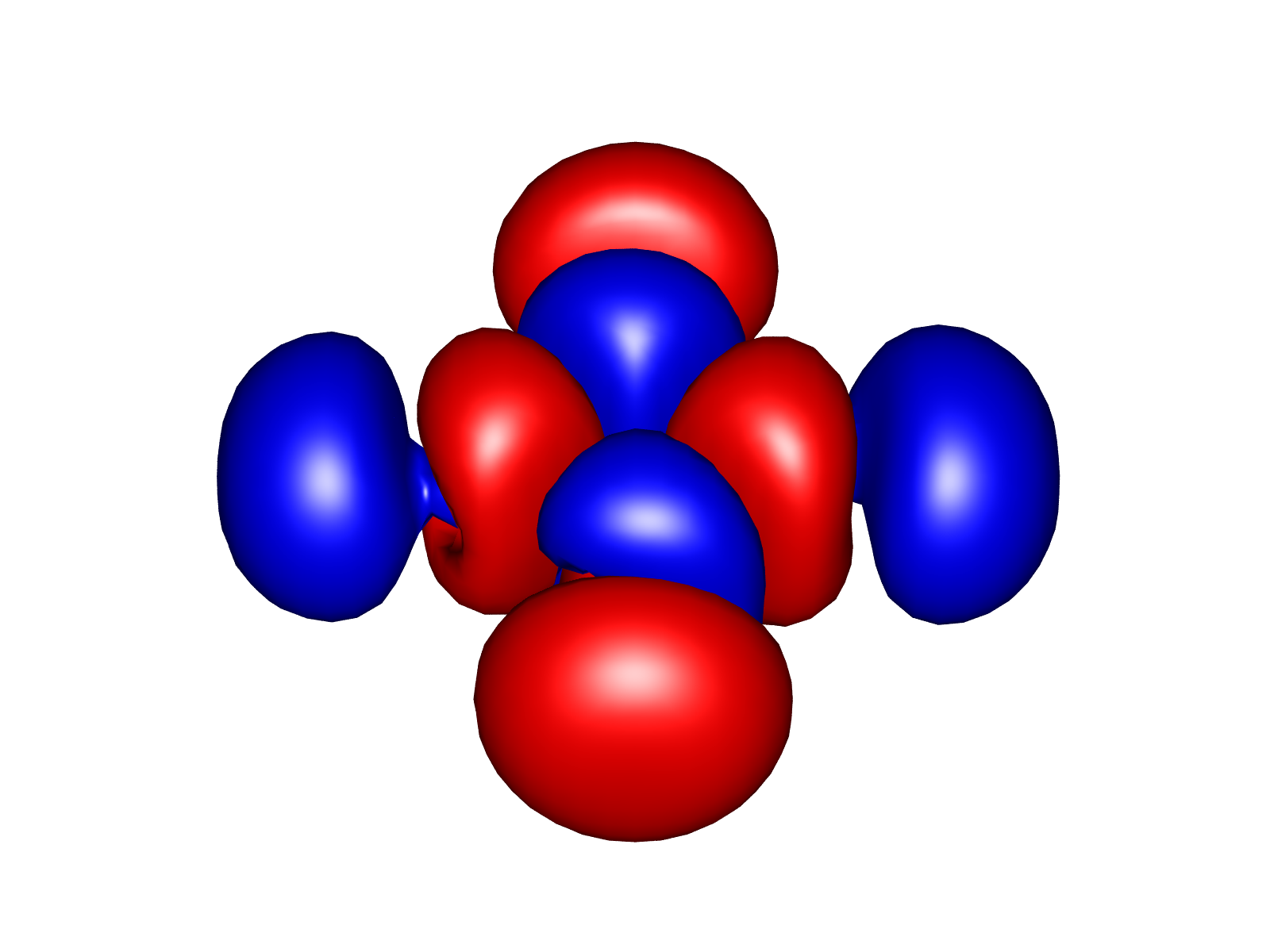}
  }
  \hfill
  \subfloat[$n_{\text{occup}}$ = 0.03]{%
    \includegraphics[width=0.22\textwidth]{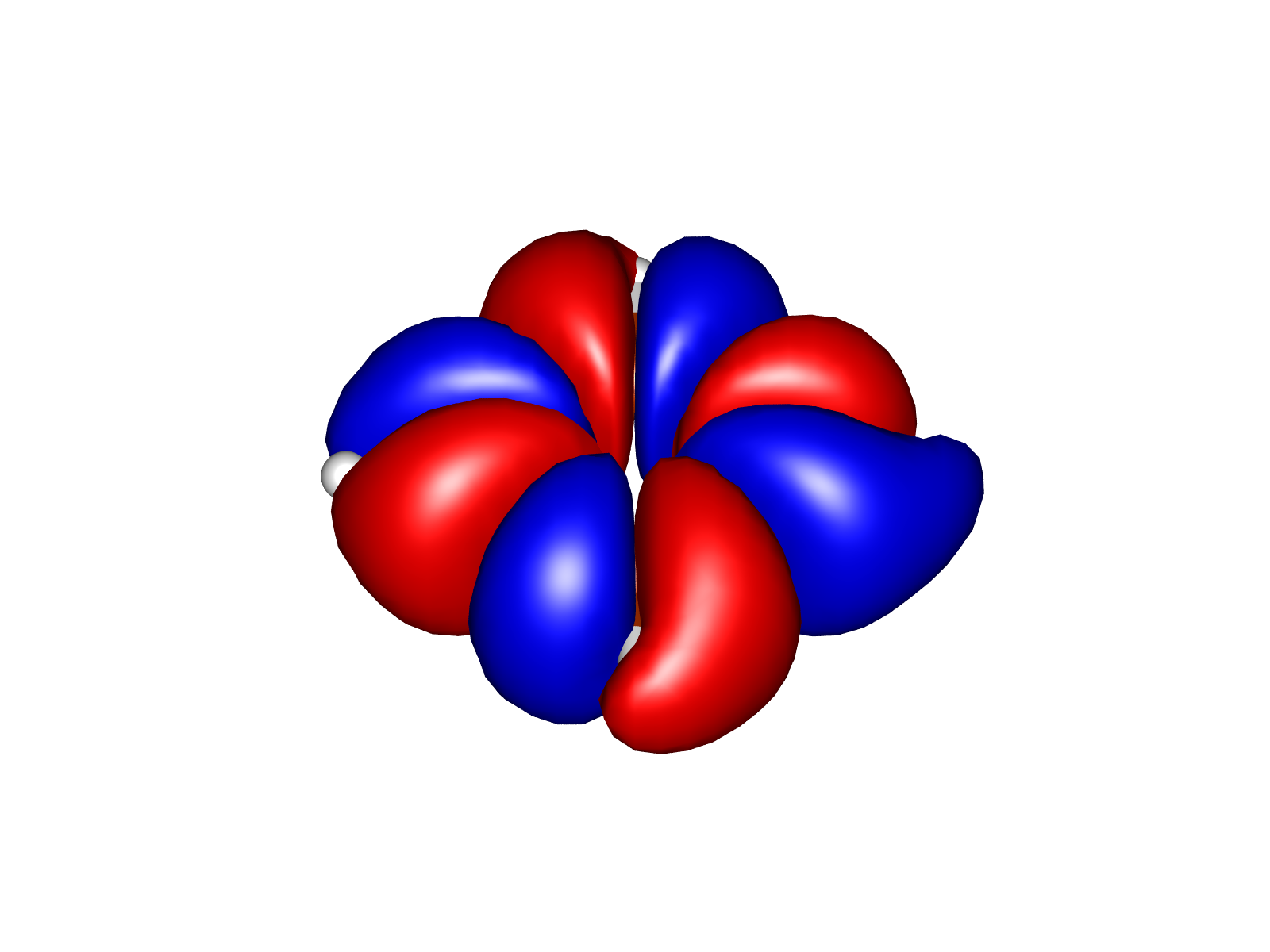}
  }
  \\
  \subfloat[$n_{\text{occup}}$ = 0.02]{%
    \includegraphics[width=0.22\textwidth]{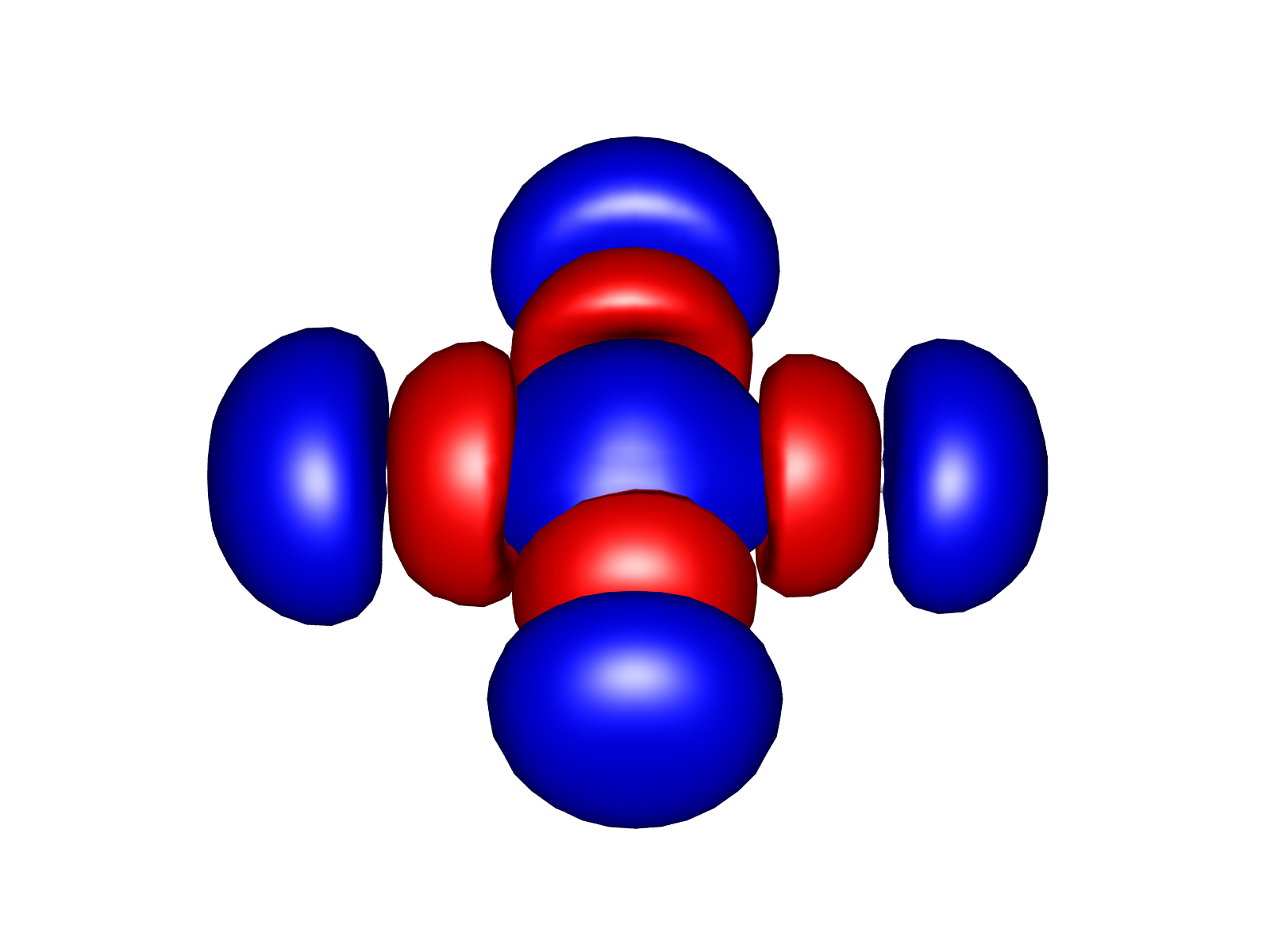}
  }
  \hfill
  \subfloat[$n_{\text{occup}}$ = 0.02]{%
    \includegraphics[width=0.22\textwidth]{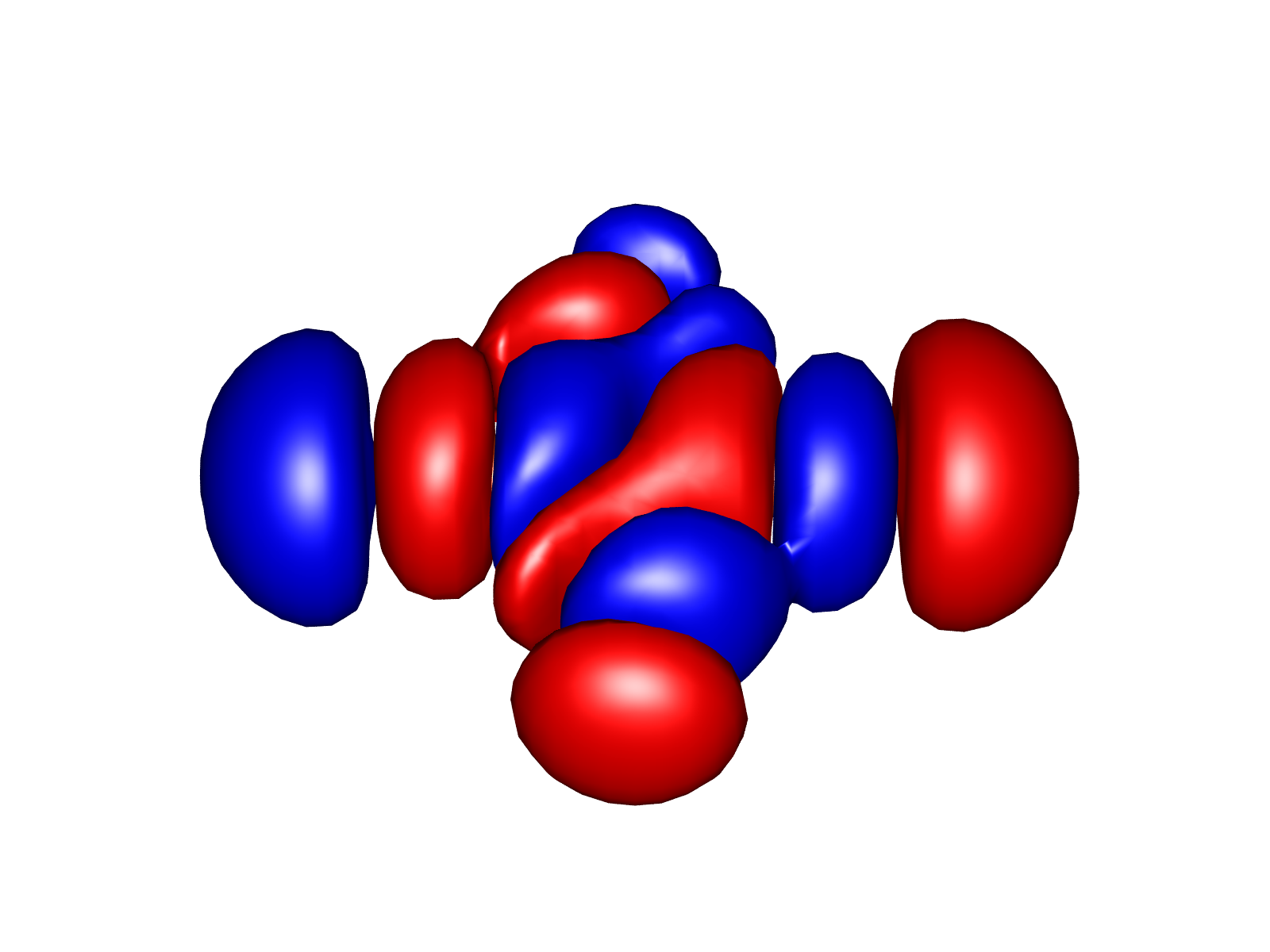}
  }
  \hfill
  \subfloat[$n_{\text{occup}}$ = 0.02]{%
    \includegraphics[width=0.22\textwidth]{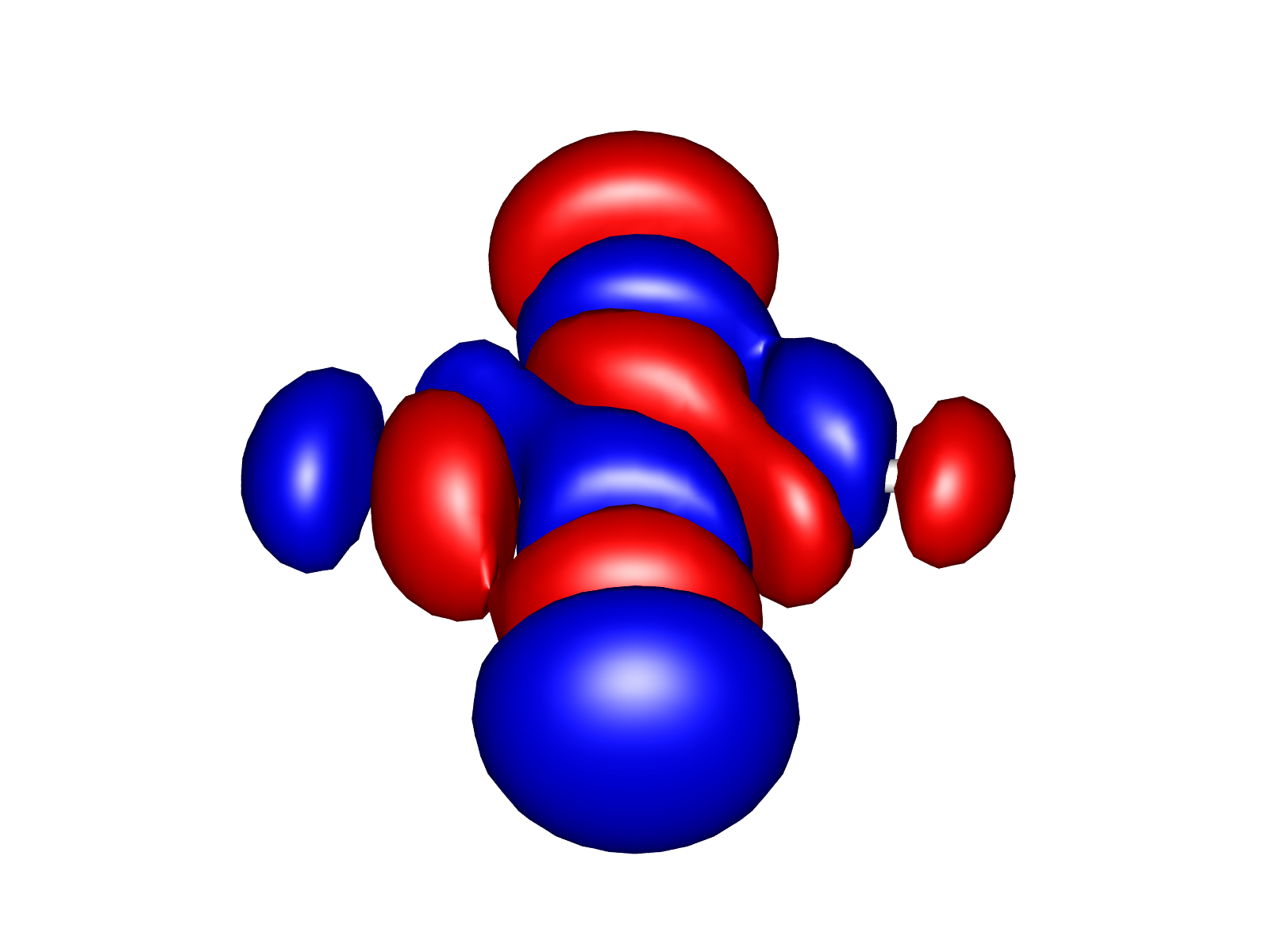}
  }
  \hfill
  \subfloat[$n_{\text{occup}}$ = 0.02]{%
    \includegraphics[width=0.22\textwidth]{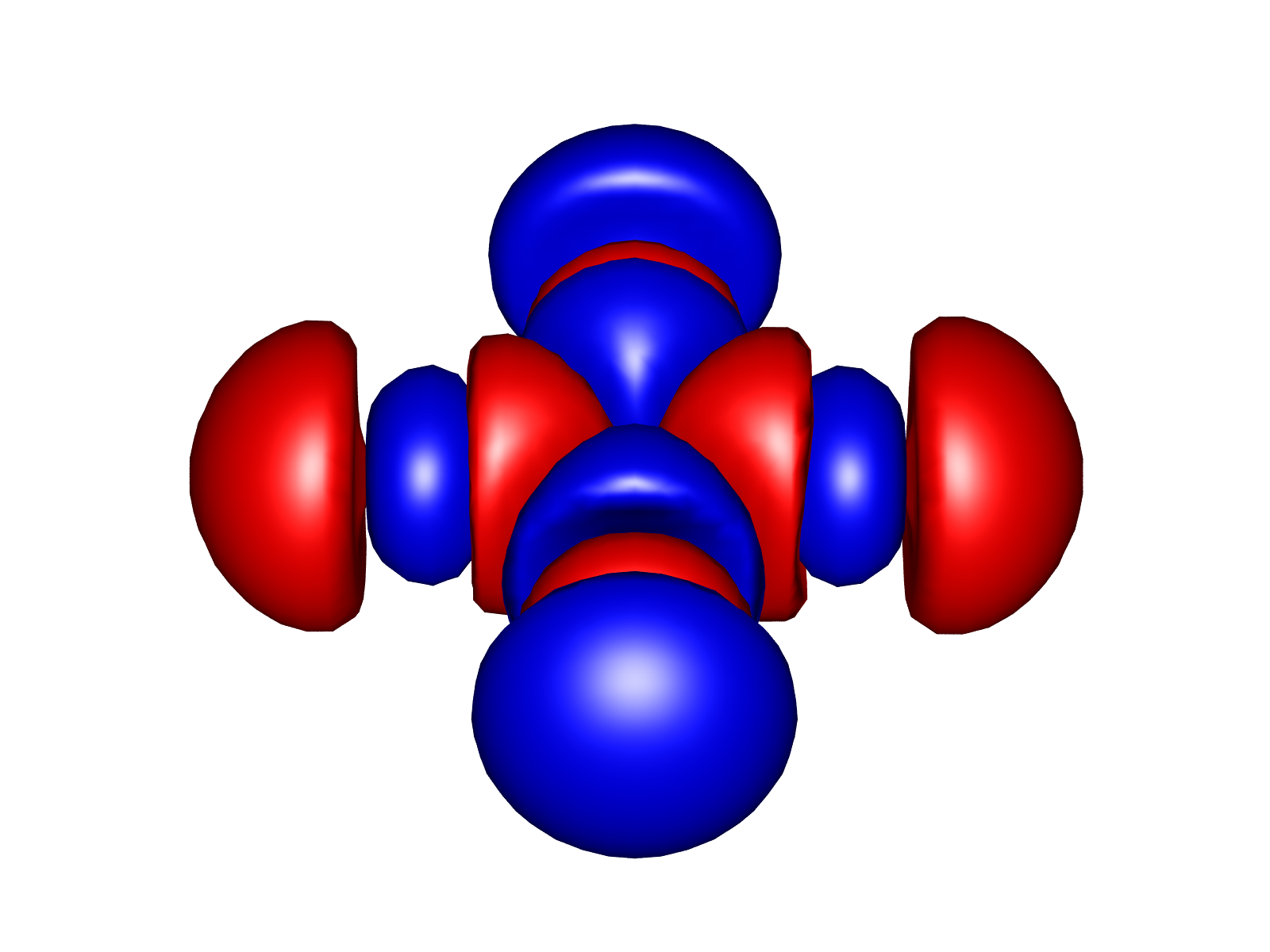}
  }
  \\
  \subfloat[$n_{\text{occup}}$ = 0.01]{%
    \includegraphics[width=0.22\textwidth]{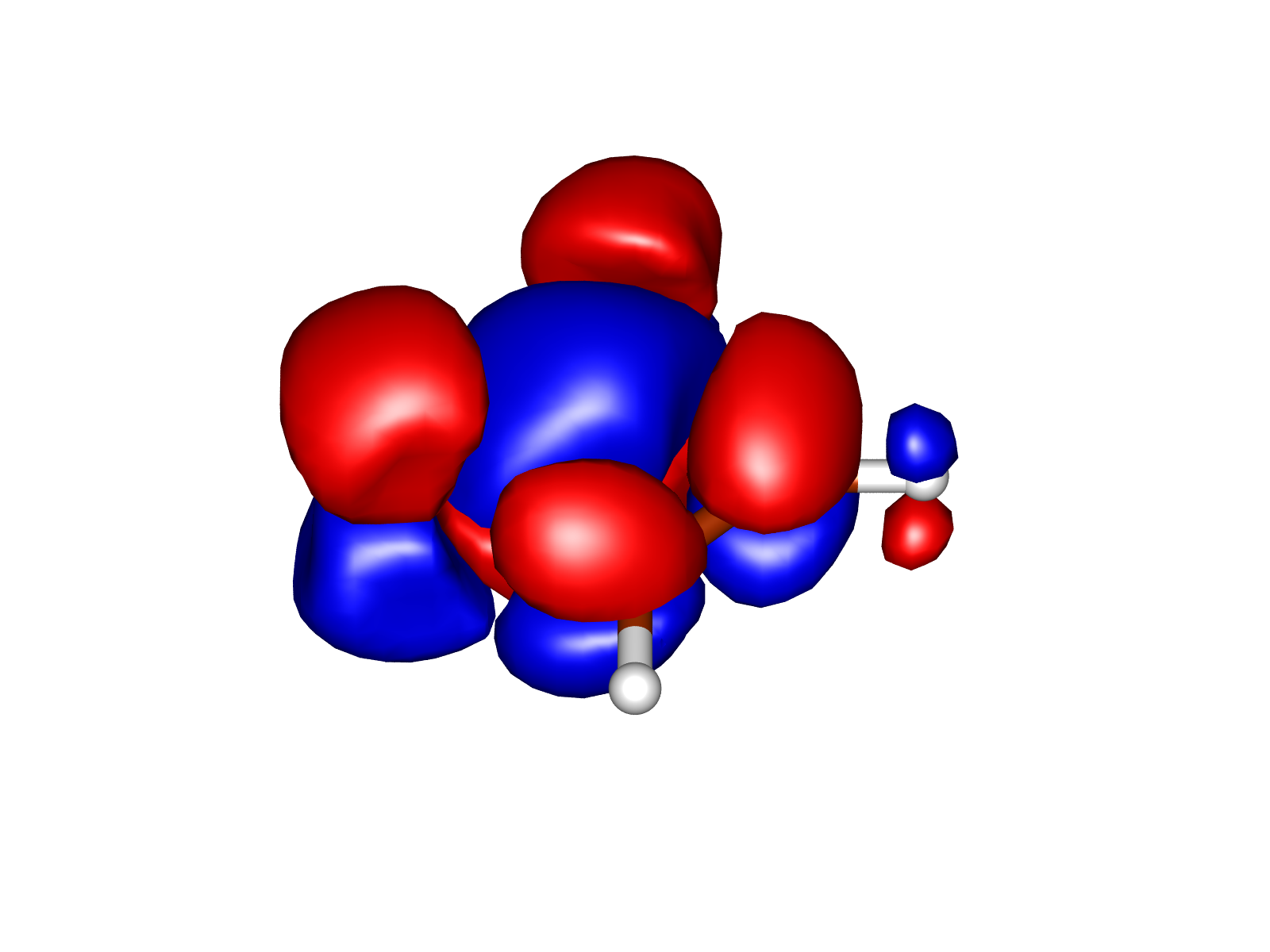}
  }
  \subfloat[$n_{\text{occup}}$ = 0.01]{%
    \includegraphics[width=0.22\textwidth]{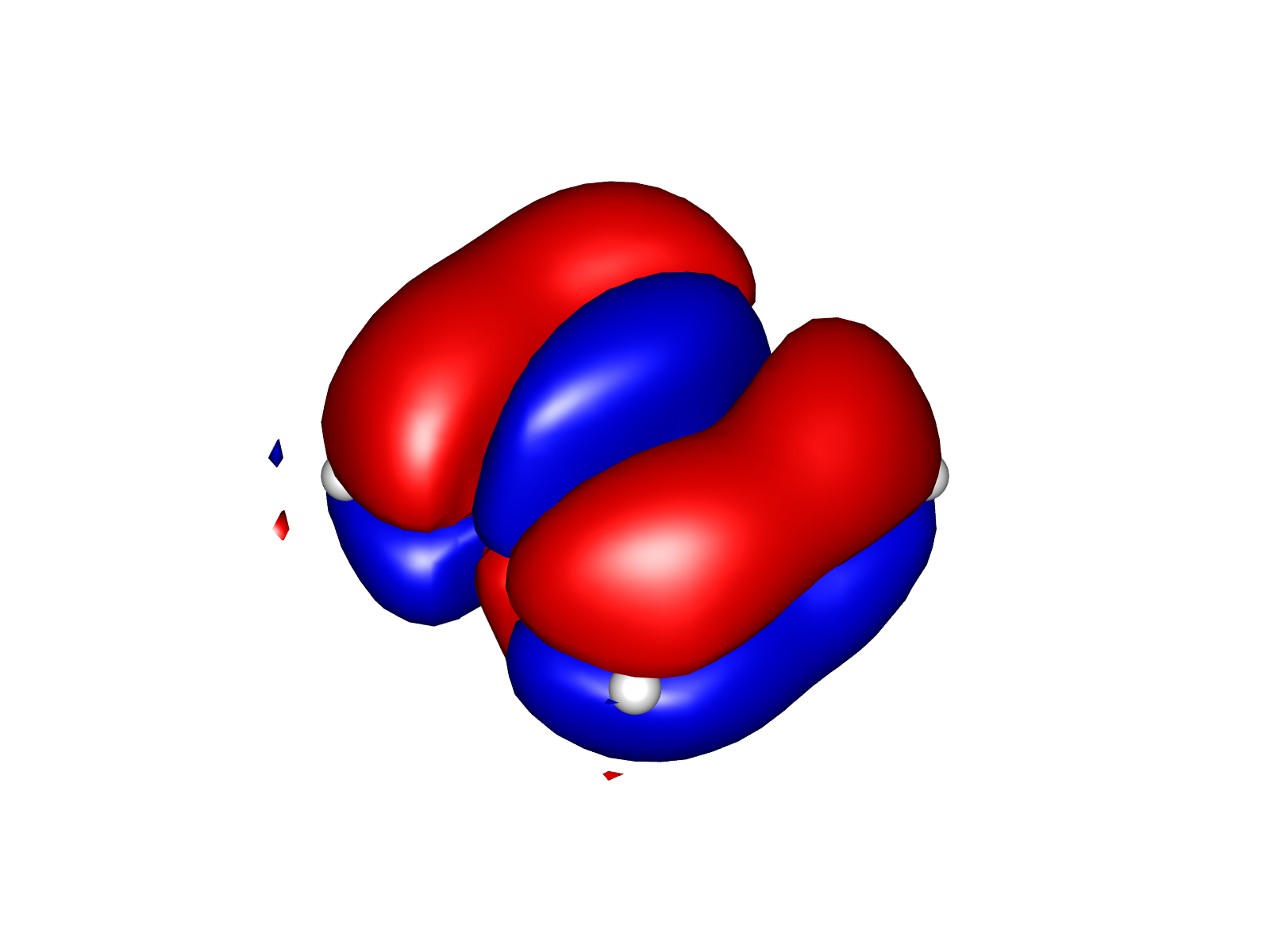}
  }
  \caption{C$_4$H$_4$, singlet state, DMRG-SCF(20, 22) \label{orbs_c4h4_s}}
\end{figure}

\begin{figure}[!h]
  \subfloat[$n_{\text{occup}}$ = 1.99]{%
    \includegraphics[width=0.22\textwidth]{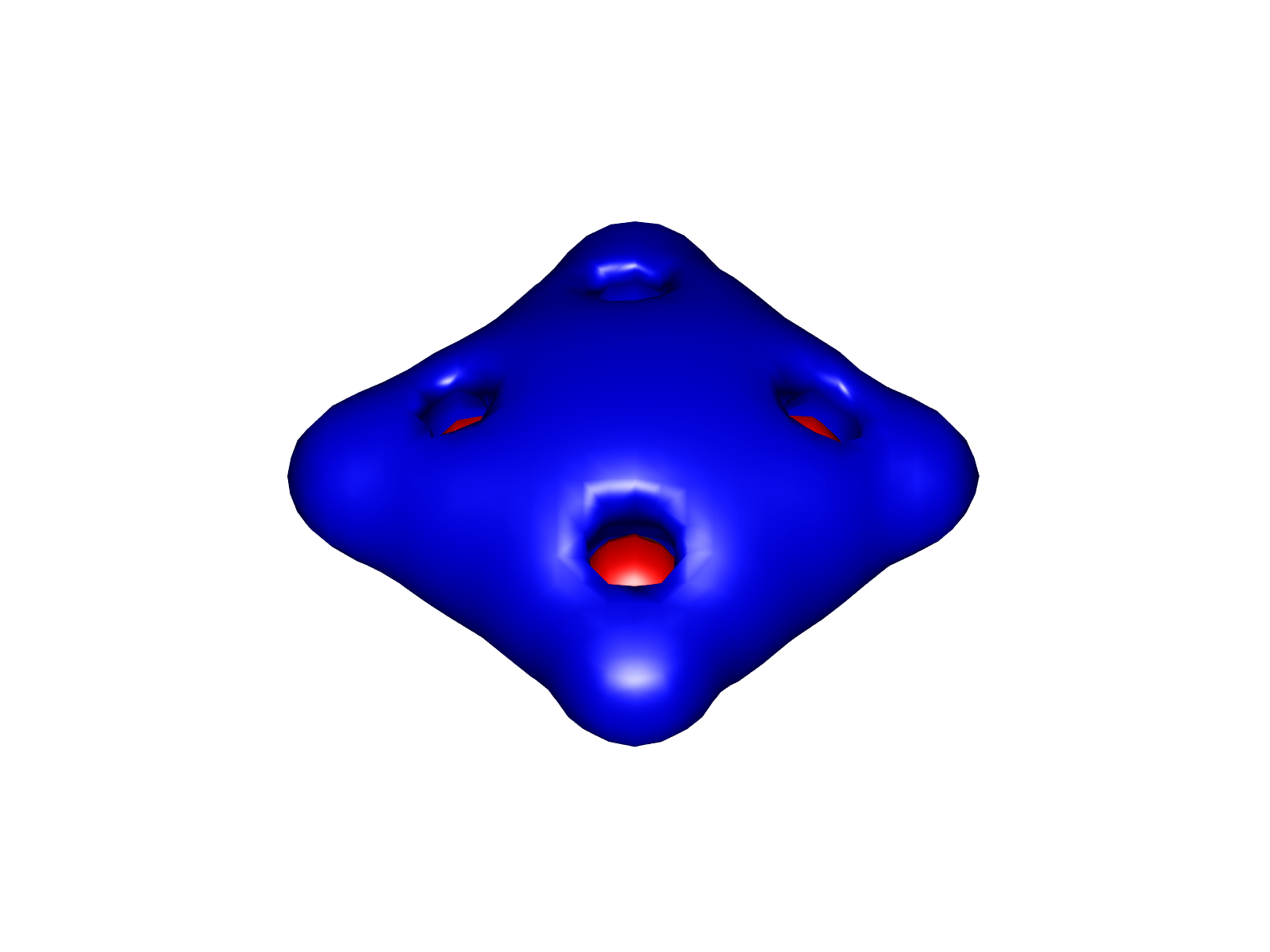}
  }
  \hfill
  \subfloat[$n_{\text{occup}}$ = 1.98]{%
    \includegraphics[width=0.22\textwidth]{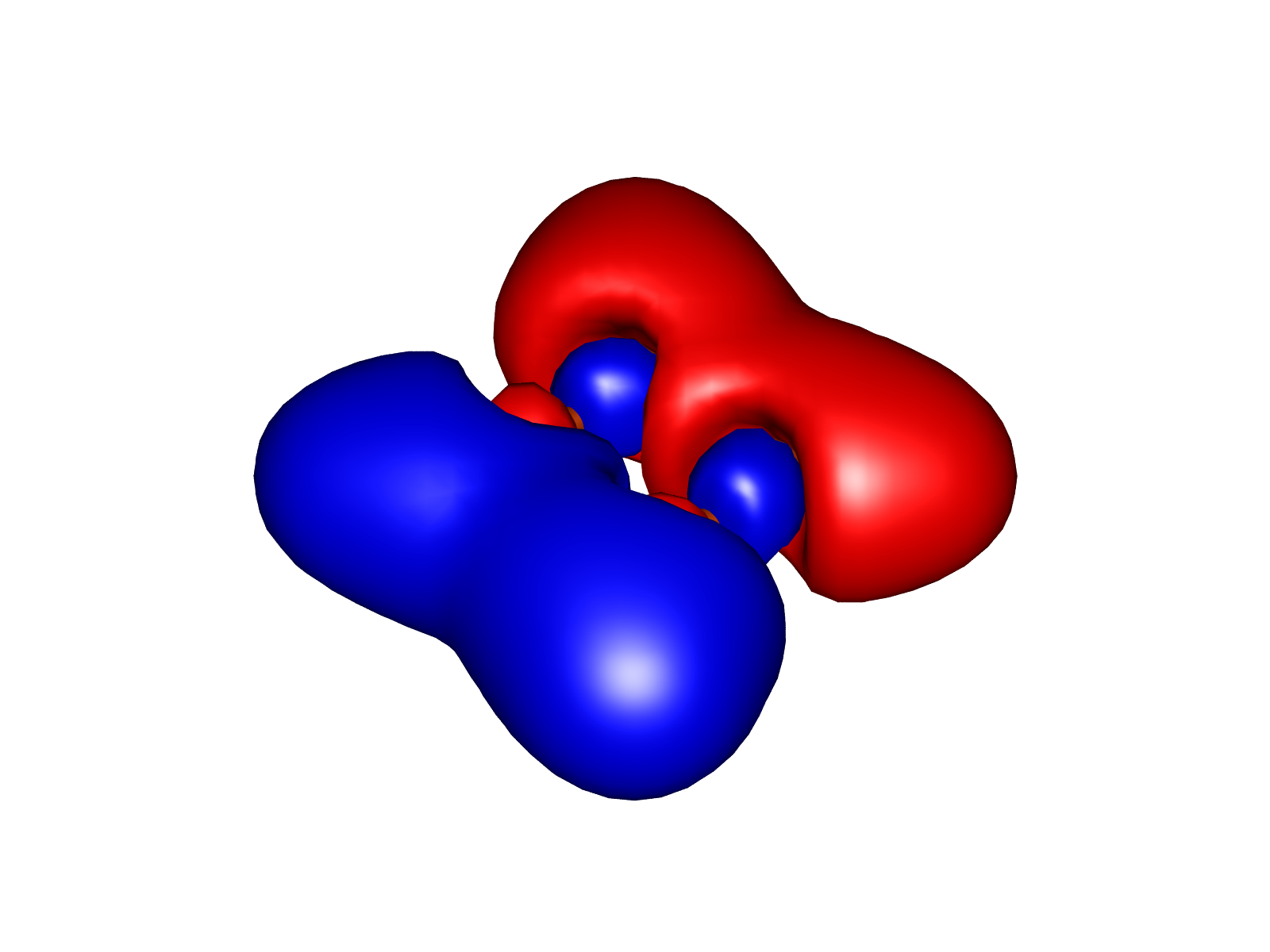}
  }
  \hfill
  \subfloat[$n_{\text{occup}}$ = 1.98]{%
    \includegraphics[width=0.22\textwidth]{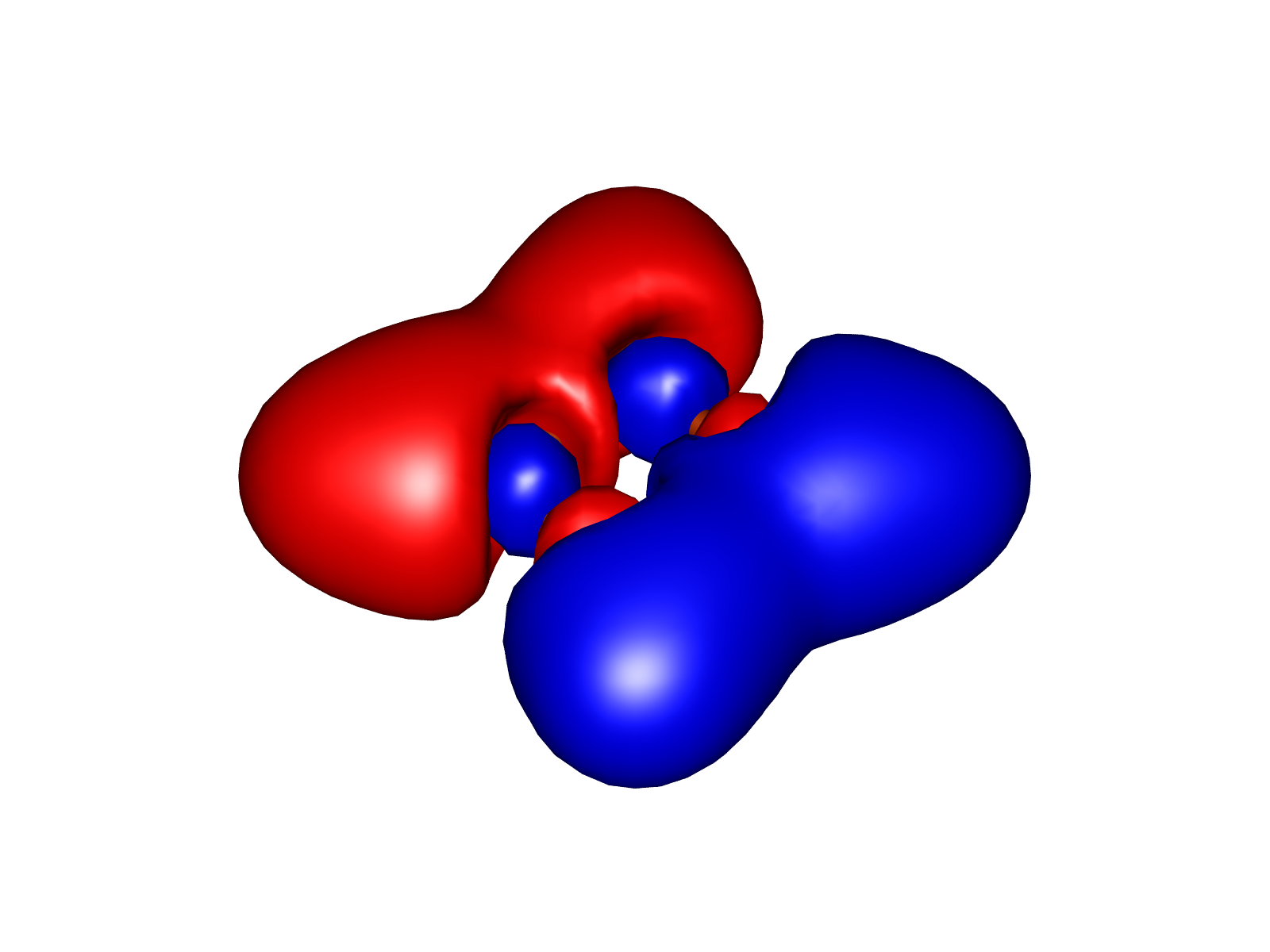}
  }
  \hfill
  \subfloat[$n_{\text{occup}}$ = 1.98]{%
    \includegraphics[width=0.22\textwidth]{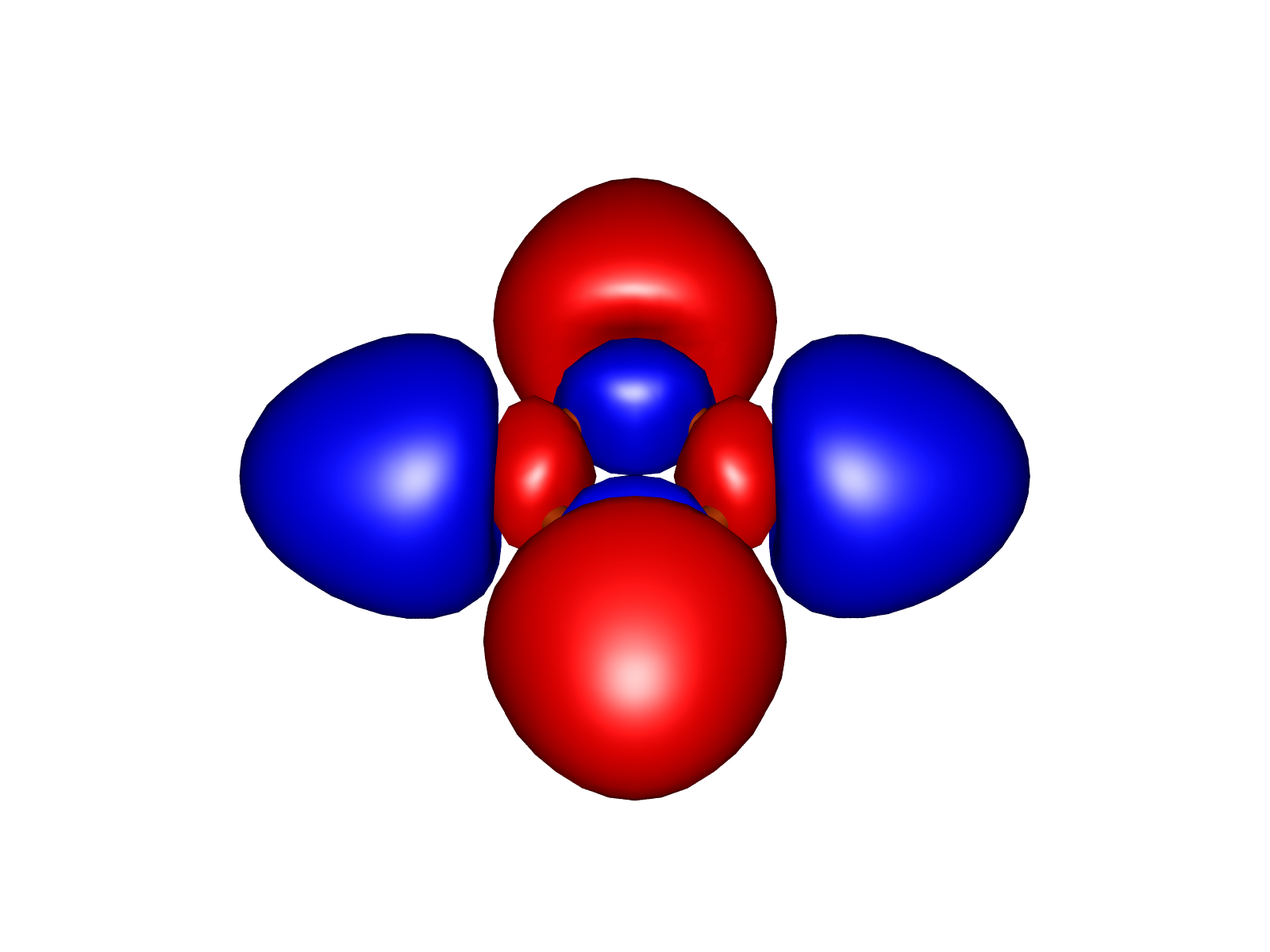}
  } \\
  \subfloat[$n_{\text{occup}}$ = 1.97]{%
    \includegraphics[width=0.22\textwidth]{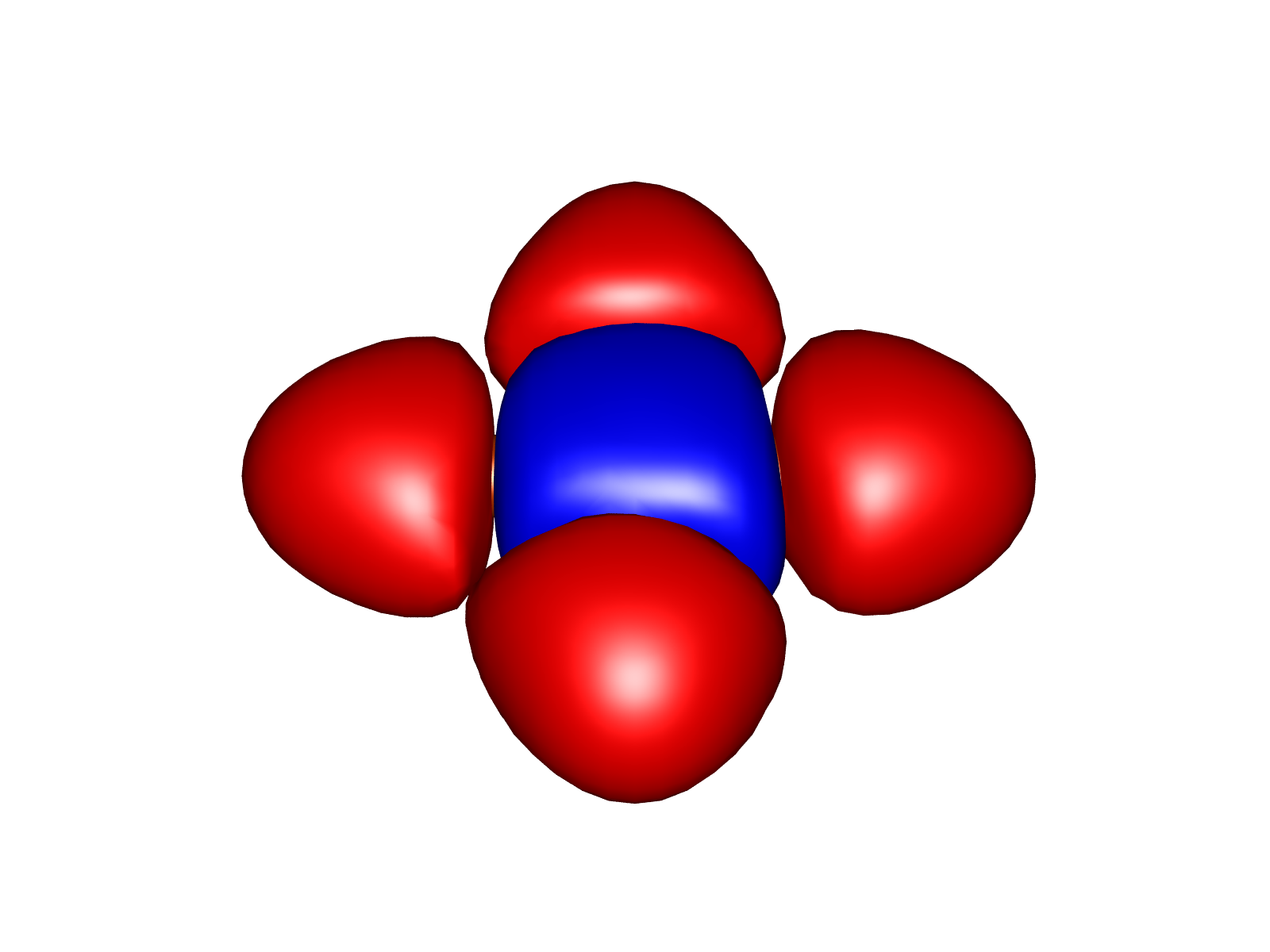}
  }
  \hfill
  \subfloat[$n_{\text{occup}}$ = 1.97]{%
    \includegraphics[width=0.22\textwidth]{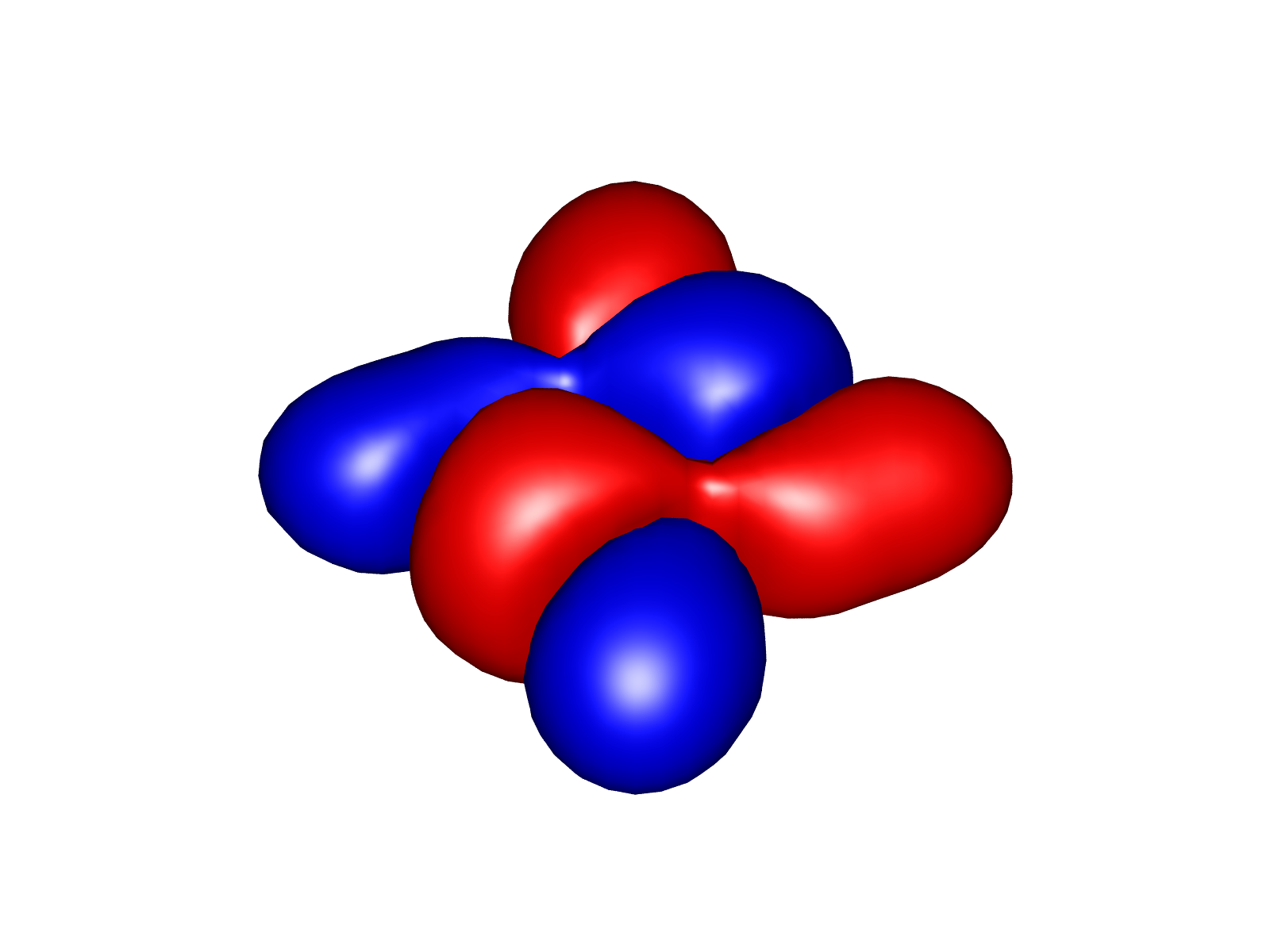}
  }
  \hfill
  \subfloat[$n_{\text{occup}}$ = 1.97]{%
    \includegraphics[width=0.22\textwidth]{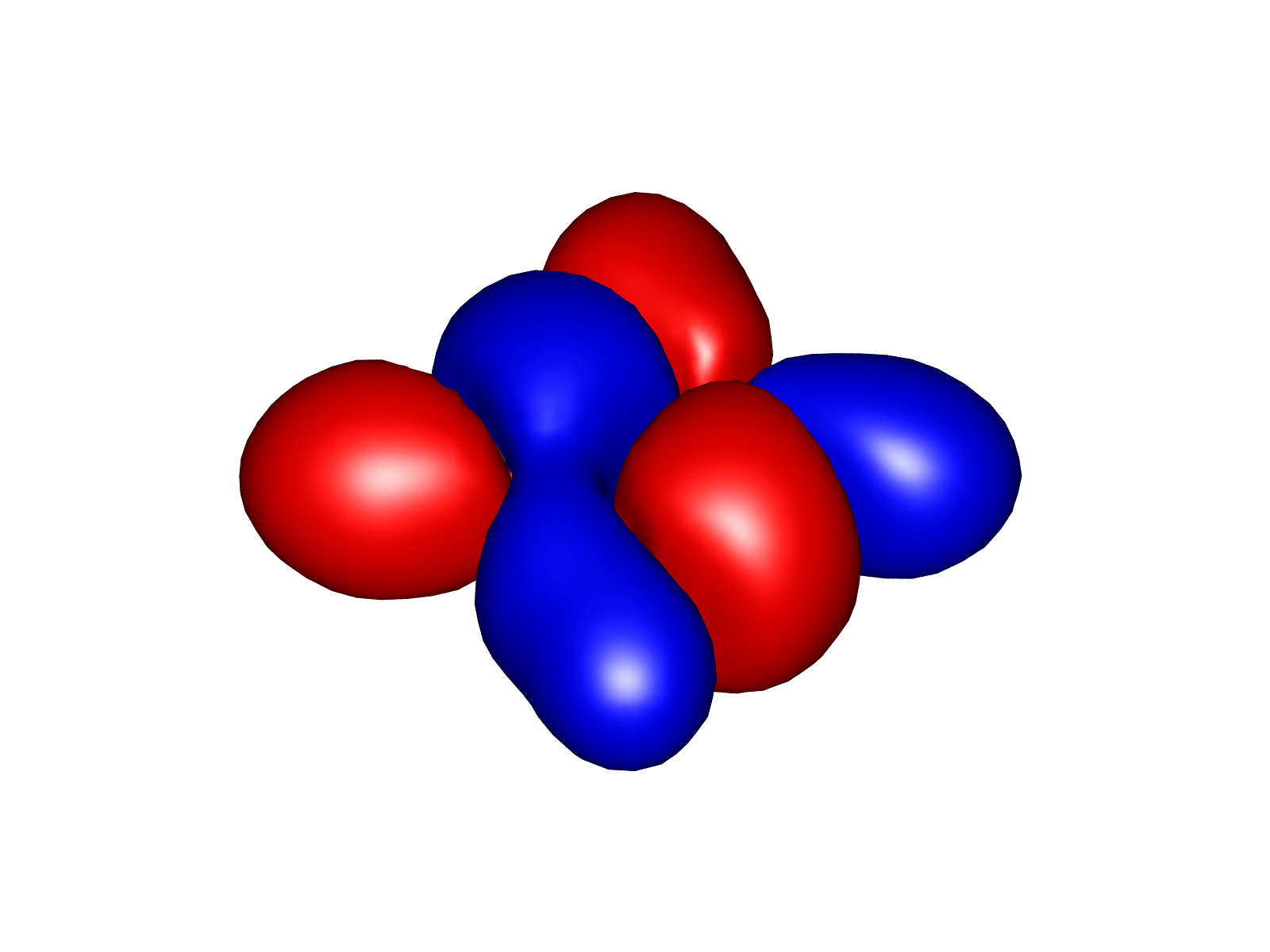}
  }
  \hfill
  \subfloat[$n_{\text{occup}}$ = 1.96]{%
    \includegraphics[width=0.22\textwidth]{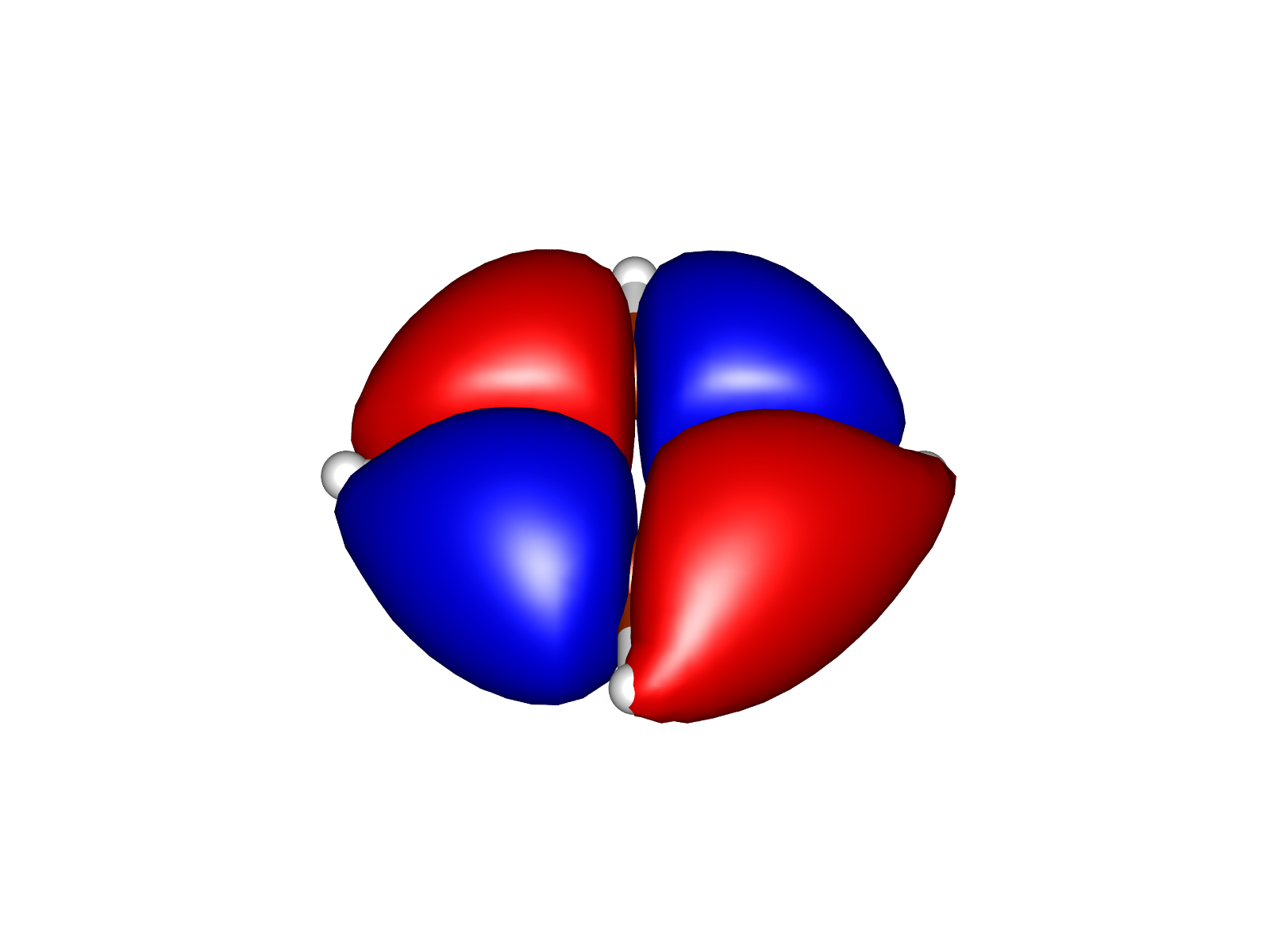}
  }
  \\
  \subfloat[$n_{\text{occup}}$ = 1.93]{%
    \includegraphics[width=0.22\textwidth]{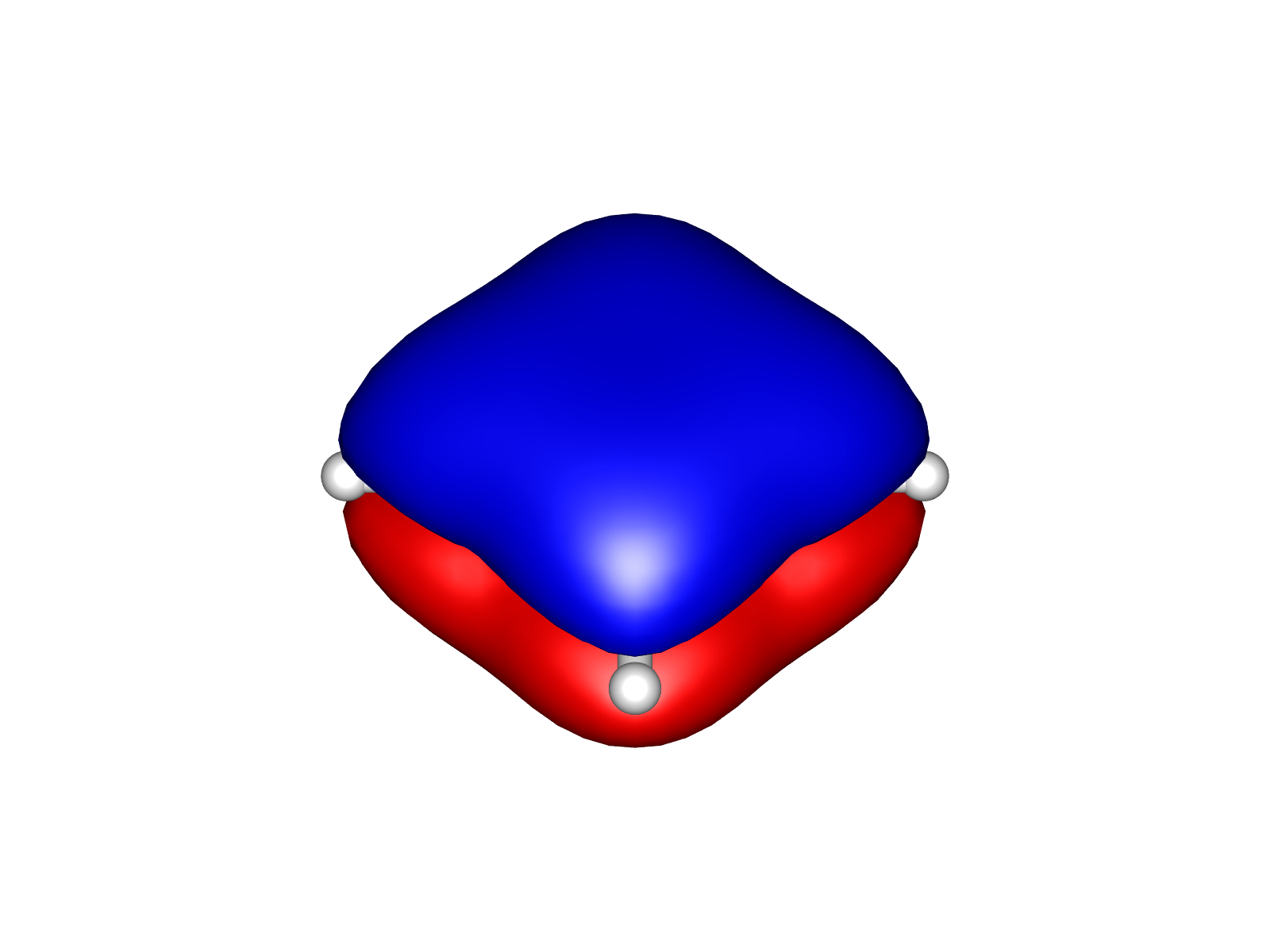}
  }
  \hfill
  \subfloat[$n_{\text{occup}}$ = 1.00]{%
    \includegraphics[width=0.22\textwidth]{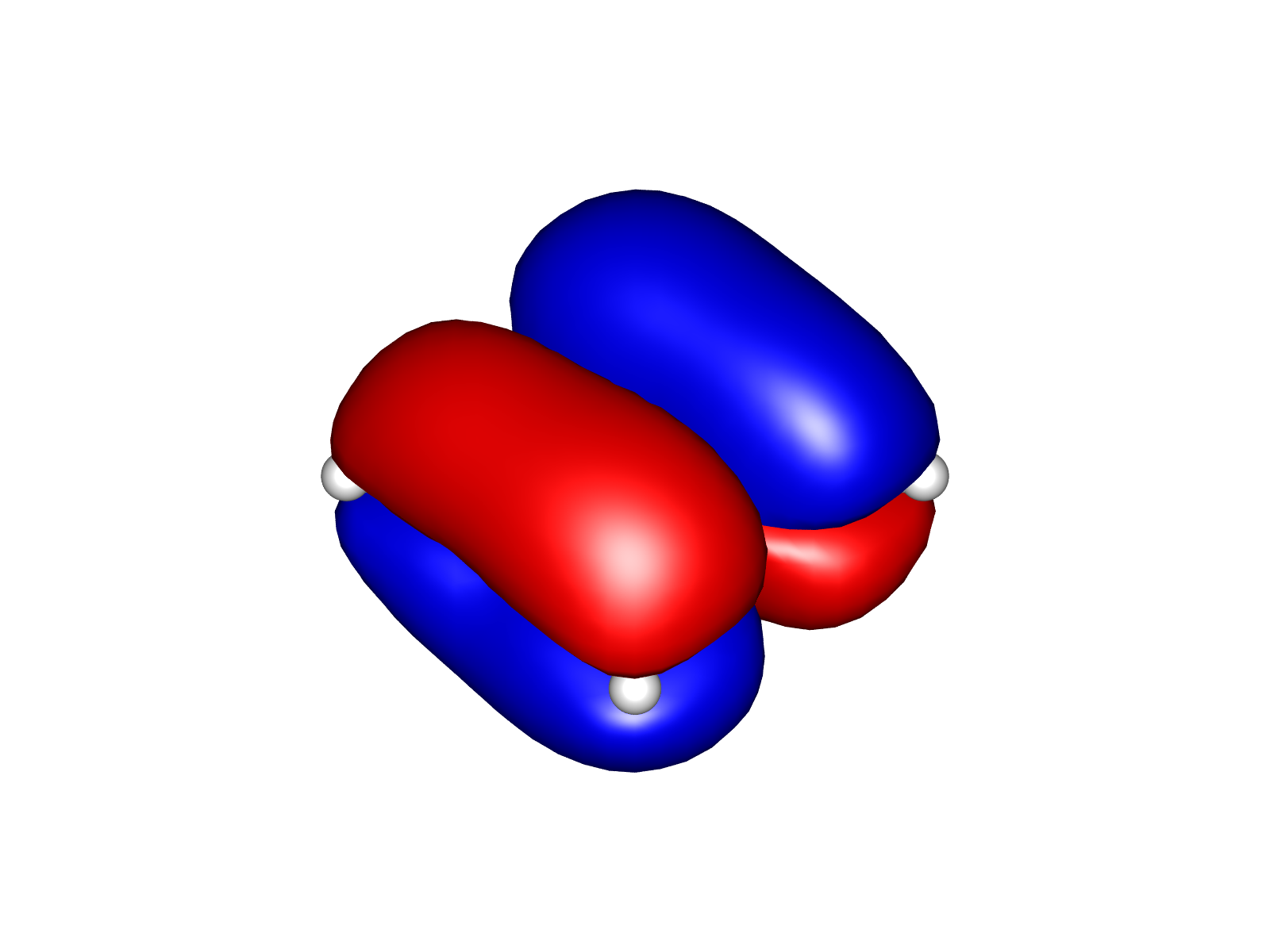}
  }
  \hfill
  \subfloat[$n_{\text{occup}}$ = 1.00]{%
    \includegraphics[width=0.22\textwidth]{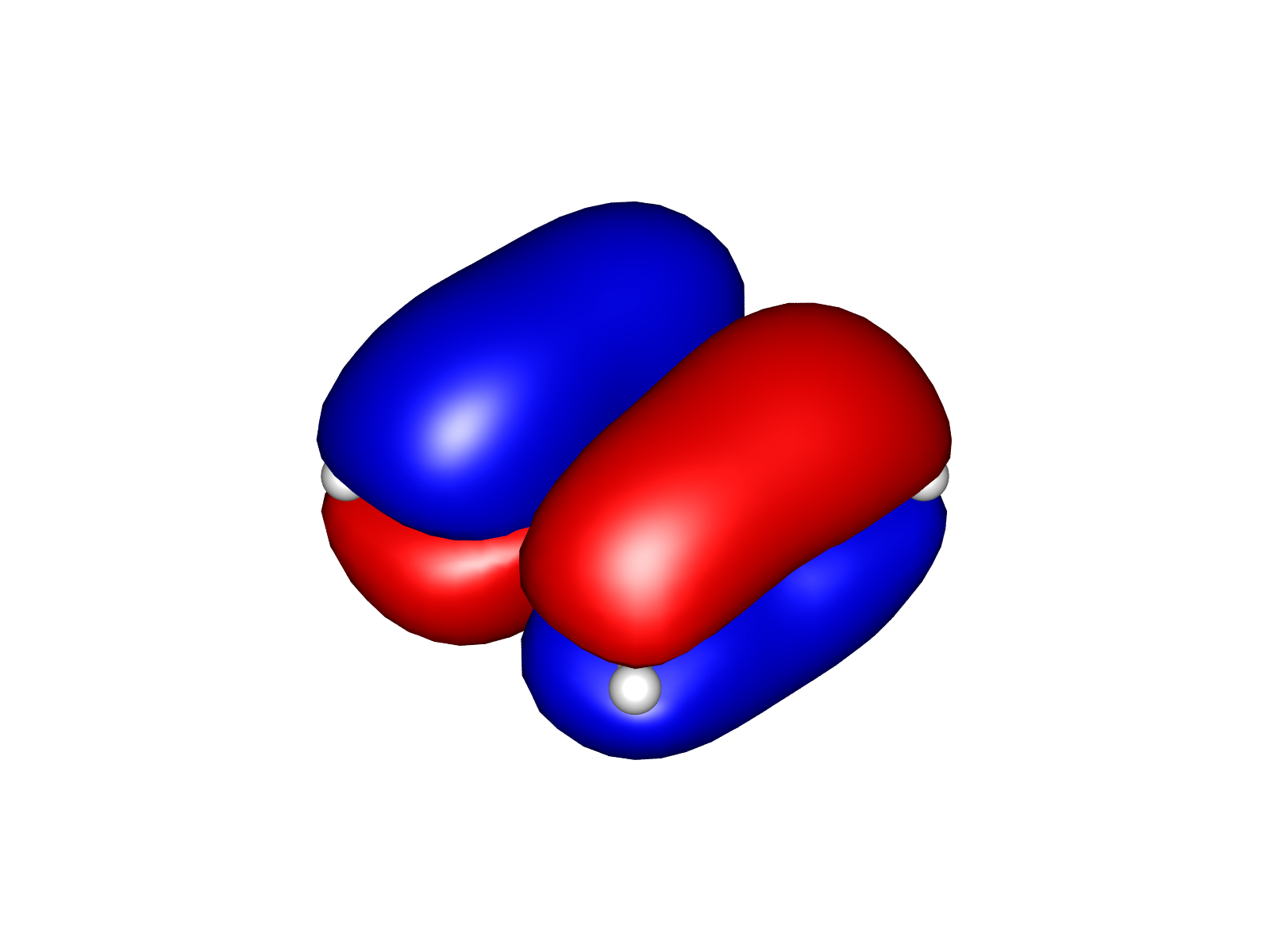}
  }
  \hfill
  \subfloat[$n_{\text{occup}}$ = 0.07]{%
    \includegraphics[width=0.22\textwidth]{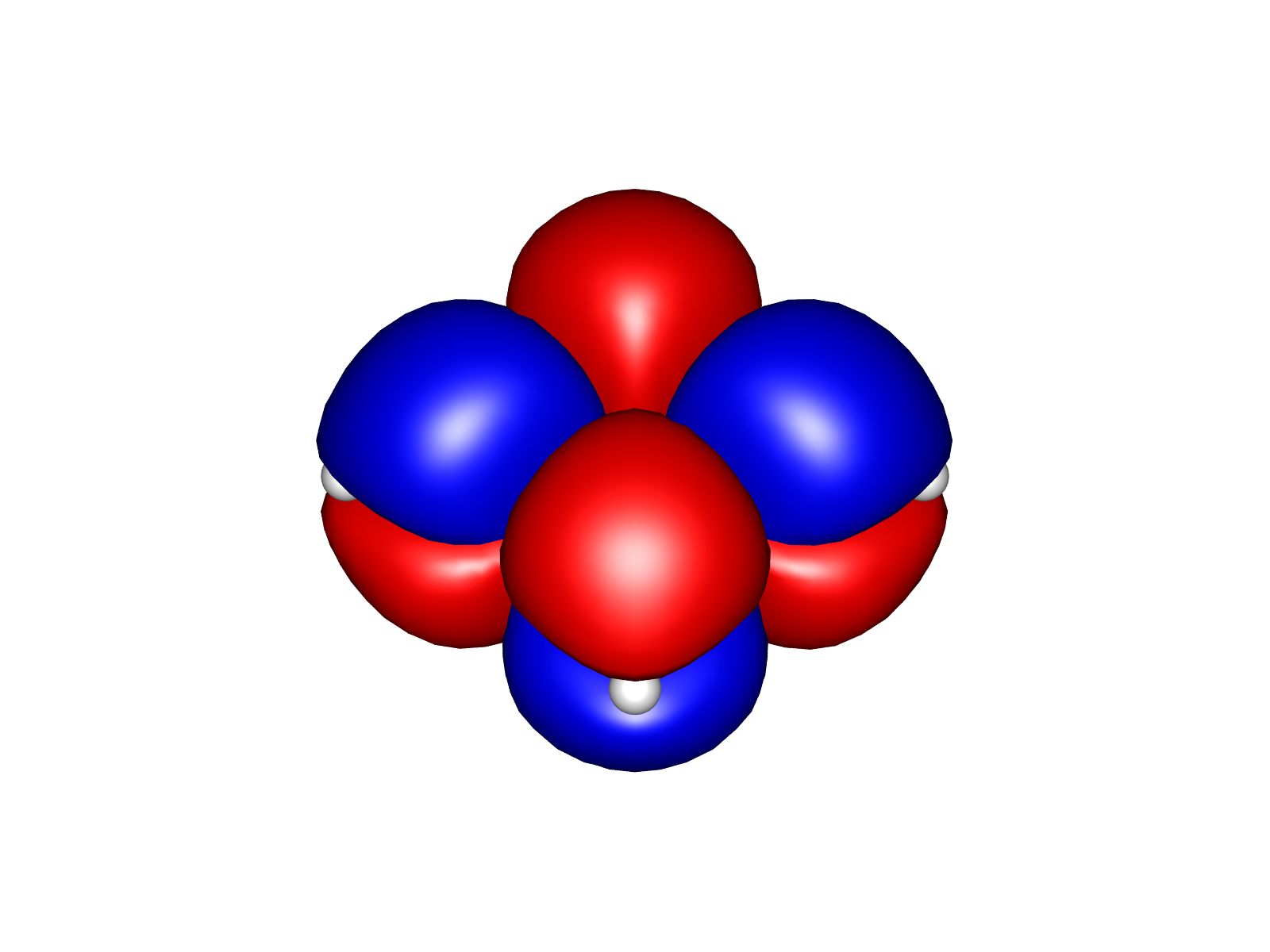}
  }
  \\
  \subfloat[$n_{\text{occup}}$ = 0.03]{%
    \includegraphics[width=0.22\textwidth]{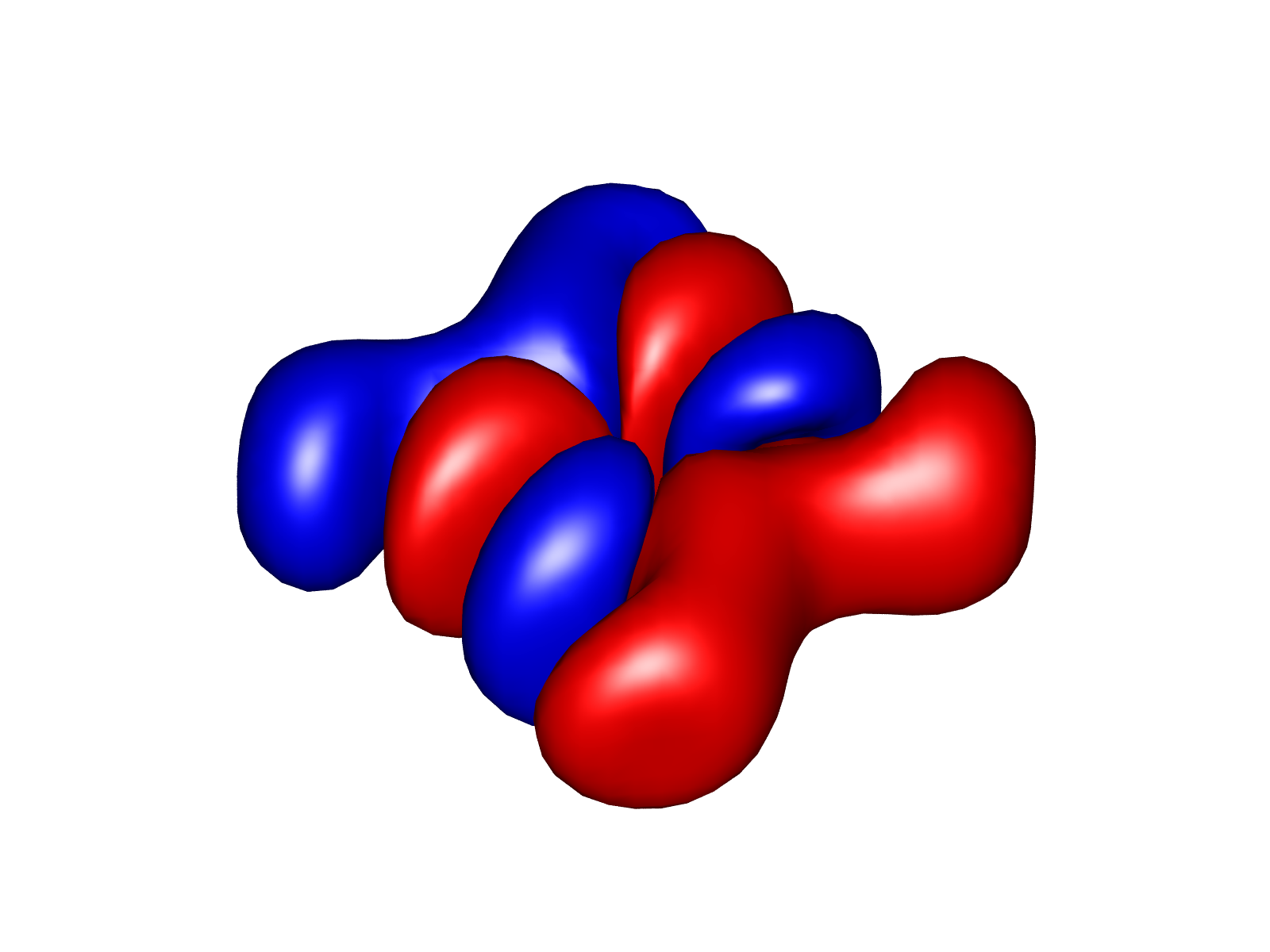}
  }
  \hfill
  \subfloat[$n_{\text{occup}}$ = 0.03]{%
    \includegraphics[width=0.22\textwidth]{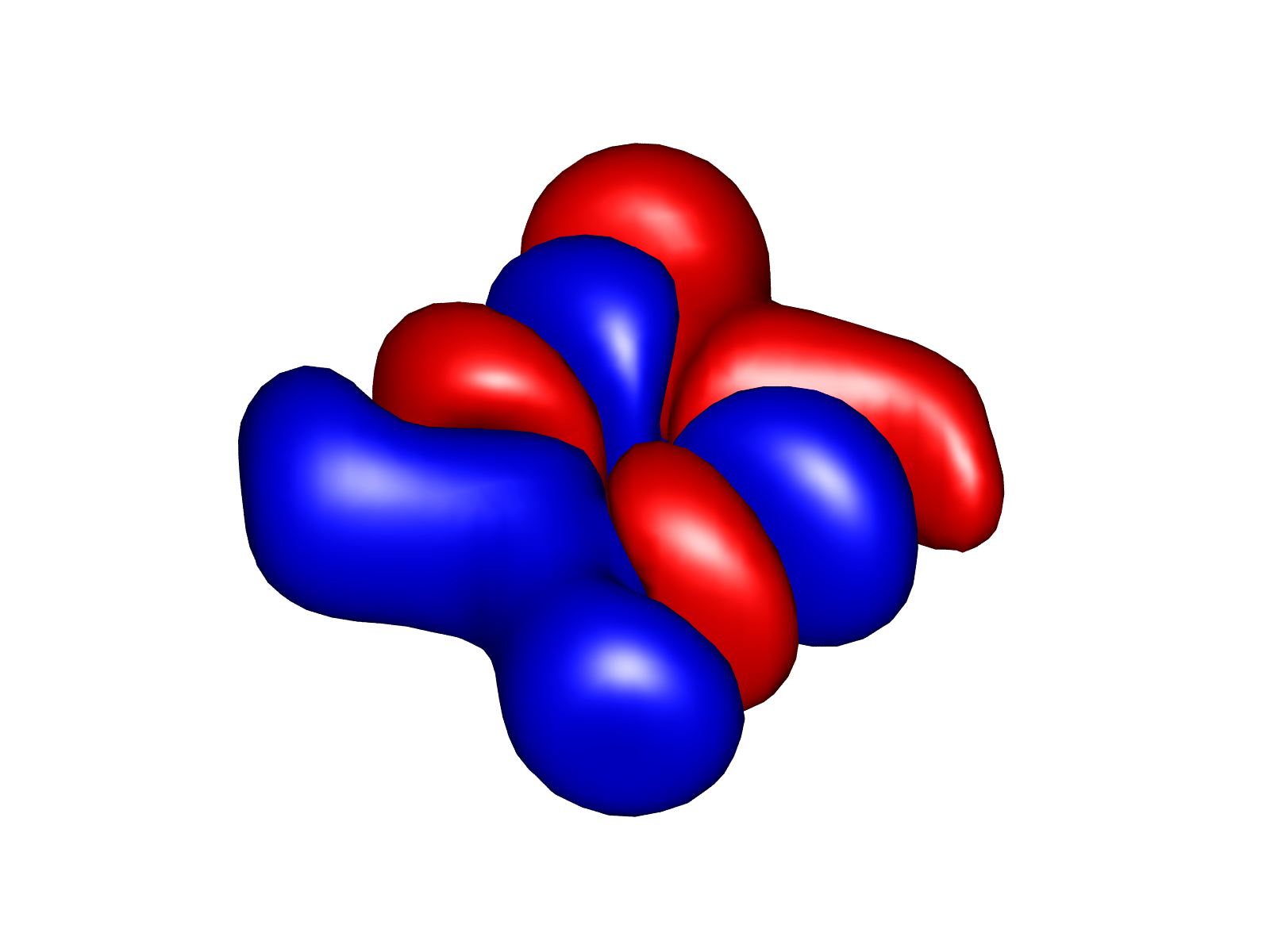}
  }
  \hfill
  \subfloat[$n_{\text{occup}}$ = 0.03]{%
    \includegraphics[width=0.22\textwidth]{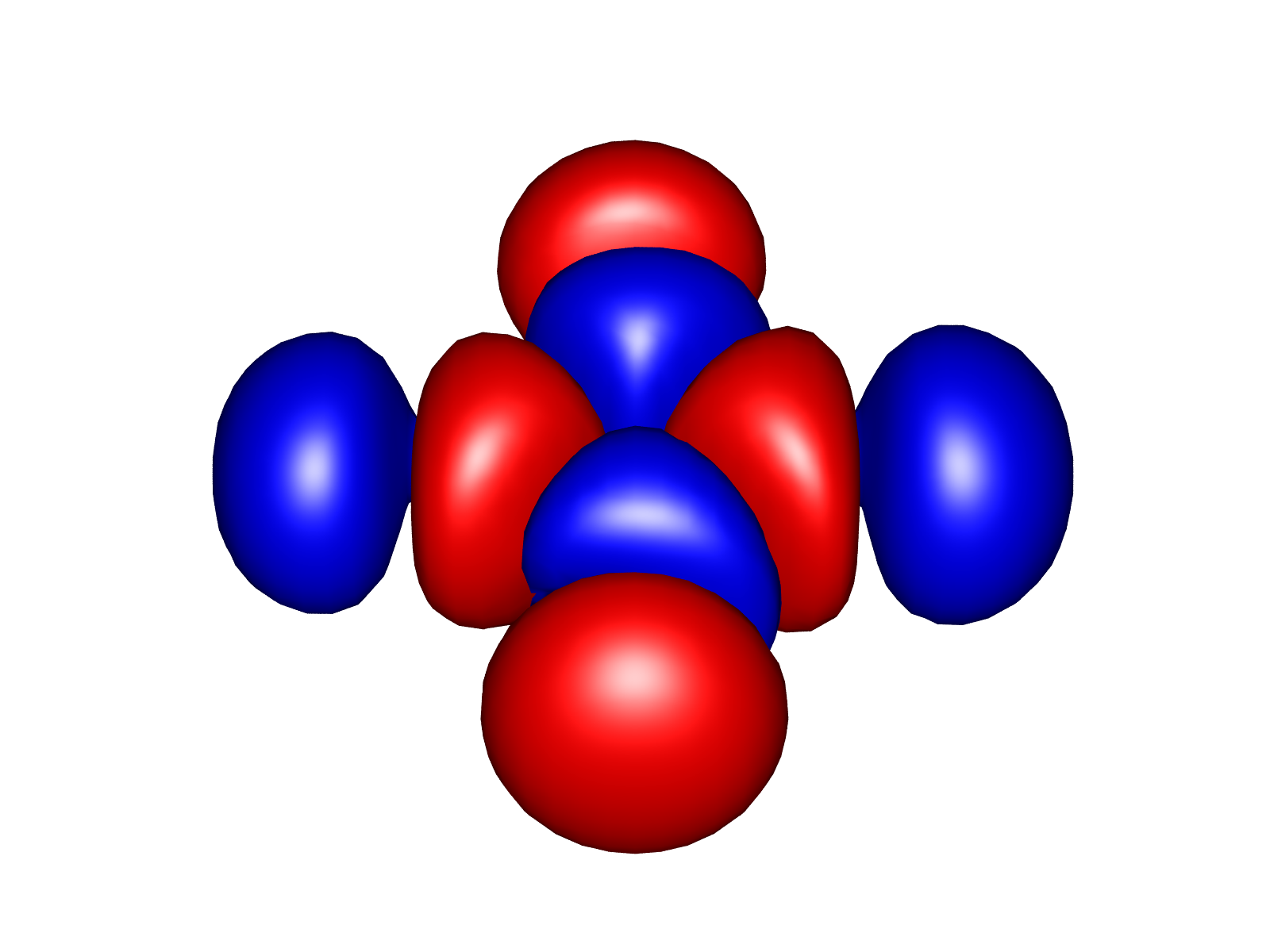}
  }
  \hfill
  \subfloat[$n_{\text{occup}}$ = 0.03]{%
    \includegraphics[width=0.22\textwidth]{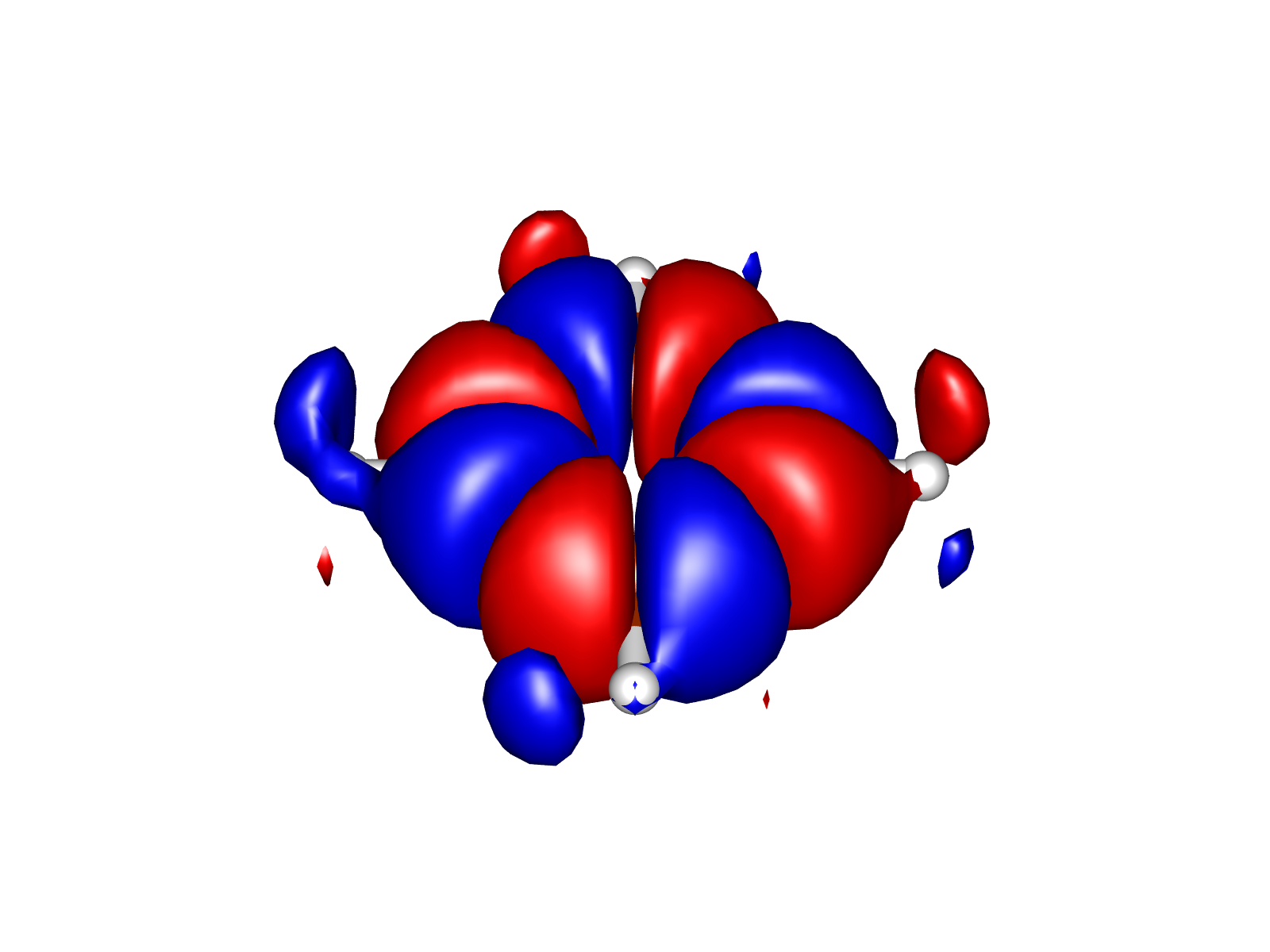}
  }
  \\
  \subfloat[$n_{\text{occup}}$ = 0.02]{%
    \includegraphics[width=0.22\textwidth]{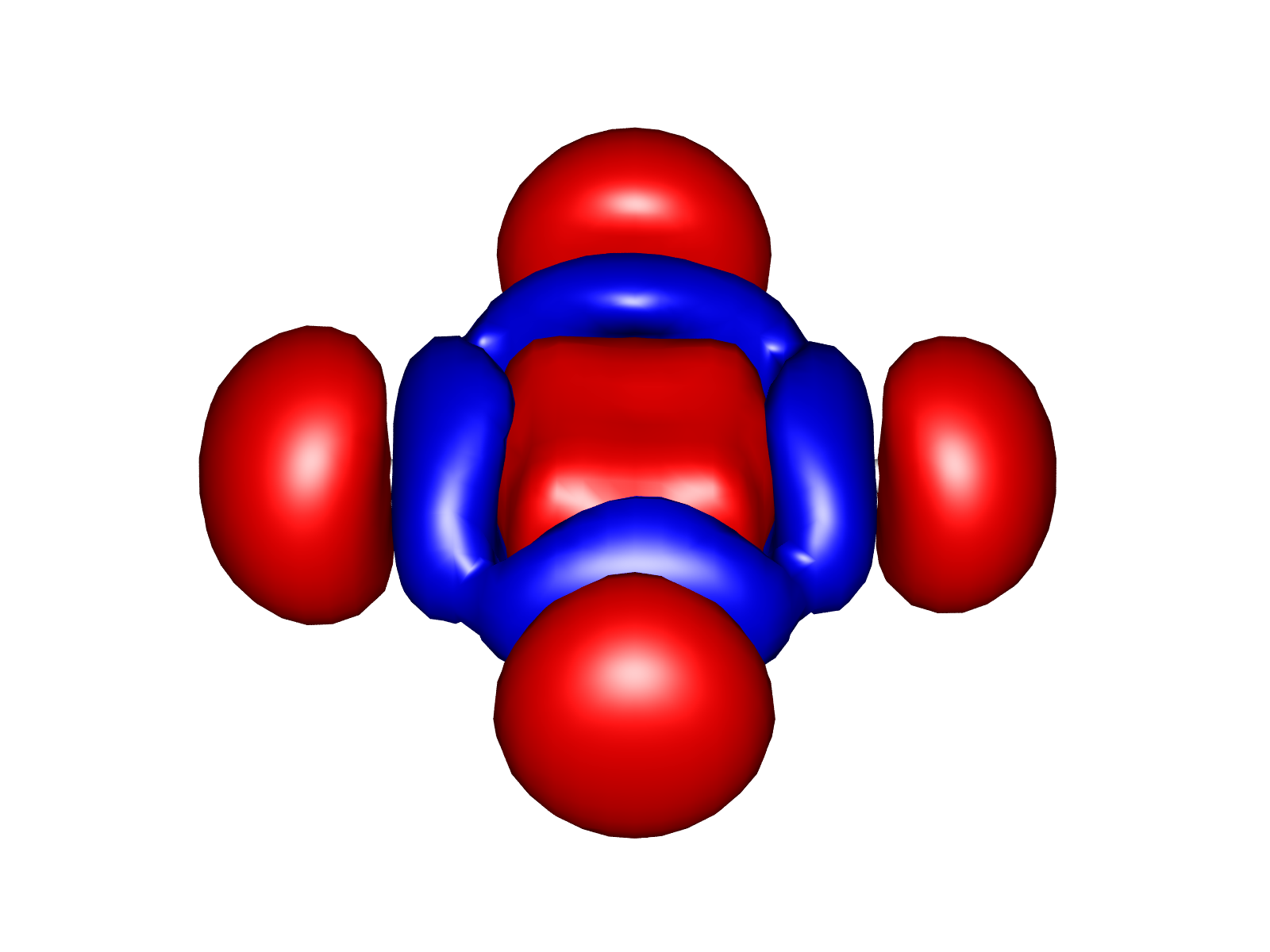}
  }
  \hfill
  \subfloat[$n_{\text{occup}}$ = 0.02]{%
    \includegraphics[width=0.22\textwidth]{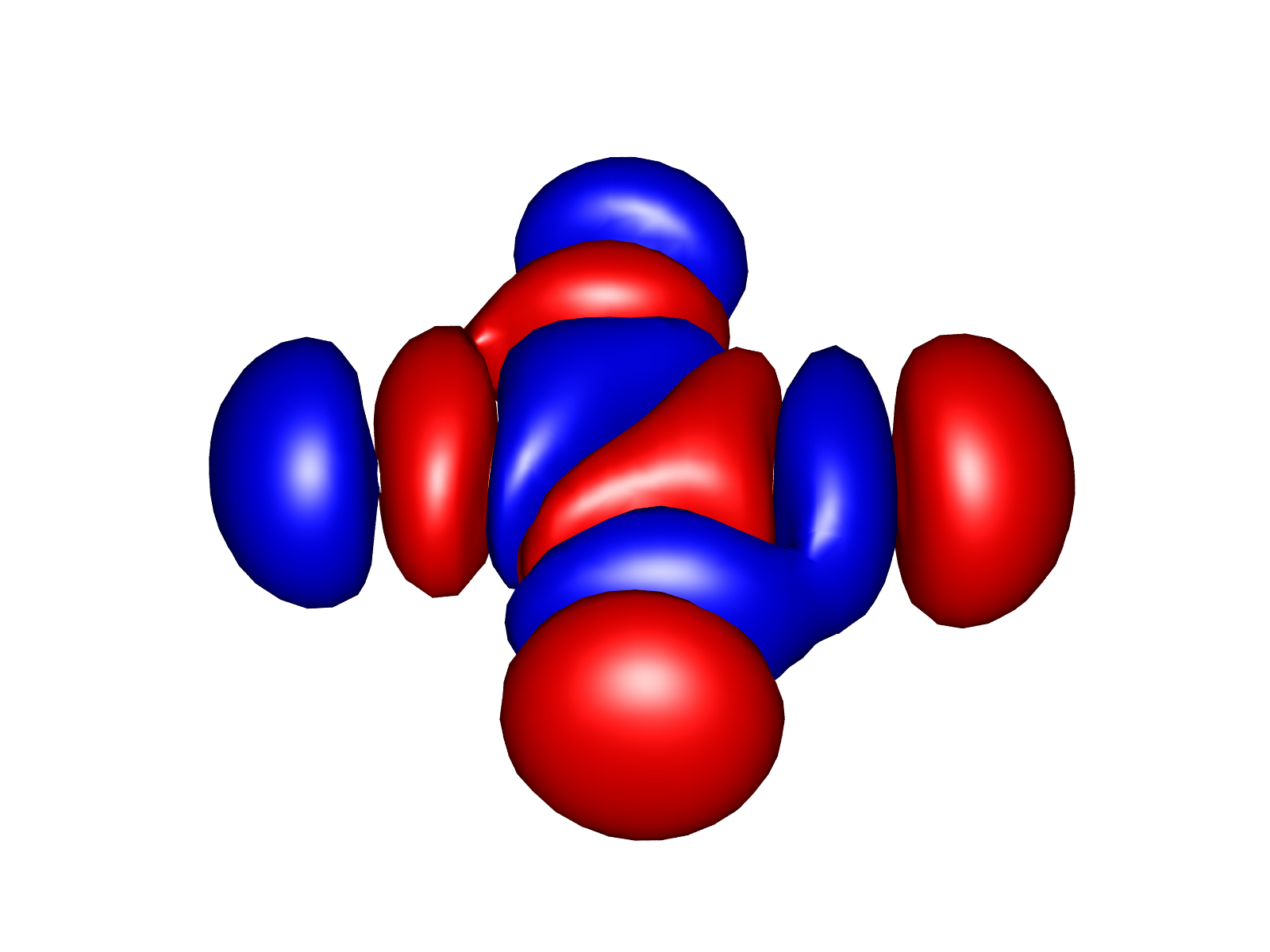}
  }
  \hfill
  \subfloat[$n_{\text{occup}}$ = 0.02]{%
    \includegraphics[width=0.22\textwidth]{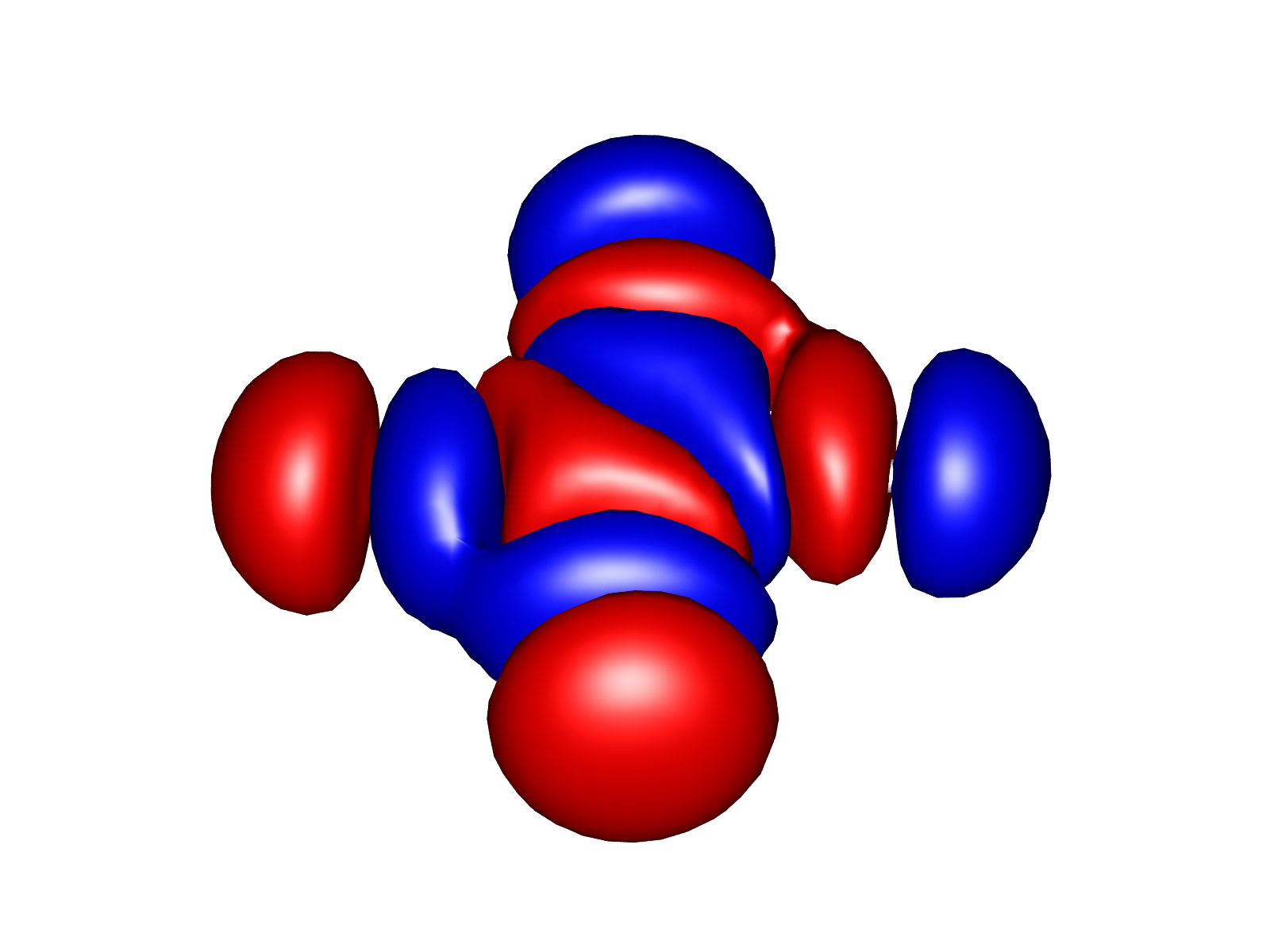}
  }
  \hfill
  \subfloat[$n_{\text{occup}}$ = 0.02]{%
    \includegraphics[width=0.22\textwidth]{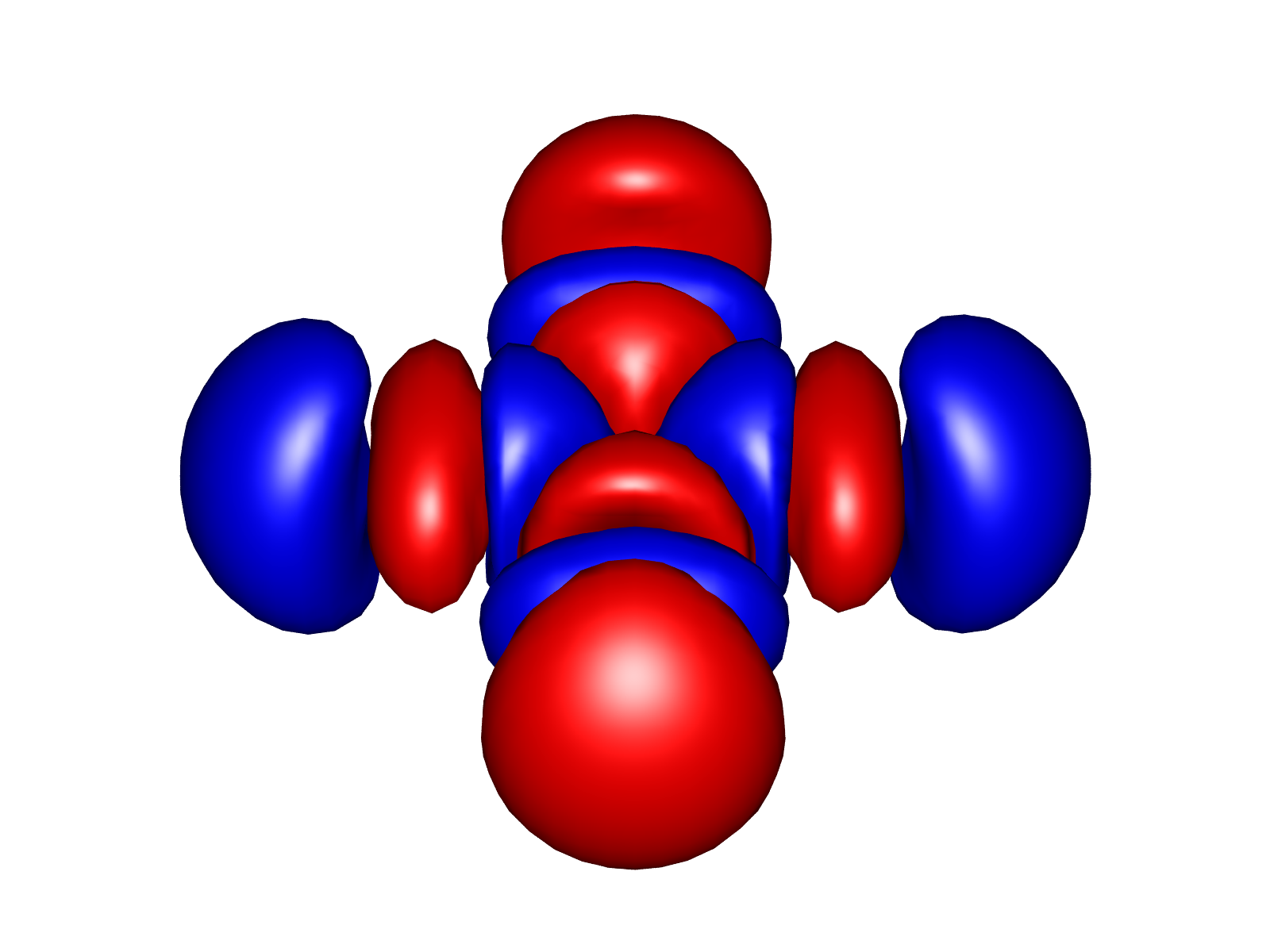}
  }
  \\
  \subfloat[$n_{\text{occup}}$ = 0.01]{%
    \includegraphics[width=0.22\textwidth]{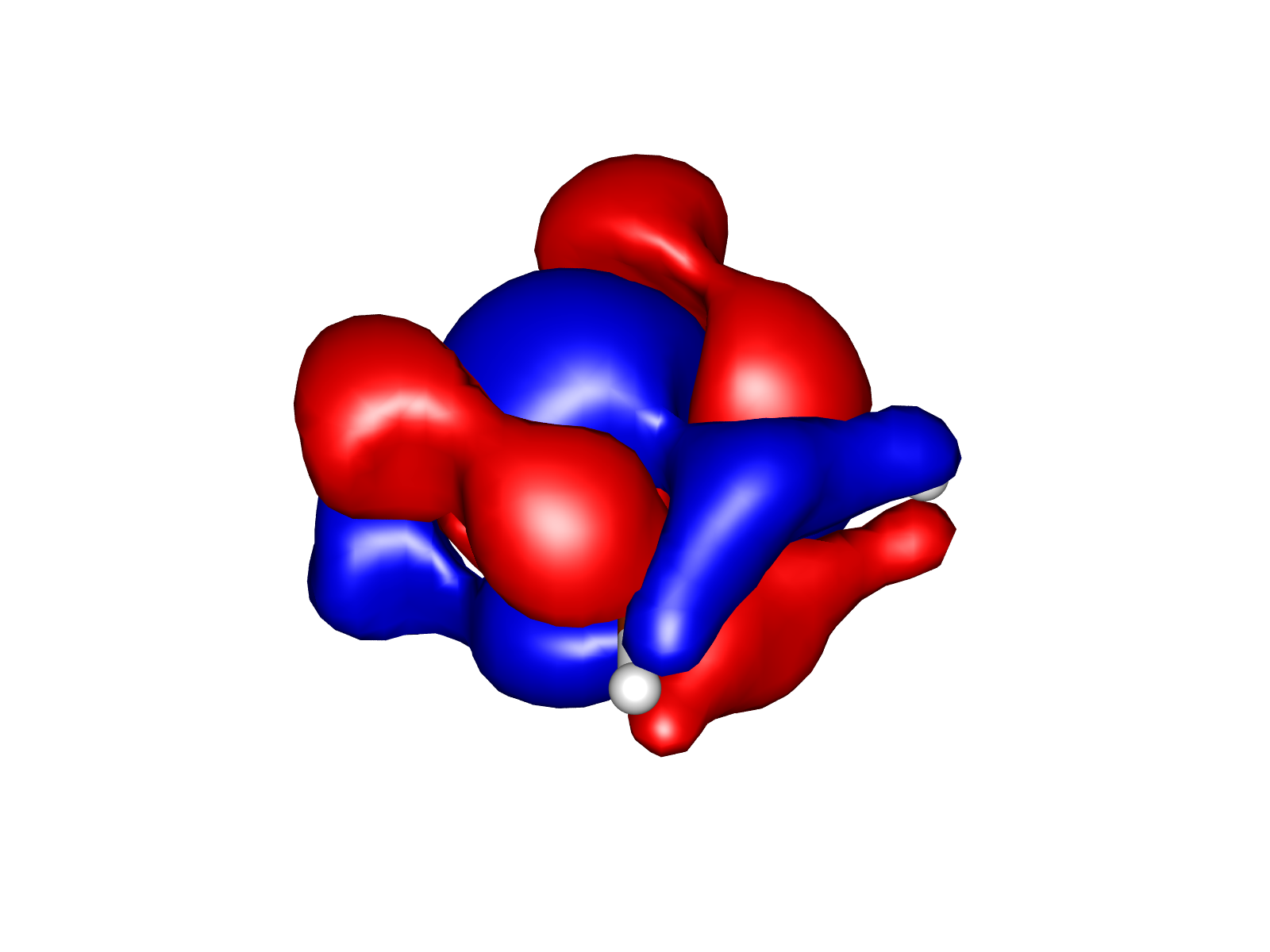}
  }
  \subfloat[$n_{\text{occup}}$ = 0.01]{%
    \includegraphics[width=0.22\textwidth]{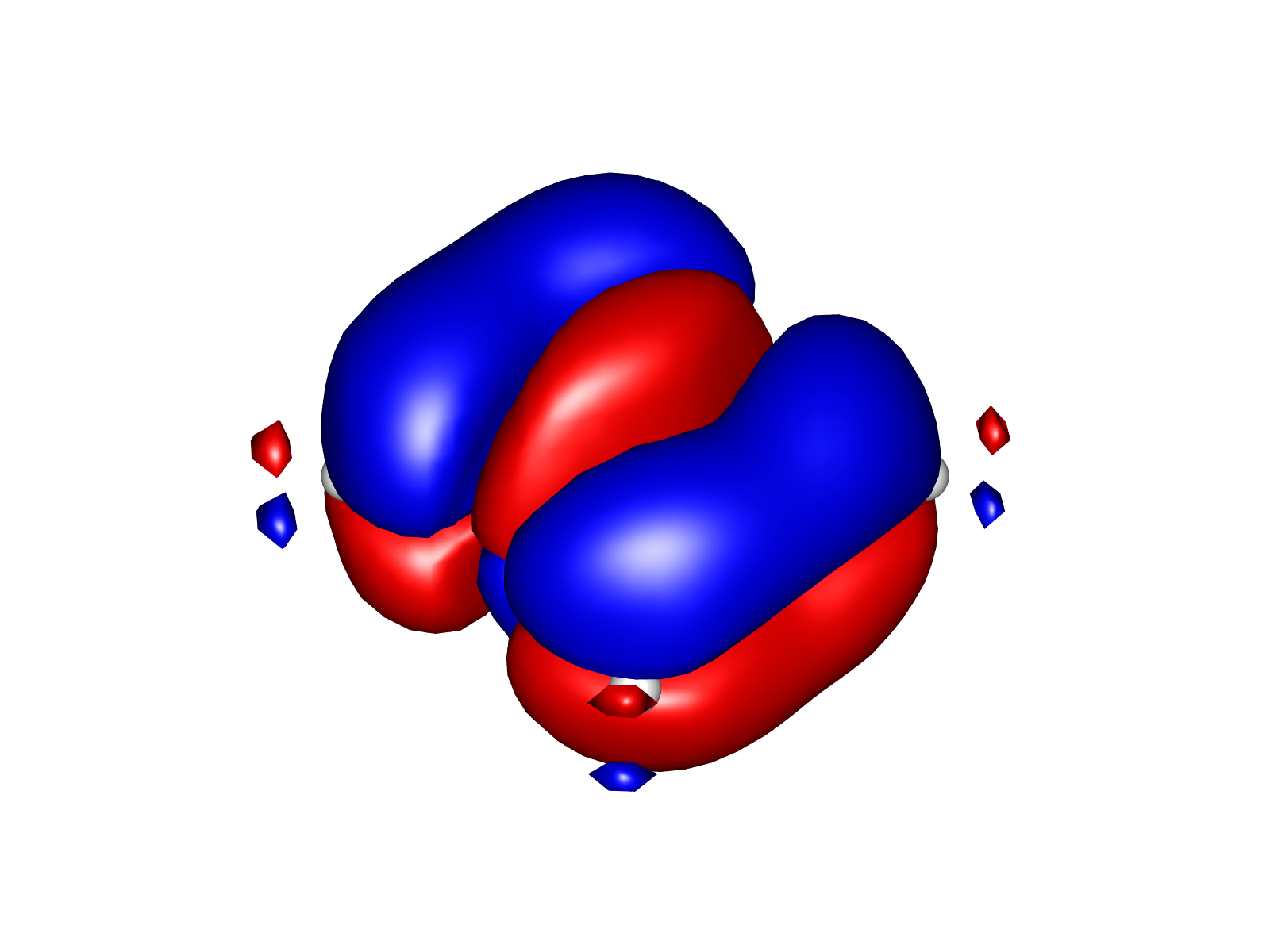}
  }
  \caption{C$_4$H$_4$, triplet state, DMRG-SCF(20, 22) \label{orbs_c4h4_t}}
\end{figure}

\begin{figure}[!h]
  \subfloat[$n_{\text{occup}}$ = 1.99]{%
    \includegraphics[width=0.22\textwidth]{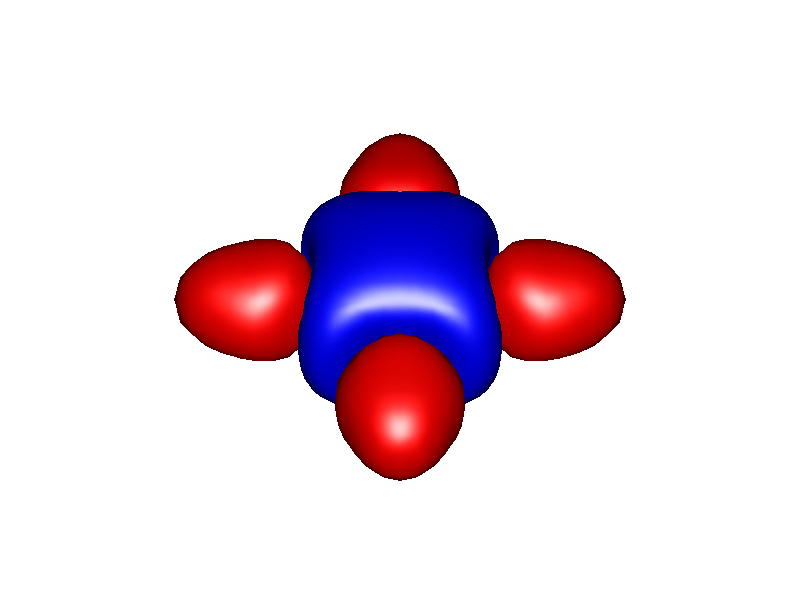}
  }
  \hfill
  \subfloat[$n_{\text{occup}}$ = 1.98]{%
    \includegraphics[width=0.22\textwidth]{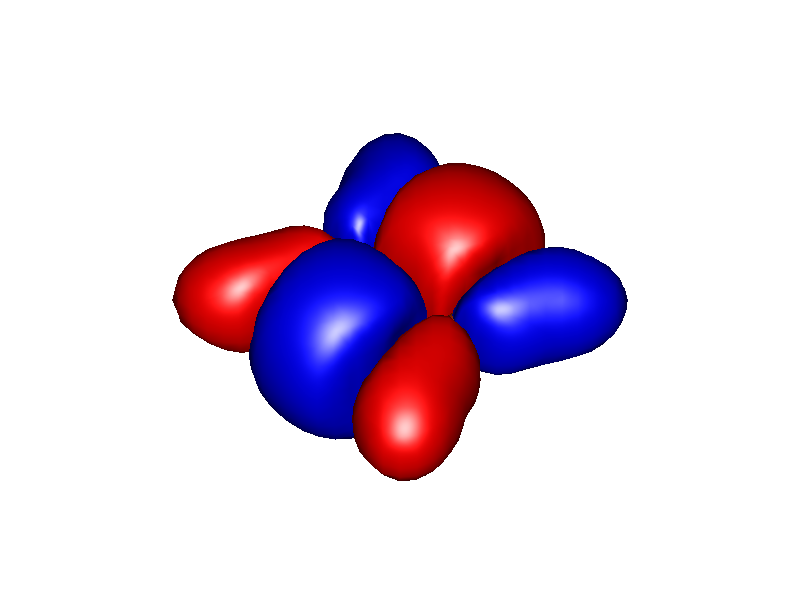}
  }
  \hfill
  \subfloat[$n_{\text{occup}}$ = 1.98]{%
    \includegraphics[width=0.22\textwidth]{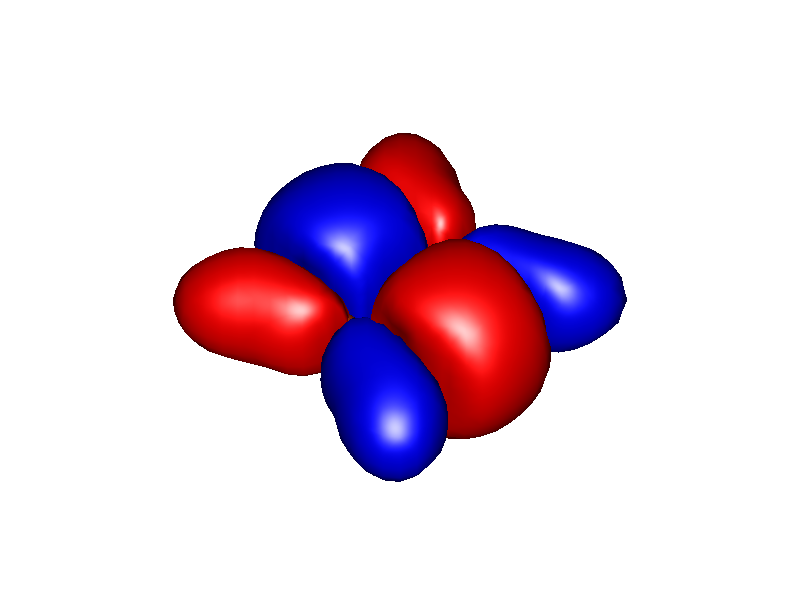}
  }
  \hfill
  \subfloat[$n_{\text{occup}}$ = 1.97]{%
    \includegraphics[width=0.22\textwidth]{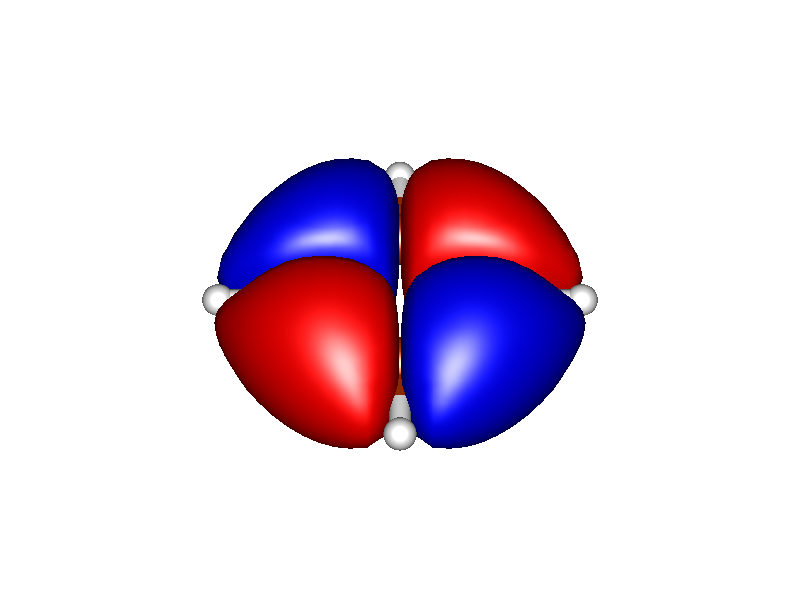}
  } \\
  \subfloat[$n_{\text{occup}}$ = 1.91]{%
    \includegraphics[width=0.22\textwidth]{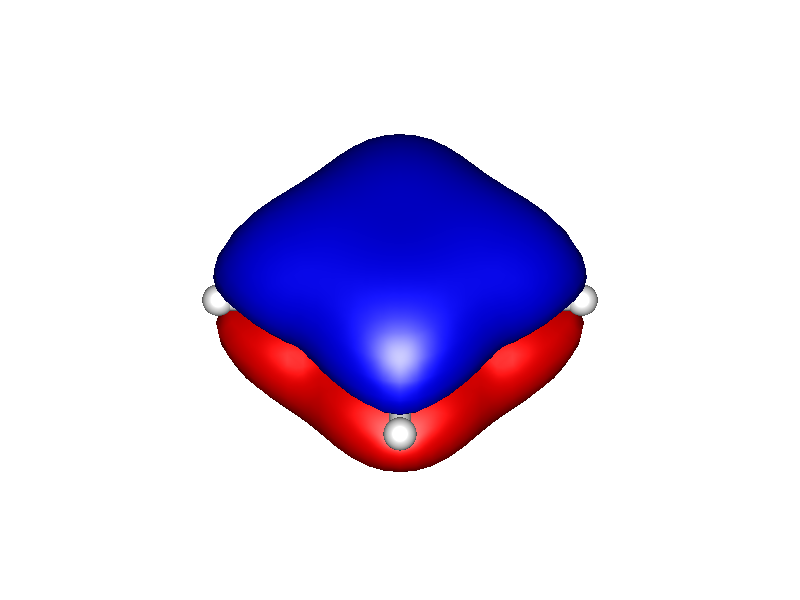}
  }
  \hfill
  \subfloat[$n_{\text{occup}}$ = 1.00]{%
    \includegraphics[width=0.22\textwidth]{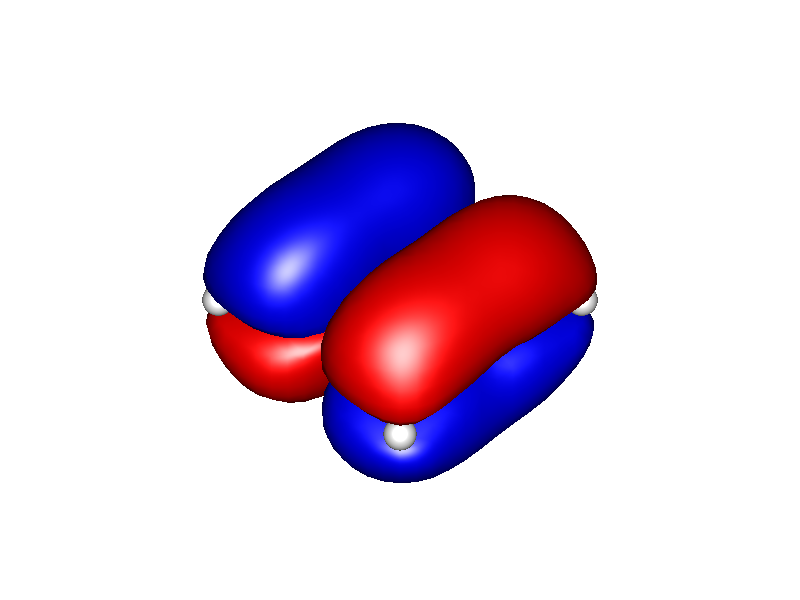}
  }
  \hfill
  \subfloat[$n_{\text{occup}}$ = 1.00]{%
    \includegraphics[width=0.22\textwidth]{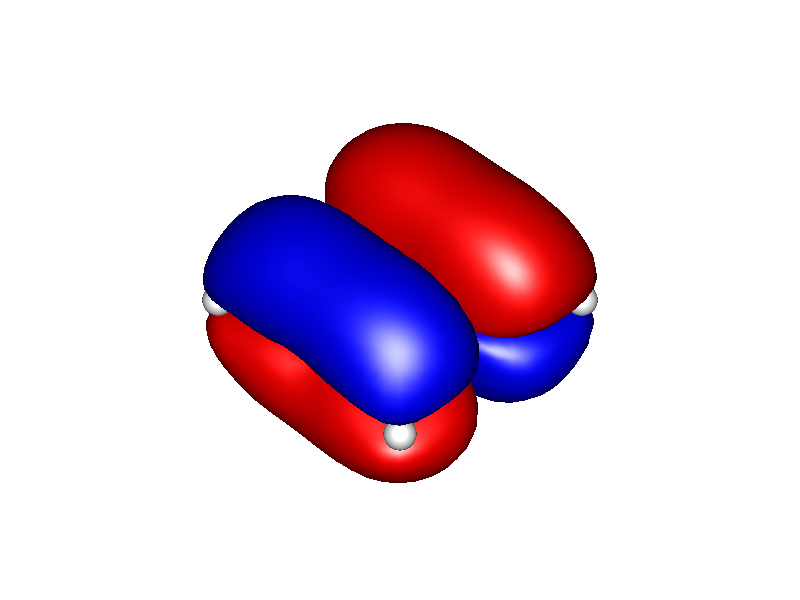}
  }
  \hfill
  \subfloat[$n_{\text{occup}}$ = 0.08]{%
    \includegraphics[width=0.22\textwidth]{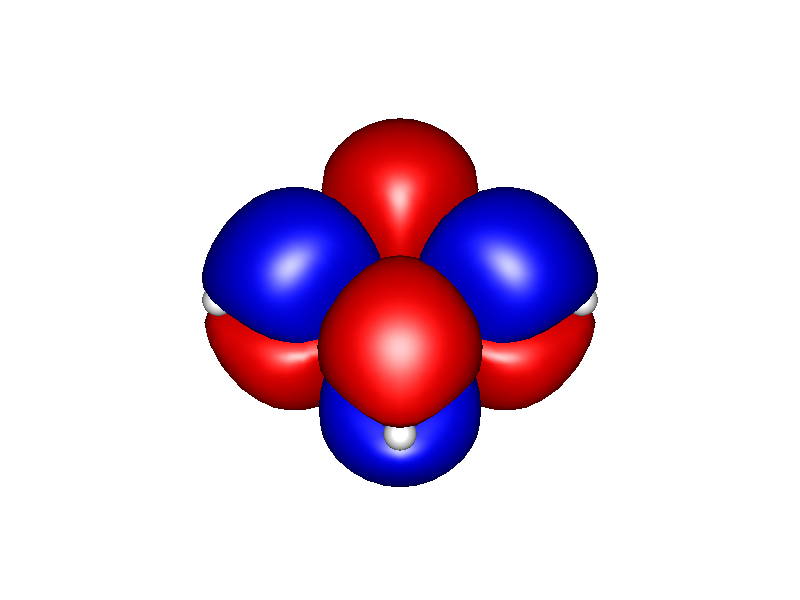}
  }
  \\
  \subfloat[$n_{\text{occup}}$ = 0.02]{%
    \includegraphics[width=0.22\textwidth]{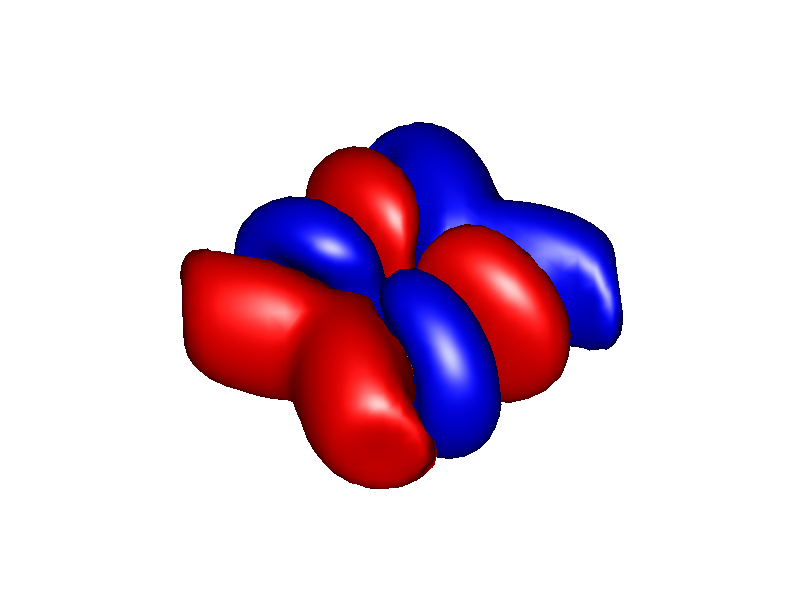}
  }
  \hfill
  \subfloat[$n_{\text{occup}}$ = 0.02]{%
    \includegraphics[width=0.22\textwidth]{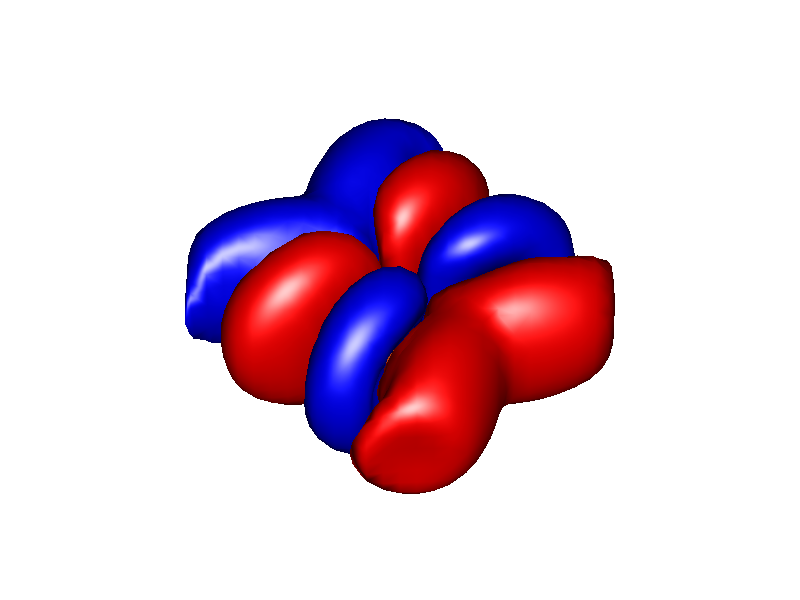}
  }
  \hfill
  \subfloat[$n_{\text{occup}}$ = 0.02]{%
    \includegraphics[width=0.22\textwidth]{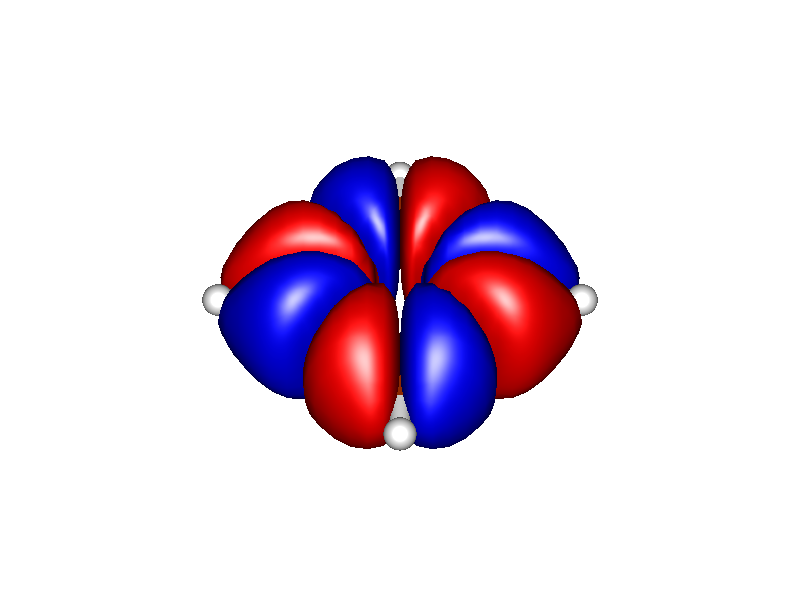}
  }
  \hfill
  \subfloat[$n_{\text{occup}}$ = 0.01]{%
    \includegraphics[width=0.22\textwidth]{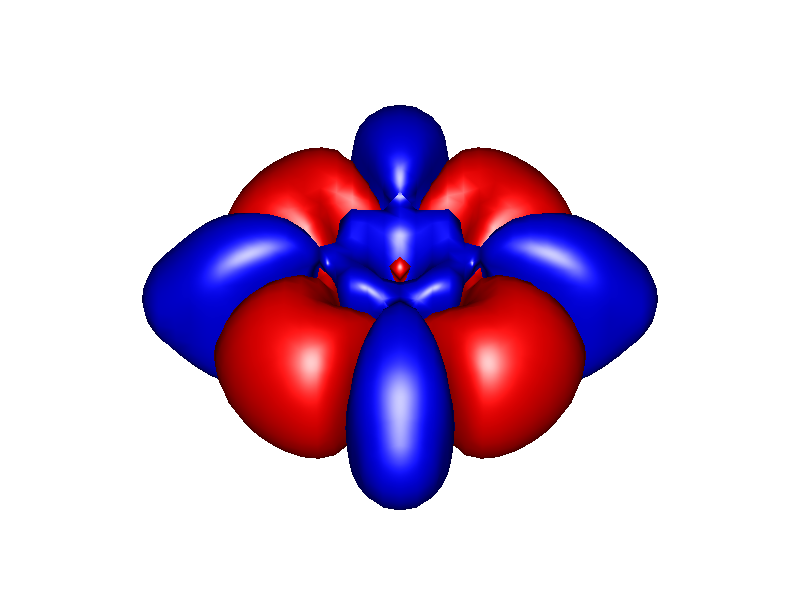}
  }
  \\
  \subfloat[$n_{\text{occup}}$ = 0.004]{%
    \includegraphics[width=0.22\textwidth]{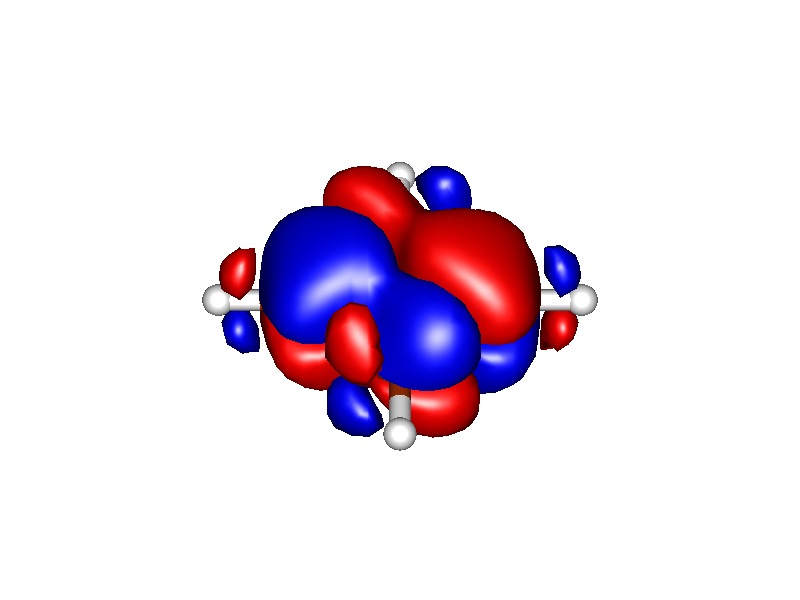}
  }
  \subfloat[$n_{\text{occup}}$ = 0.004]{%
    \includegraphics[width=0.22\textwidth]{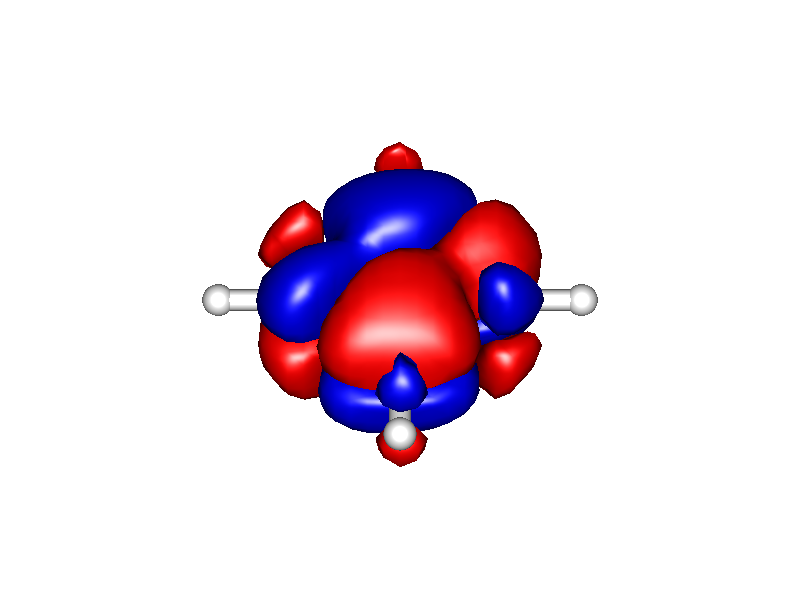}
  }
  \caption{C$_4$H$_4$, singlet state, CASSCF(12, 14) \label{orbs_c4h4small_s}}
\end{figure}

\begin{figure}[!h]
  \subfloat[$n_{\text{occup}}$ = 1.99]{%
    \includegraphics[width=0.22\textwidth]{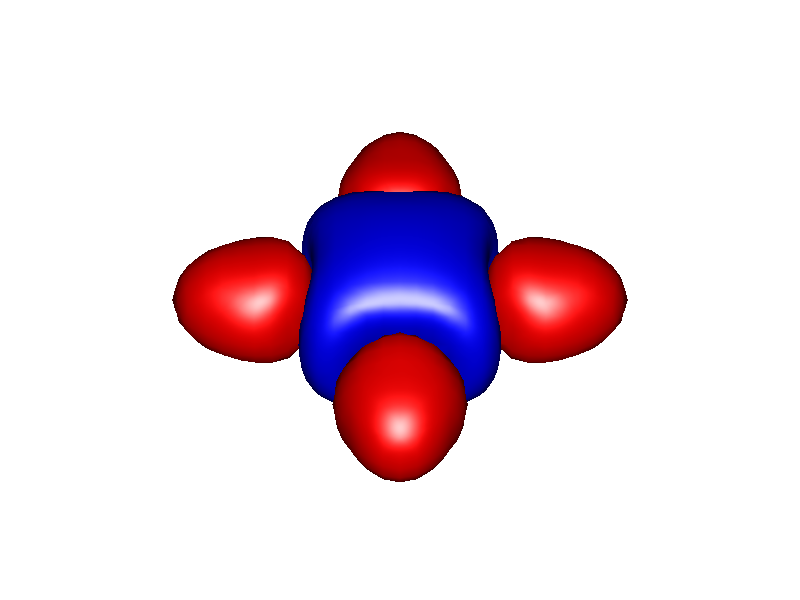}
  }
  \hfill
  \subfloat[$n_{\text{occup}}$ = 1.98]{%
    \includegraphics[width=0.22\textwidth]{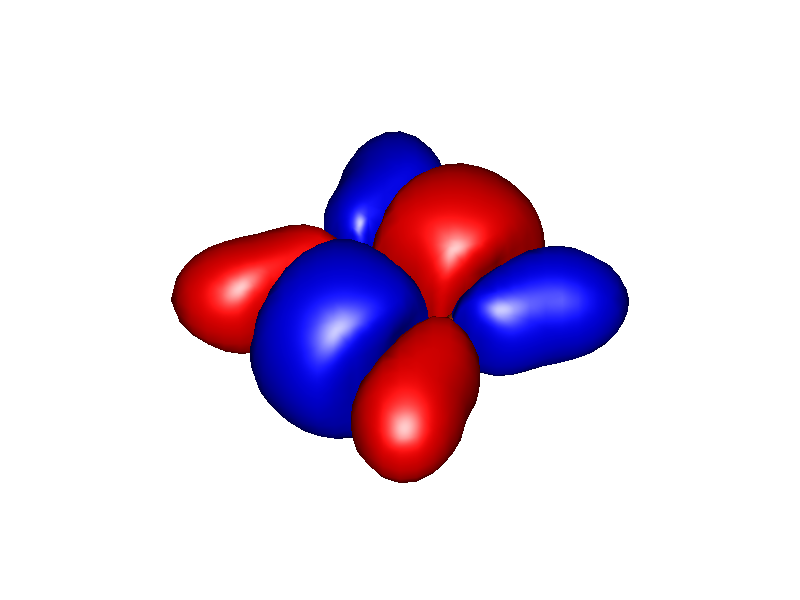}
  }
  \hfill
  \subfloat[$n_{\text{occup}}$ = 1.98]{%
    \includegraphics[width=0.22\textwidth]{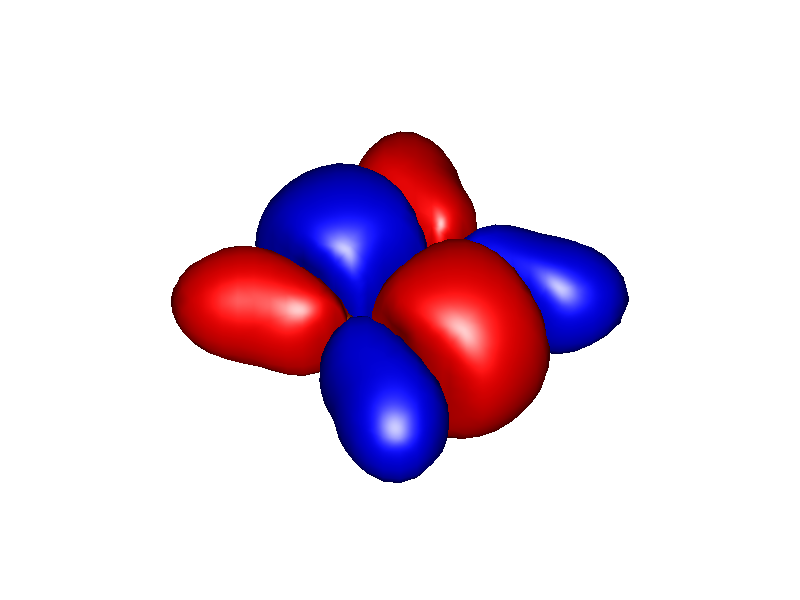}
  }
  \hfill
  \subfloat[$n_{\text{occup}}$ = 1.97]{%
    \includegraphics[width=0.22\textwidth]{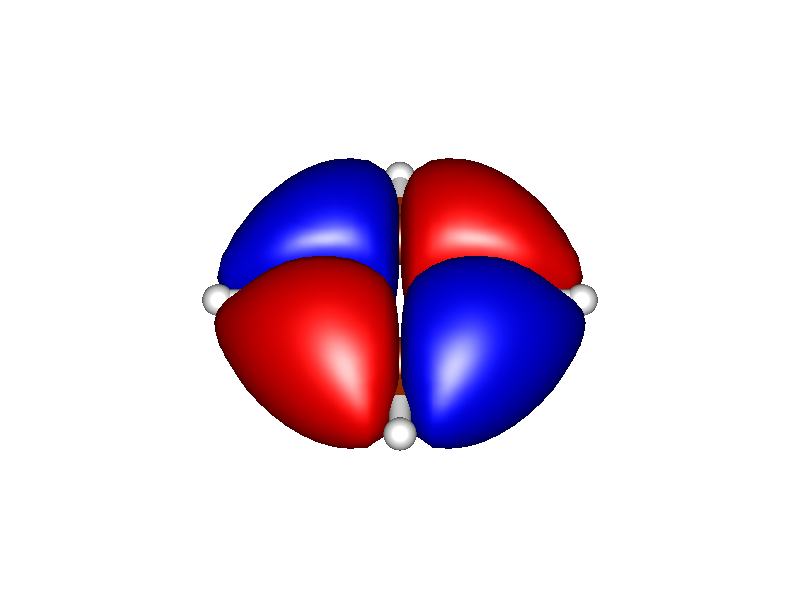}
  } \\
  \subfloat[$n_{\text{occup}}$ = 1.92]{%
    \includegraphics[width=0.22\textwidth]{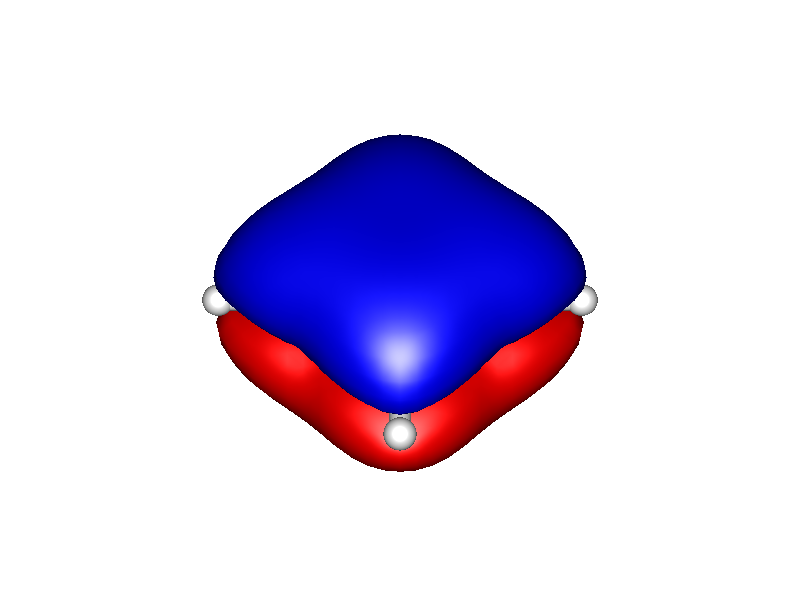}
  }
  \hfill
  \subfloat[$n_{\text{occup}}$ = 1.00]{%
    \includegraphics[width=0.22\textwidth]{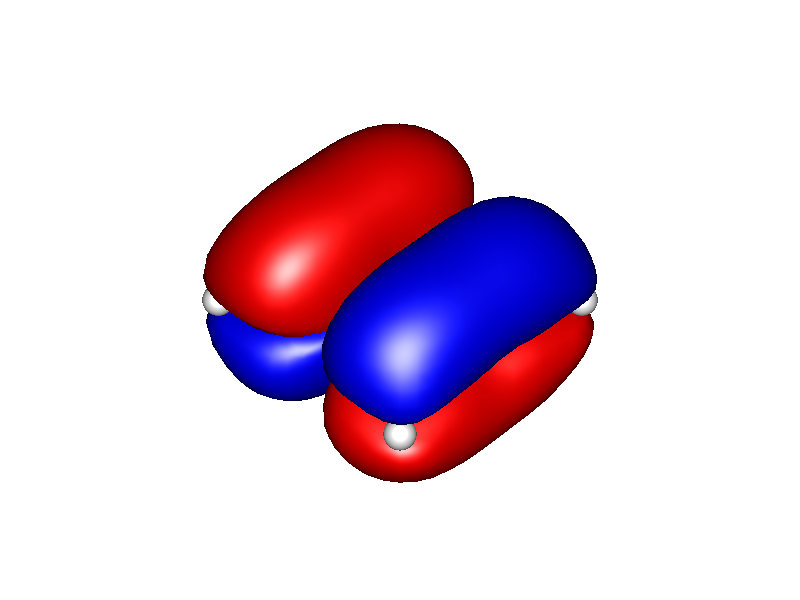}
  }
  \hfill
  \subfloat[$n_{\text{occup}}$ = 1.00]{%
    \includegraphics[width=0.22\textwidth]{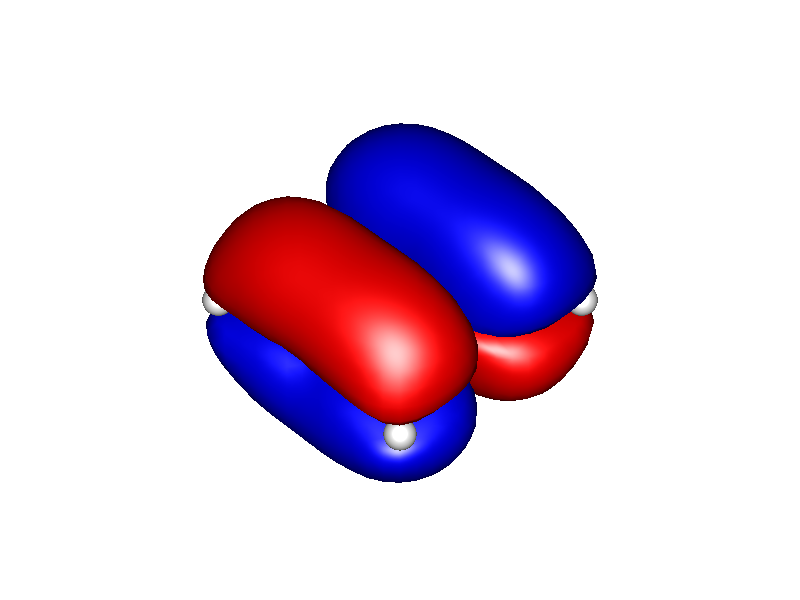}
  }
  \hfill
  \subfloat[$n_{\text{occup}}$ = 0.07]{%
    \includegraphics[width=0.22\textwidth]{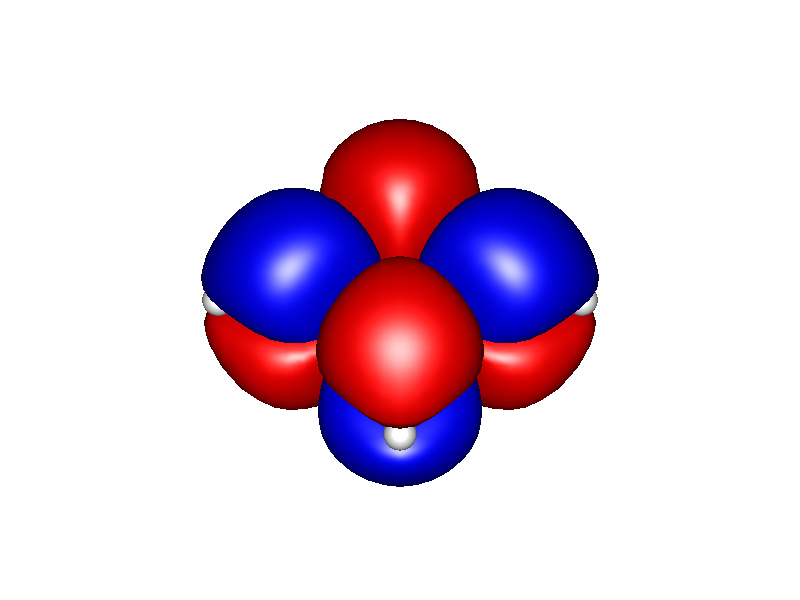}
  }
  \\
  \subfloat[$n_{\text{occup}}$ = 0.02]{%
    \includegraphics[width=0.22\textwidth]{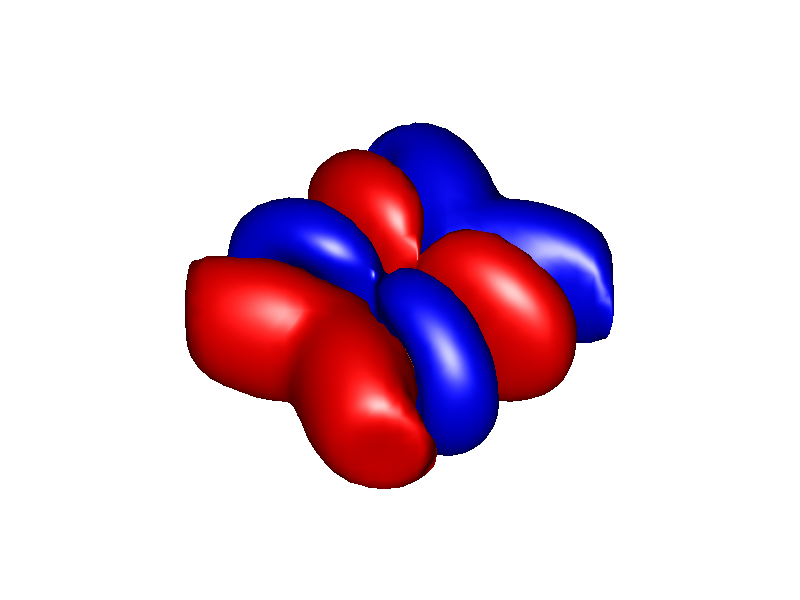}
  }
  \hfill
  \subfloat[$n_{\text{occup}}$ = 0.02]{%
    \includegraphics[width=0.22\textwidth]{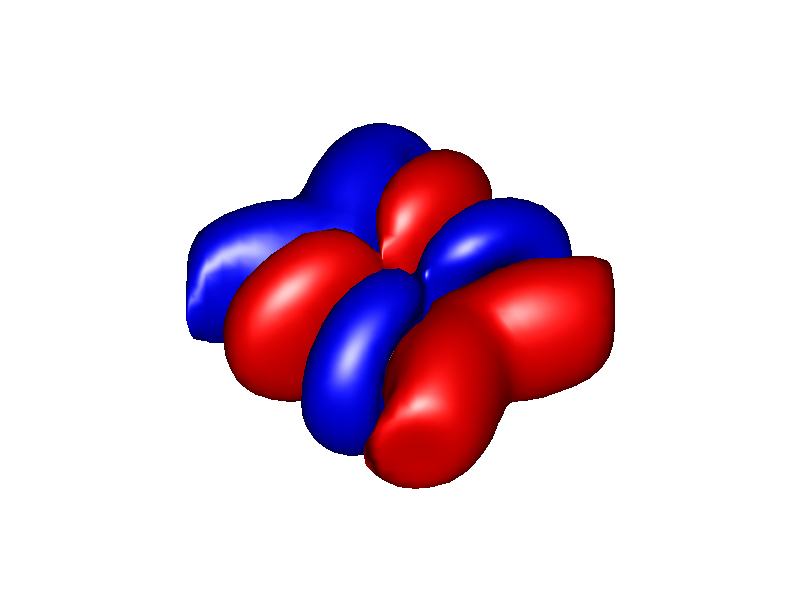}
  }
  \hfill
  \subfloat[$n_{\text{occup}}$ = 0.02]{%
    \includegraphics[width=0.22\textwidth]{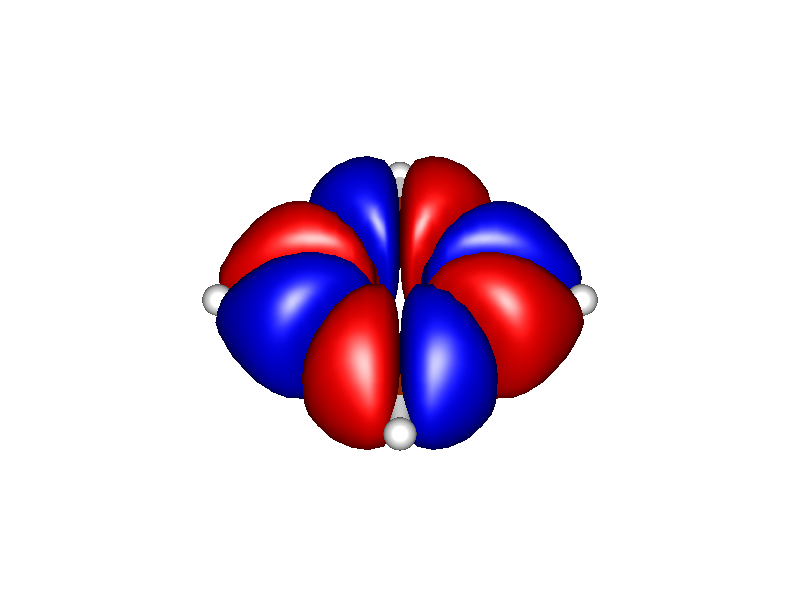}
  }
  \hfill
  \subfloat[$n_{\text{occup}}$ = 0.01]{%
    \includegraphics[width=0.22\textwidth]{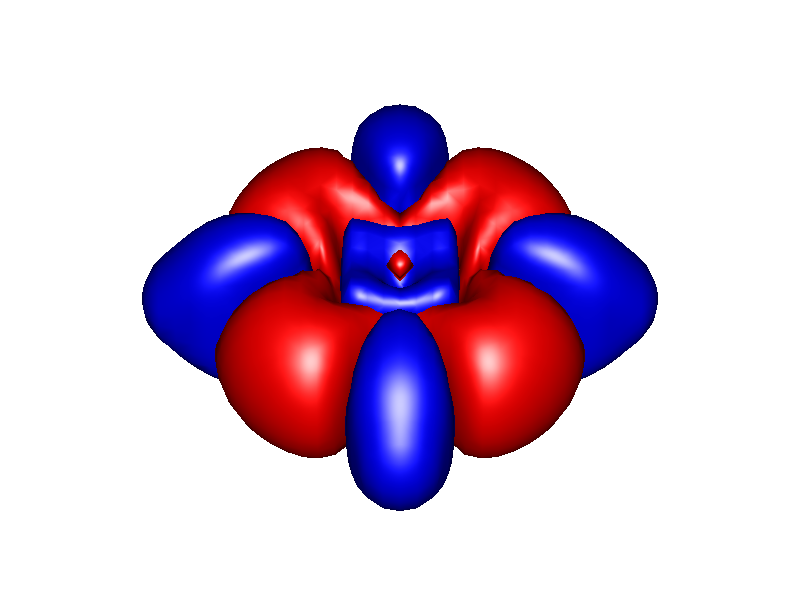}
  }
  \\
  \subfloat[$n_{\text{occup}}$ = 0.004]{%
    \includegraphics[width=0.22\textwidth]{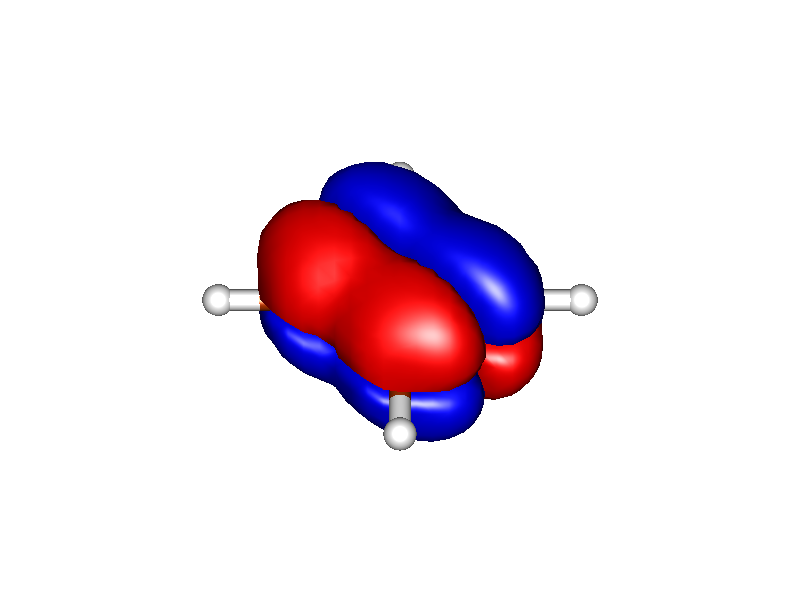}
  }
  \subfloat[$n_{\text{occup}}$ = 0.004]{%
    \includegraphics[width=0.22\textwidth]{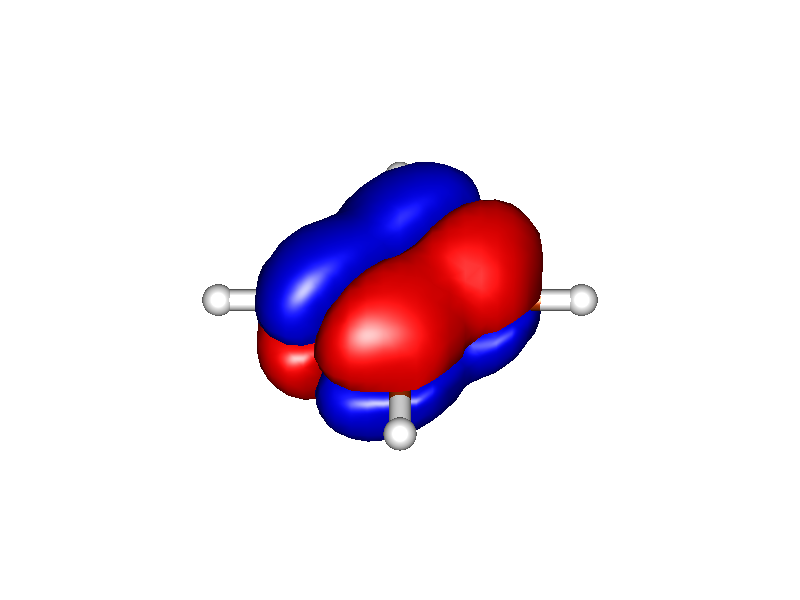}
  }
  \caption{C$_4$H$_4$, triplet state, CASSCF(12, 14) \label{orbs_c4h4small_t}}
\end{figure}

\begin{figure}[!h]
  \subfloat[$n_{\text{occup}}$ = 1.99]{%
    \includegraphics[width=0.22\textwidth]{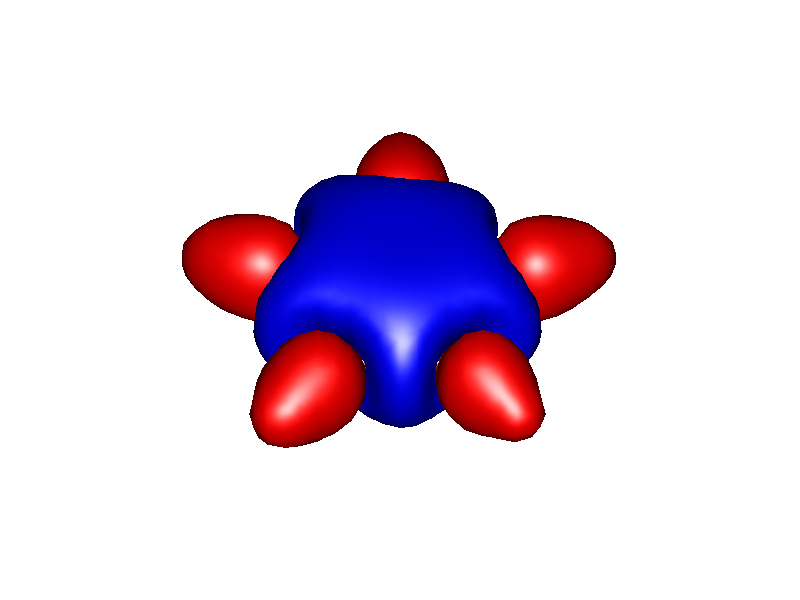}
  }
  \hfill
  \subfloat[$n_{\text{occup}}$ = 1.98]{%
    \includegraphics[width=0.22\textwidth]{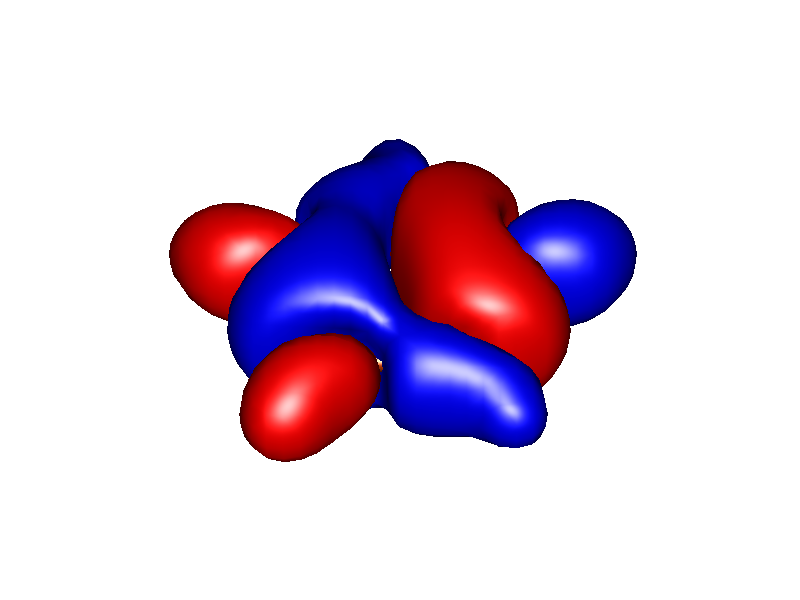}
  }
  \hfill
  \subfloat[$n_{\text{occup}}$ = 1.98]{%
    \includegraphics[width=0.22\textwidth]{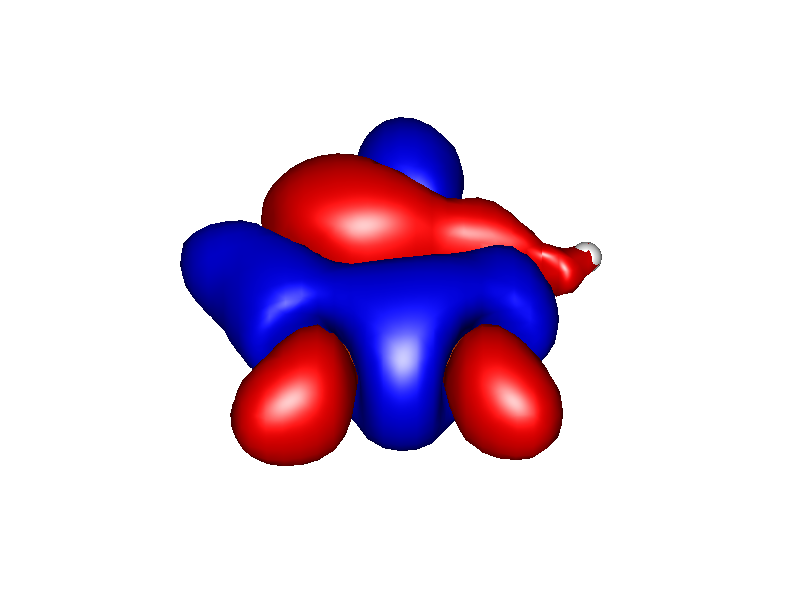}
  }
  \hfill
  \subfloat[$n_{\text{occup}}$ = 1.97]{%
    \includegraphics[width=0.22\textwidth]{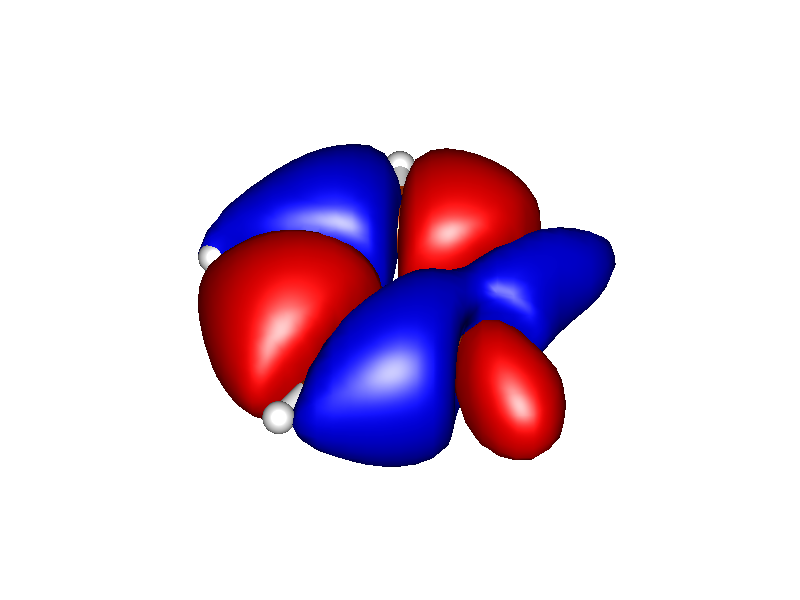}
  } \\
  \subfloat[$n_{\text{occup}}$ = 1.97]{%
    \includegraphics[width=0.22\textwidth]{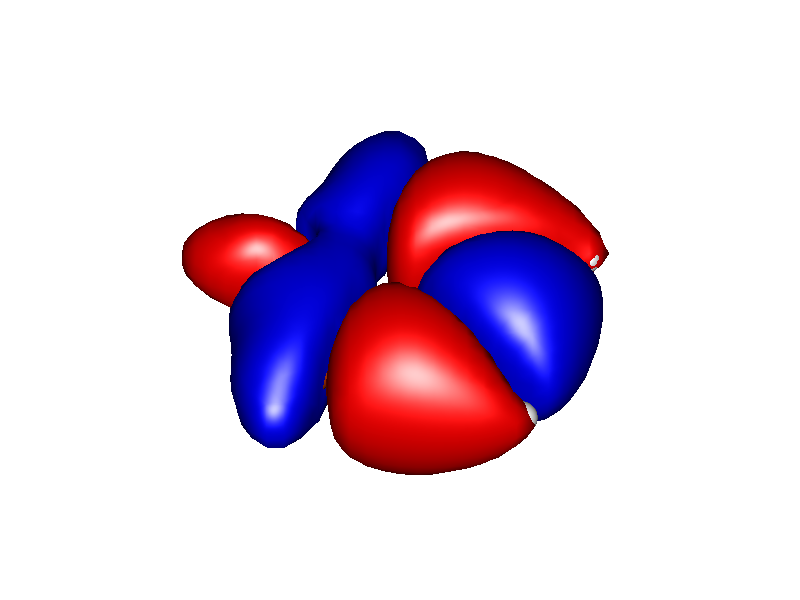}
  }
  \hfill
  \subfloat[$n_{\text{occup}}$ = 1.91]{%
    \includegraphics[width=0.22\textwidth]{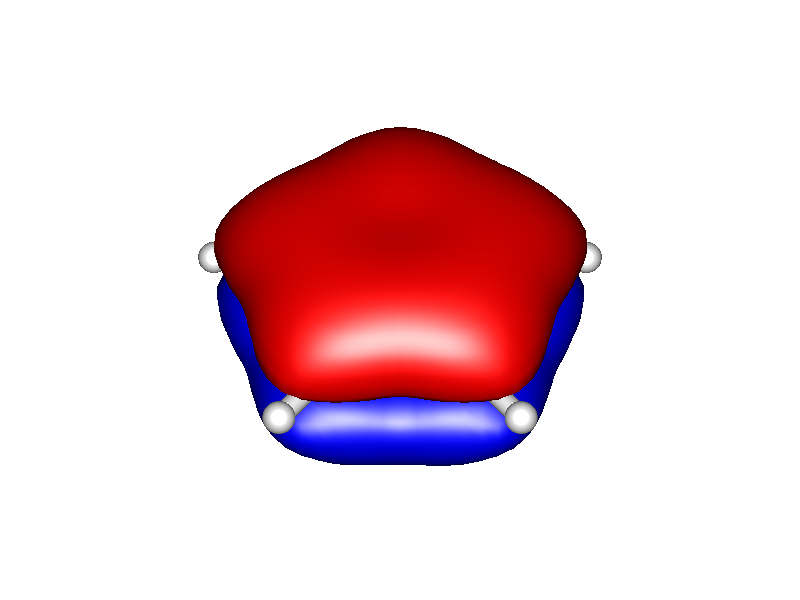}
  }
  \hfill
  \subfloat[$n_{\text{occup}}$ = 1.37]{%
    \includegraphics[width=0.22\textwidth]{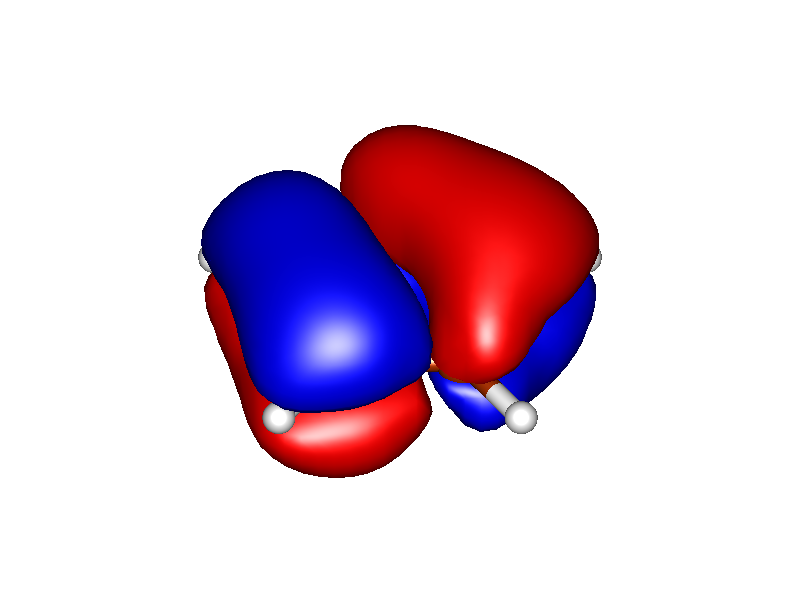}
  }
  \hfill
  \subfloat[$n_{\text{occup}}$ = 0.63]{%
    \includegraphics[width=0.22\textwidth]{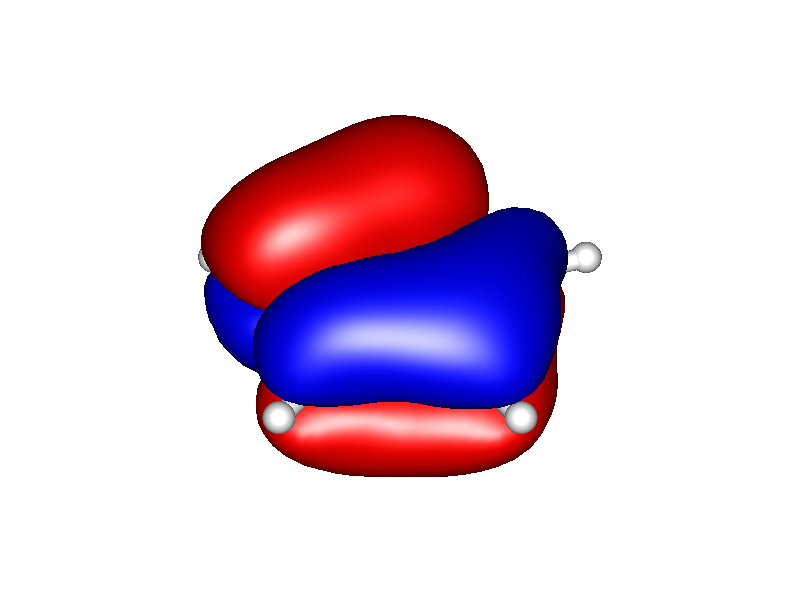}
  }
  \\
  \subfloat[$n_{\text{occup}}$ = 0.07]{%
    \includegraphics[width=0.22\textwidth]{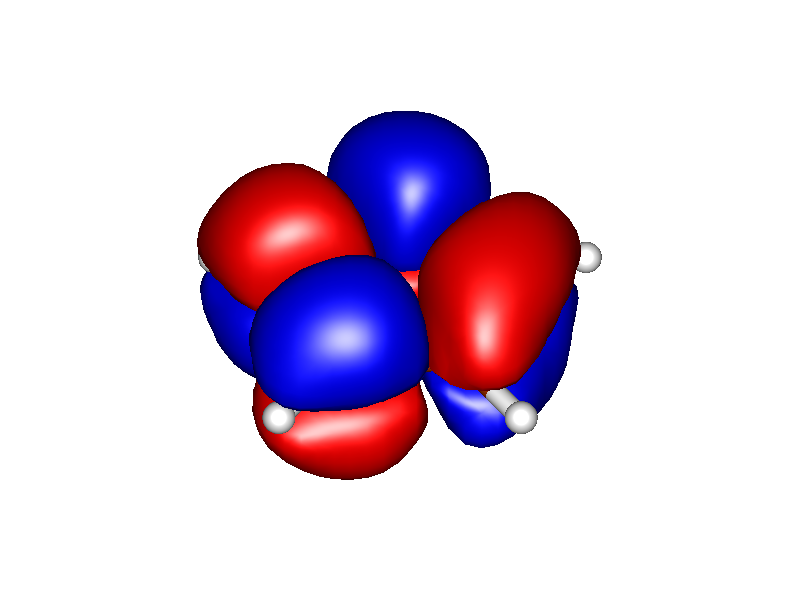}
  }
  \hfill
  \subfloat[$n_{\text{occup}}$ = 0.02]{%
    \includegraphics[width=0.22\textwidth]{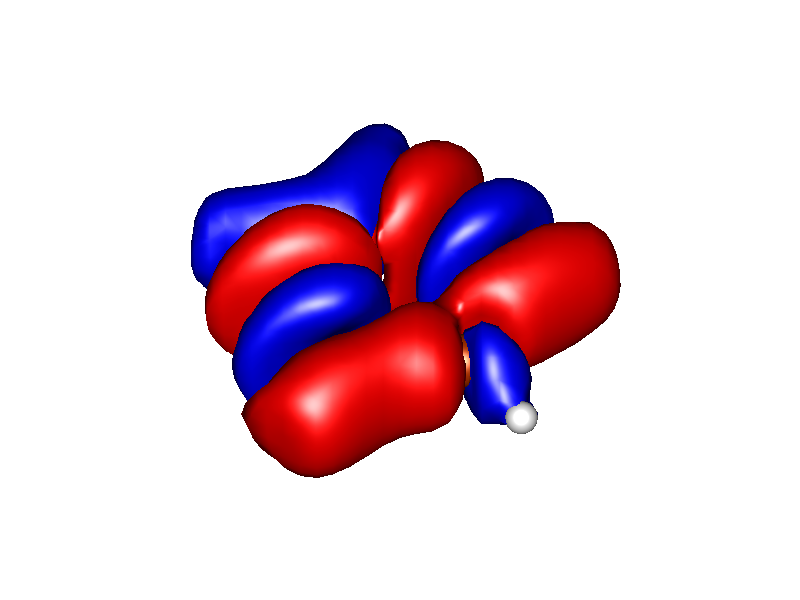}
  }
  \hfill
  \subfloat[$n_{\text{occup}}$ = 0.02]{%
    \includegraphics[width=0.22\textwidth]{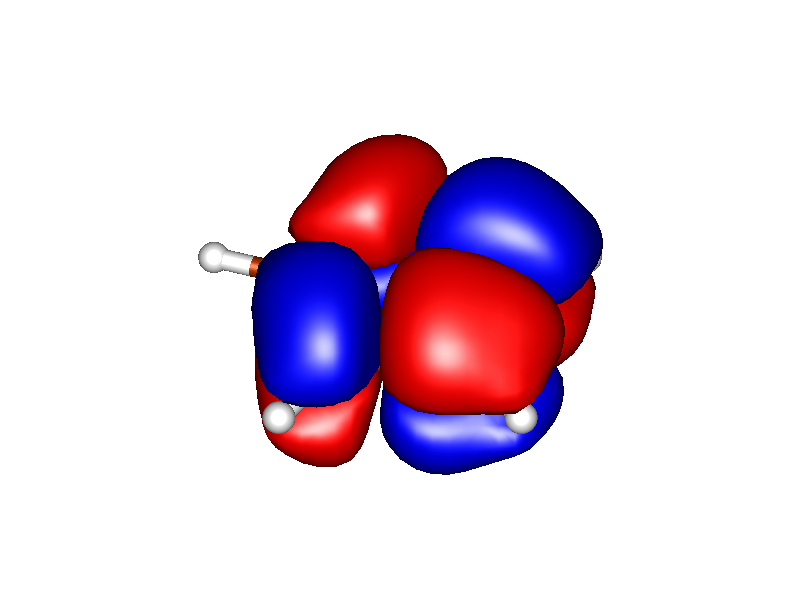}
  }
  \hfill
  \subfloat[$n_{\text{occup}}$ = 0.02]{%
    \includegraphics[width=0.22\textwidth]{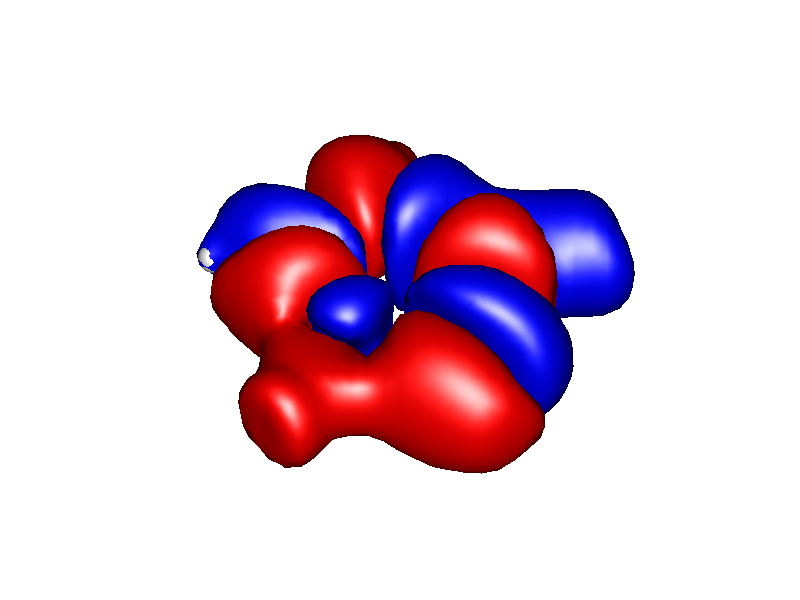}
  }
  \\
  \subfloat[$n_{\text{occup}}$ = 0.02]{%
    \includegraphics[width=0.22\textwidth]{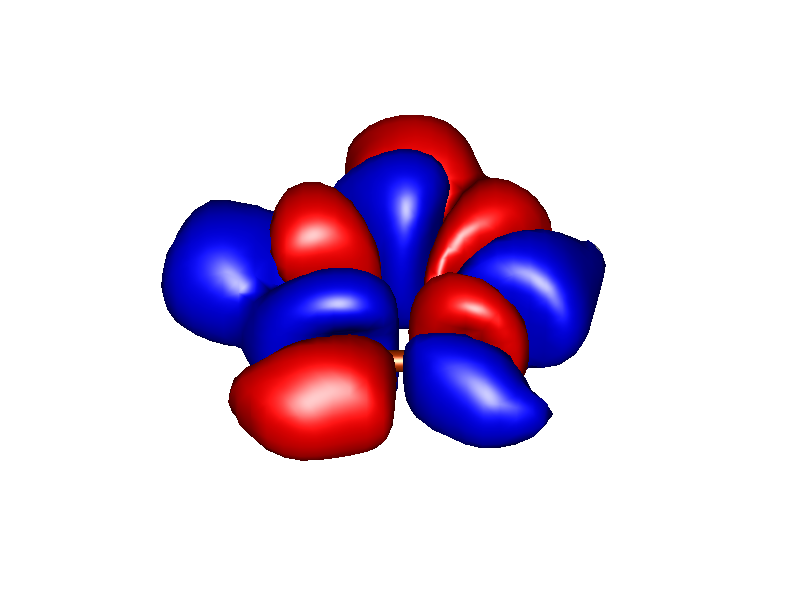}
  }
  \hfill
  \subfloat[$n_{\text{occup}}$ = 0.02]{%
    \includegraphics[width=0.22\textwidth]{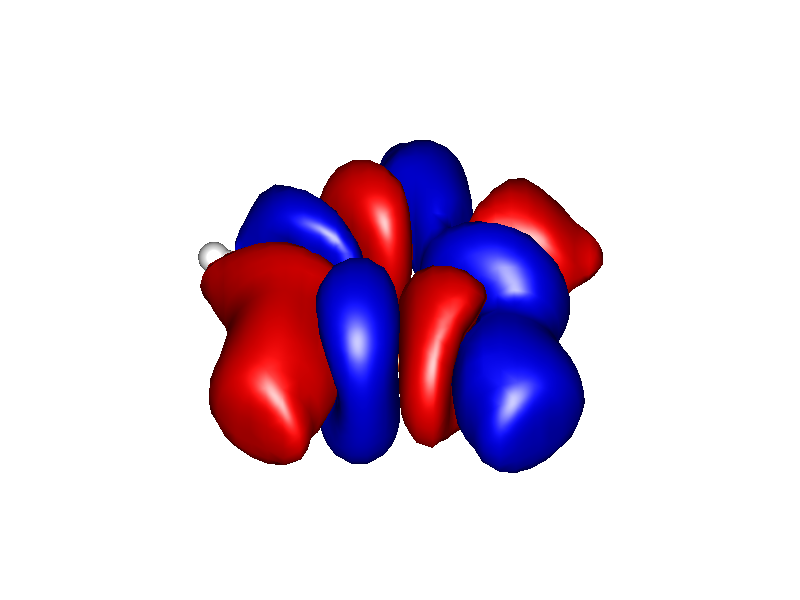}
  }
  \hfill
  \subfloat[$n_{\text{occup}}$ = 0.02]{%
    \includegraphics[width=0.22\textwidth]{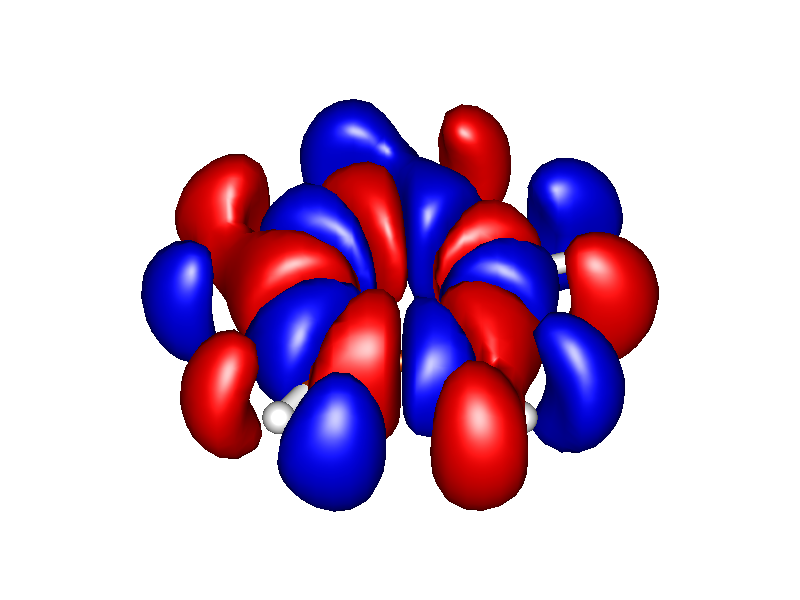}
  }
  \hfill
  \subfloat[$n_{\text{occup}}$ = 0.004]{%
    \includegraphics[width=0.22\textwidth]{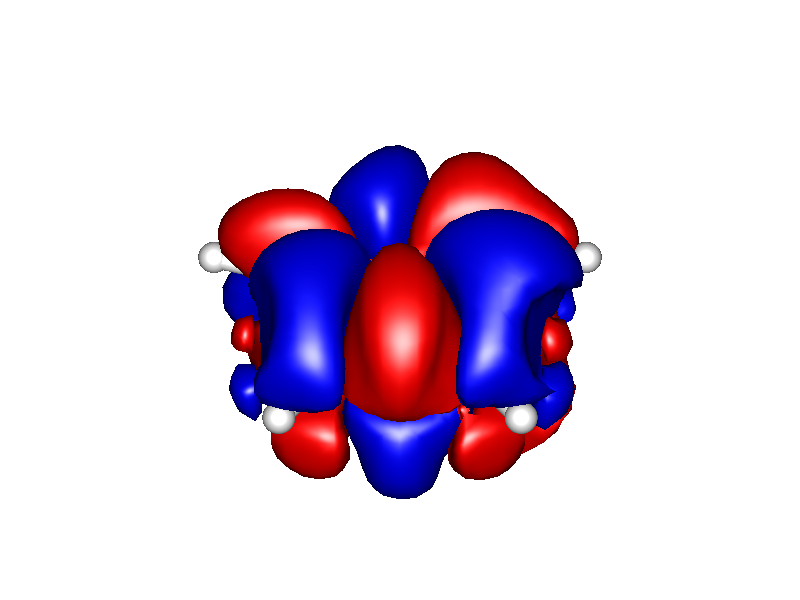}
  }
  \caption{C$_5$H$_5$, singlet state, CASSCF(14, 16) \label{orbs_c5h5_s}}
\end{figure}

\begin{figure}[!h]
  \subfloat[$n_{\text{occup}}$ = 1.99]{%
    \includegraphics[width=0.23\textwidth]{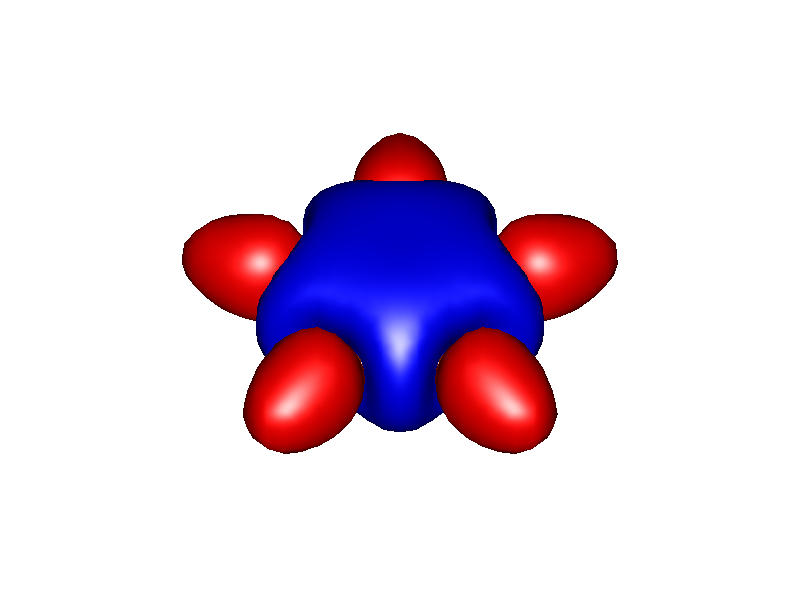}
  }
  \hfill
  \subfloat[$n_{\text{occup}}$ = 1.98]{%
    \includegraphics[width=0.23\textwidth]{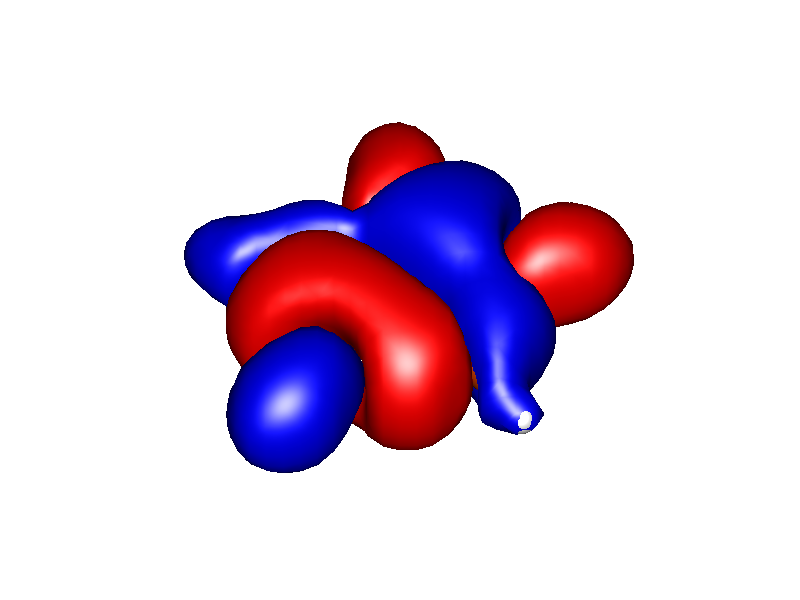}
  }
  \hfill
  \subfloat[$n_{\text{occup}}$ = 1.98]{%
    \includegraphics[width=0.23\textwidth]{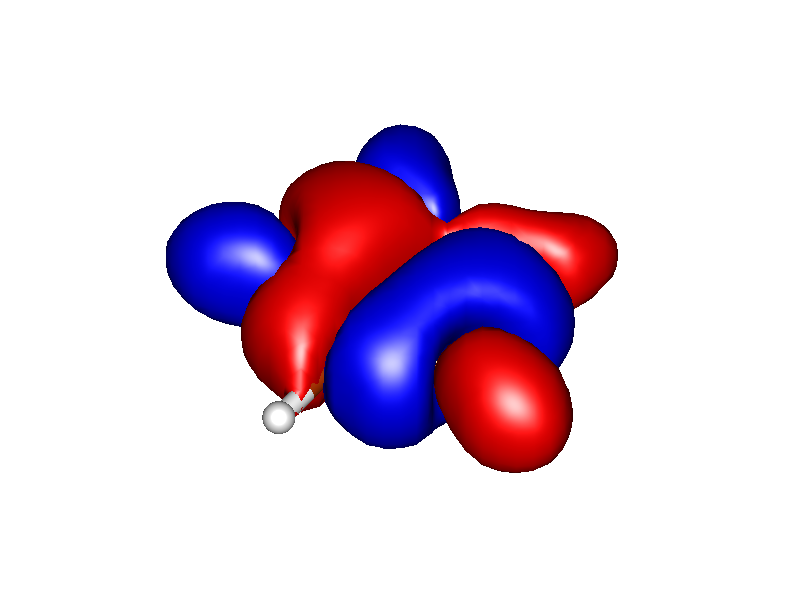}
  }
  \hfill
  \subfloat[$n_{\text{occup}}$ = 1.97]{%
    \includegraphics[width=0.23\textwidth]{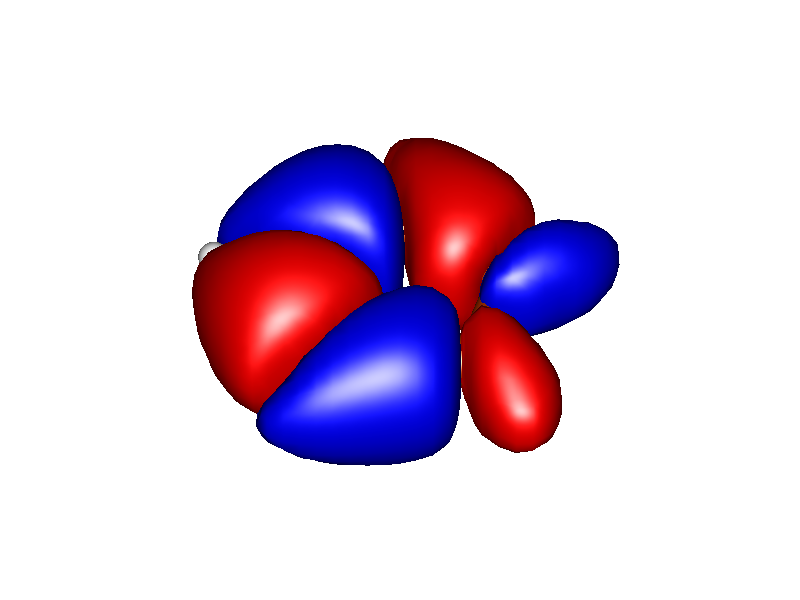}
  } \\
  \subfloat[$n_{\text{occup}}$ = 1.97]{%
    \includegraphics[width=0.23\textwidth]{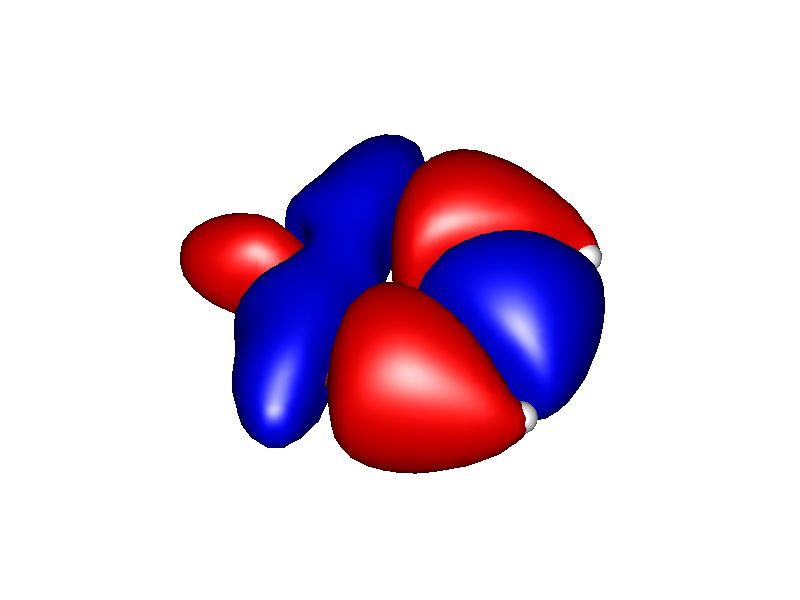}
  }
  \hfill
  \subfloat[$n_{\text{occup}}$ = 1.93]{%
    \includegraphics[width=0.23\textwidth]{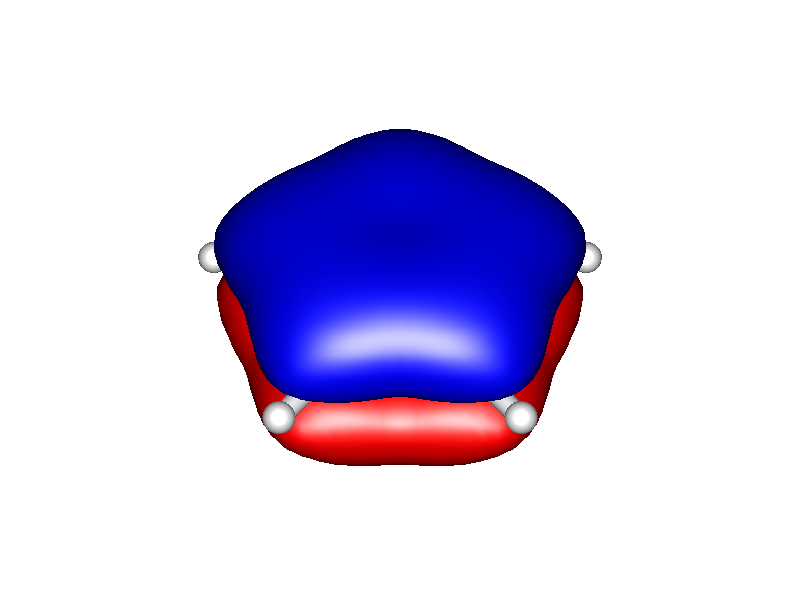}
  }
  \hfill
  \subfloat[$n_{\text{occup}}$ = 0.99]{%
    \includegraphics[width=0.23\textwidth]{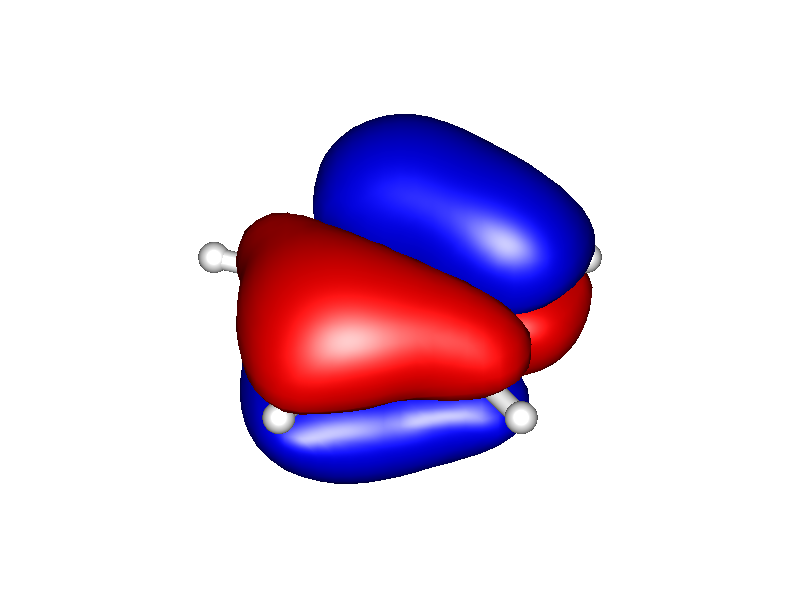}
  }
  \hfill
  \subfloat[$n_{\text{occup}}$ = 0.99]{%
    \includegraphics[width=0.23\textwidth]{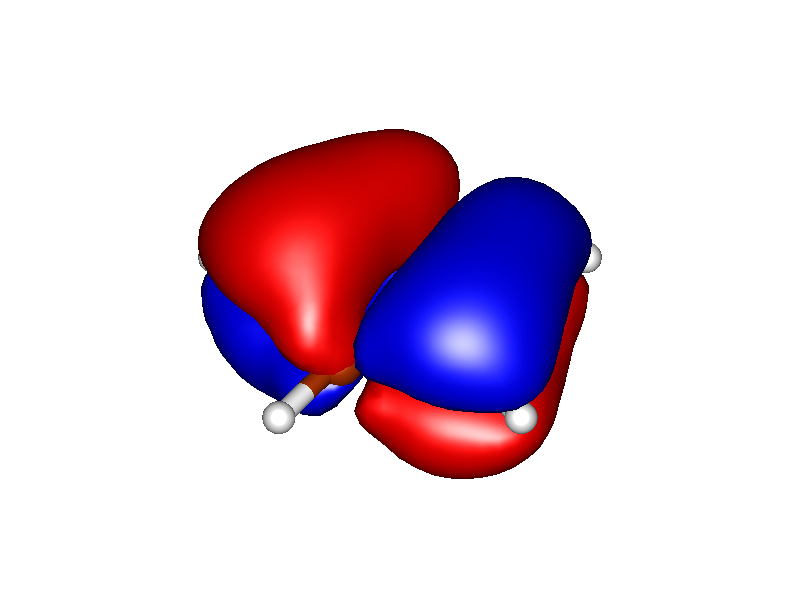}
  }
  \\
  \subfloat[$n_{\text{occup}}$ = 0.04]{%
    \includegraphics[width=0.23\textwidth]{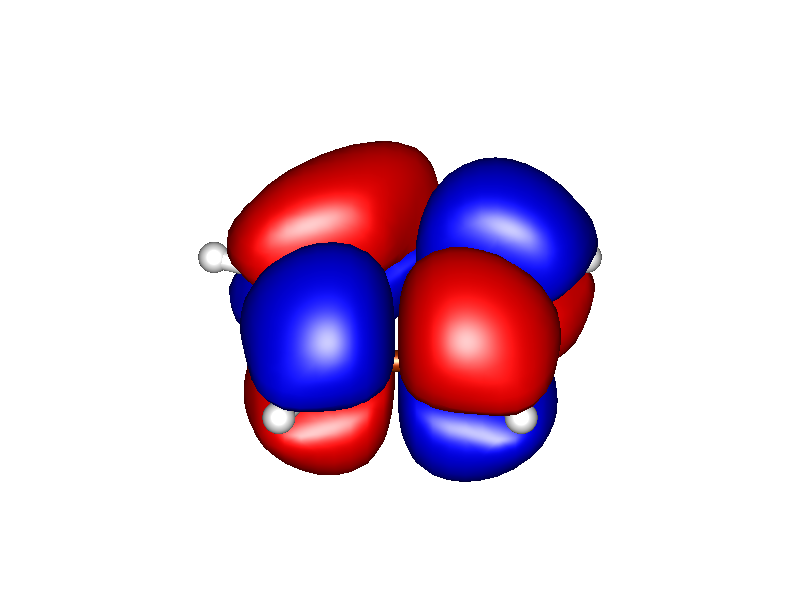}
  }
  \hfill
  \subfloat[$n_{\text{occup}}$ = 0.04]{%
    \includegraphics[width=0.23\textwidth]{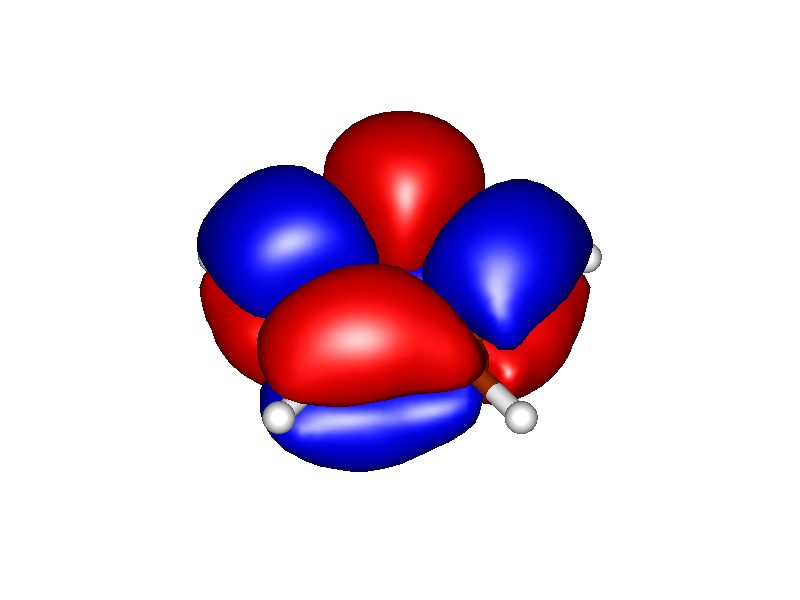}
  }
  \hfill
  \subfloat[$n_{\text{occup}}$ = 0.02]{%
    \includegraphics[width=0.23\textwidth]{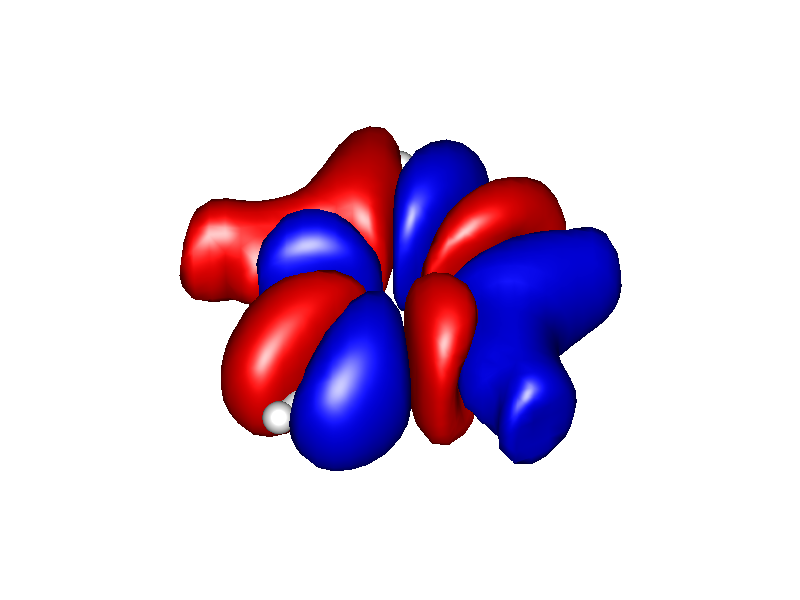}
  }
  \hfill
  \subfloat[$n_{\text{occup}}$ = 0.02]{%
    \includegraphics[width=0.23\textwidth]{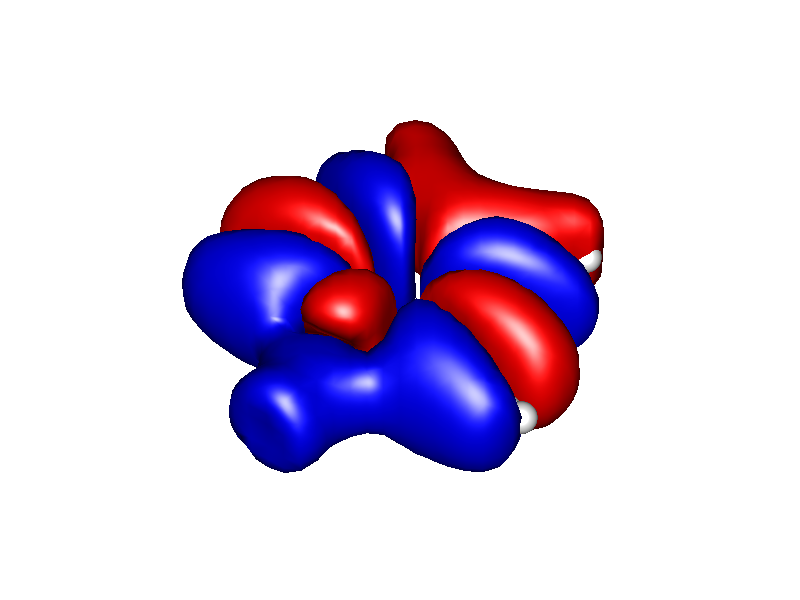}
  }
  \\
  \subfloat[$n_{\text{occup}}$ = 0.02]{%
    \includegraphics[width=0.23\textwidth]{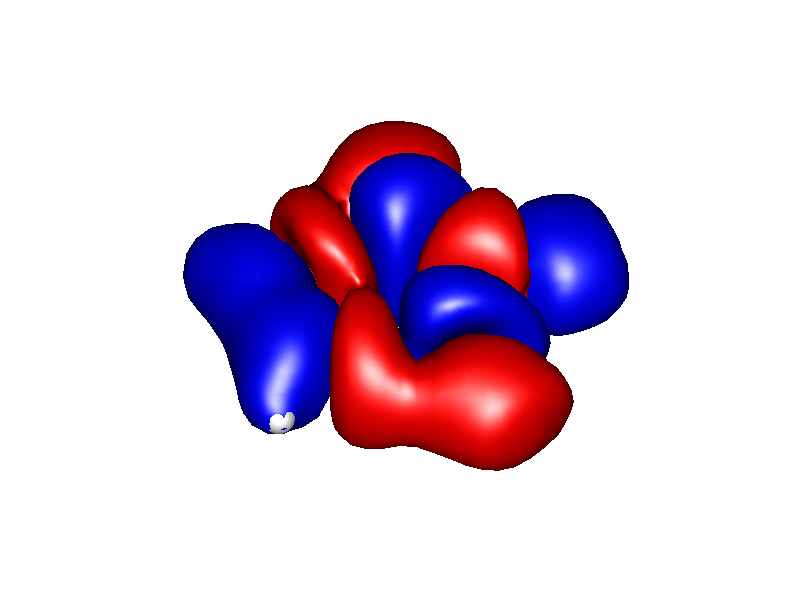}
  }
  \hfill
  \subfloat[$n_{\text{occup}}$ = 0.02]{%
    \includegraphics[width=0.23\textwidth]{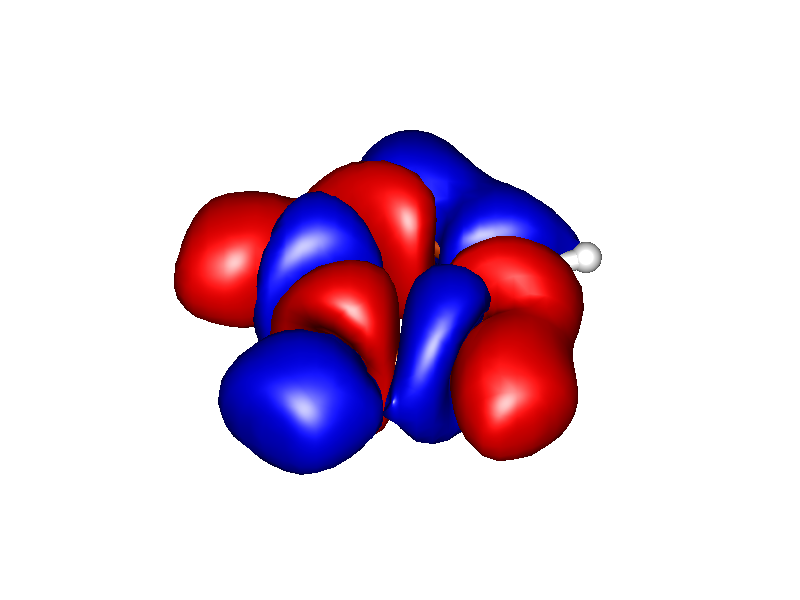}
  }
  \hfill
  \subfloat[$n_{\text{occup}}$ = 0.02]{%
    \includegraphics[width=0.23\textwidth]{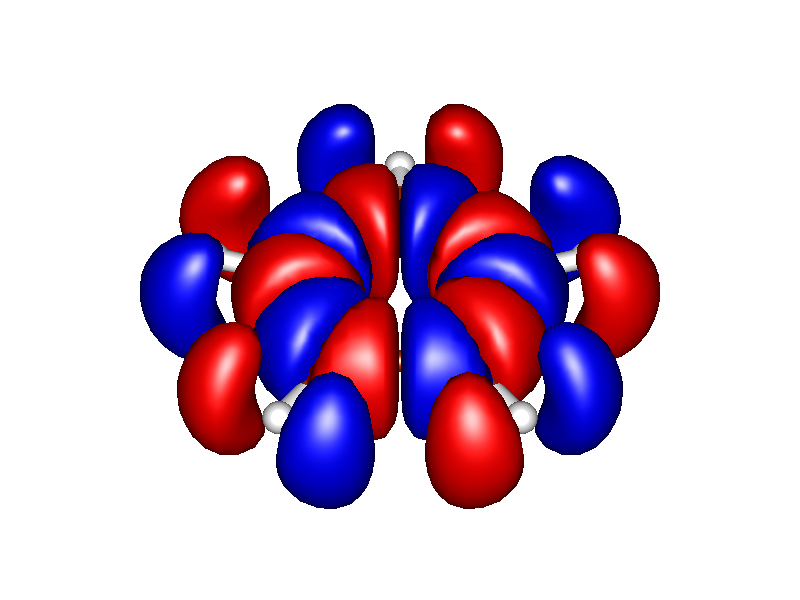}
  }
  \hfill
  \subfloat[$n_{\text{occup}}$ = 0.004]{%
    \includegraphics[width=0.23\textwidth]{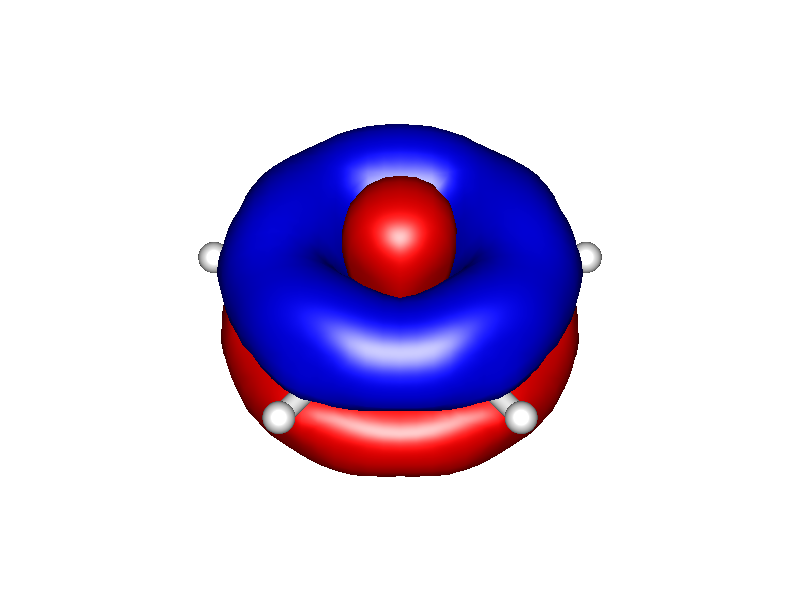}
  }
  \caption{C$_5$H$_5$, triplet state, CASSCF(14, 16) \label{orbs_c5h5_t}}
\end{figure}

The DMRG-SCF (C$_4$H$_4$, eCPO) natural orbitals with their occupation numbers are depicted in Figures \ref{orbs_c4h4_s} and \ref{orbs_c4h4_t}. CASSCF (C$_4$H$_4$, C$_5$H$_5^+$, mCPO) natural orbitals and the corresponding occupation numbers are depicted in Figures \ref{orbs_c4h4small_s}, \ref{orbs_c4h4small_t}, \ref{orbs_c5h5_s}, and \ref{orbs_c5h5_t}.

\renewcommand{\thesubfigure}{\arabic{subfigure}}

\renewcommand{\thesubfigure}{\arabic{subfigure}}

\renewcommand{\thesubfigure}{\arabic{subfigure}}

\clearpage

\bibliography{biblio_AC_CD}